\documentclass[english,11pt,oneside]{article}

\pdfoutput=1
\usepackage[T1]{fontenc}
\usepackage[latin9]{inputenc}
\usepackage[usenames,dvipsnames]{xcolor}
\usepackage{color}
\usepackage{babel}
\usepackage{setspace}
\usepackage{cancel}
\usepackage{afterpage}
\usepackage{float}
\usepackage{amstext}
\usepackage{amssymb}
\PassOptionsToPackage{normalem}{ulem}
\usepackage{ulem}
\usepackage[unicode=true,pdfusetitle,
 bookmarks=true,bookmarksnumbered=false,bookmarksopen=false,citecolor=Turquoise,
 breaklinks=false,pdfborder={0 0 1},backref=false,colorlinks=true,
linktoc=page]
 {hyperref}
\usepackage{graphicx,array}
\usepackage{mathtools}
\usepackage[small]{caption}
\usepackage{changepage}
\usepackage{slashed}
\usepackage{makecell}
\usepackage{multirow}
\usepackage{enumerate}
\usepackage{comment}
\usepackage{pdflscape}
\providecommand{\tabularnewline}{\\}
\newcolumntype{P}[1]{>{\centering\arraybackslash}m{#1}}

\definecolor{note_fontcolor}{rgb}{0.80078125, 0.80078125, 0.80078125}
\usepackage[top=3 cm, bottom=3 cm, left=3.1416 cm, right=3.1416 cm]{geometry}

\def\beq{\begin{equation}}
\def\eeq{\end{equation}}
\def\bea{\begin{eqnarray}}
\def\eea{\end{eqnarray}}
\def\no{\nonumber}
\newcommand{\nn}{\nonumber\\}
\newcommand{\al}{\alpha}
\newcommand{\vep}{\varepsilon}
\newcommand{\ep}{\epsilon}

\begin{document} 

\baselineskip=17pt

%%%%%%%%%%
%%%%%%%%%%    Title page
%%%%%%%%%%

\thispagestyle{empty}
\vspace{20pt}
\font\cmss=cmss10 \font\cmsss=cmss10 at 7pt

\begin{flushright}
UMD-PP-018-11 \\
\end{flushright}

\begin{center}
{\Large \textbf
{Natural Seesaw and Leptogenesis from Hybrid of High-Scale Type I and TeV-Scale Inverse }} \\
\end{center}

\vspace{15pt}

\begin{center}
{\large Kaustubh Agashe$\, ^{a}$, Peizhi Du$\, ^{a}$, Majid Ekhterachian$\, ^{a}$, Chee Sheng Fong$\, ^{b}$, Sungwoo Hong$\, ^{c}$, Luca Vecchi$\, ^{d}$ \\
\vspace{15pt}
$^{a}$\textit{Maryland Center for Fundamental Physics,
     Department of Physics,
     University of Maryland,
     College Park, MD 20742, U.~S.~A.} \\
   $^{b}$\textit{Centro de Ci\^{e}ncias Naturais e Humanas, 
   	Universidade Federal do ABC, Santo Andr\'{e}, 09210-580 SP, Brazil
} \\
     $^{c}$\textit{Department of Physics, LEPP, 
     	Cornell University, Ithaca NY 14853, U.~S.~A.    
   } \\
   $^{d}$\textit{Theoretical Particle Physics Laboratory, Institute of Physics, EPFL, Lausanne, Switzerland}
   
\vspace{0.3cm}
      
{\it email addresses}: kagashe@umd.edu, pdu@umd.edu, ekhtera@umd.edu, 
sheng.fong@ufabc.edu.br, sh768@cornell.edu, luca.vecchi@epfl.ch}

\end{center}

\vspace{5pt}

\begin{center}
\textbf{Abstract}
\end{center}
\vspace{5pt}

%%%%%%%%%%%%%%%%%%%
We develop an extension of the basic inverse seesaw model which addresses simultaneously two of its drawbacks, namely, the lack of explanation of the tiny Majorana mass term $\mu$ for the TeV-scale singlet fermions and the difficulty in achieving successful leptogenesis.
Firstly, we investigate systematically leptogenesis within the inverse (and the related linear) seesaw models and show that a successful scenario requires either small Yukawa couplings, implying loss of experimental signals, and/or quasi-degeneracy among singlets mass of different generations, suggesting extra structure must be invoked.
Then we move to the analysis of our new framework, which we refer to as \emph{hybrid seesaw}. This combines the TeV degrees of freedom of the inverse seesaw with those of a high-scale ($M_N\gg$ TeV) seesaw module in such a way as to retain the main features of both pictures: naturally small neutrino masses, successful leptogenesis, and accessible experimental signatures.  
We show how the required structure can arise from a more fundamental theory with a gauge symmetry or from warped extra dimensions/composite Higgs. 
We provide a detailed derivation of all the analytical formulae necessary to analyze leptogenesis in this new framework, and discuss the entire gamut of possibilities our scenario encompasses---including scenarios with singlet masses in the enlarged range $M_N \sim 10^6 - 10^{16}$ GeV.
The idea of hybrid seesaw was proposed by us in arXiv:1804.06847; here, we substantially elaborate upon and extend earlier results.

%%%%%%%%%%%%%%%%%%%

\vfill\eject
\noindent

%%%%%%%%%%%%%%%%%%%%%%%%%%%%%%%%%%%%%%%%%%%%%%%%%%%%%
%%%%%  Introduction
%%%%%%%%%%%%%%%%%%%%%%%%%%%%%%%%%%%%%%%%%%%%%%%%%%%%%

\newpage

{
	\hypersetup{linkcolor=black}
	\tableofcontents
}

\section{Introduction}

Viewing the Standard Model (SM) as an effective field theory, Majorana neutrino masses $m_\nu$ dominantly arise from the unique dimension-five Weinberg operator:
%%%
\bea
C \frac{y^2}{m_{\rm NP}}\ell H \ell H ~~~~~~\to~~~~~~ m_\nu = C \frac{y^2v^2}{m_{\rm NP}},
\label{eq:Weinberg_operator}
\eea
%%%
where $\ell$ and $H$ are respectively the SM lepton and Higgs doublets with vacuum expectation value (VEV) $v$. Within our conventions, the new degrees of freedom responsible for generating the operator in eq.~(\ref{eq:Weinberg_operator}) are assumed to be characterized by a mass scale $m_{\rm NP}$ and a leading coupling $y$ to the SM lepton (and the Higgs) (couplings among new states are measured by other couplings in general). We next elaborate on the ``$C$'' parameter.

The operator in eq.~(\ref{eq:Weinberg_operator}) violates $U(1)_{B-L}$ by two units. Such a violation may be induced directly from $y^2/m_{\rm NP}$, as in ordinary type I seesaw scenarios~\cite{original}.
In all those cases we conventionally say $U(1)_{B-L}$ breaking is {\emph{maximal}} and set $C \equiv 1$ to mean that no further parameter is necessary to generate neutrino masses. On the other hand, in all UV completions in which $y^2/m_{\rm NP}$ does not have spurious $U(1)_{B-L}$ charge $2$, eq.~(\ref{eq:Weinberg_operator}) will have to be proportional to some additional $U(1)_{B-L}$-breaking parameter $C$. In particular, when $m_{\rm NP}\ll10^{14}$ GeV and $y=\mathcal{O} (1)$ such $U(1)_{B-L}$-breaking parameter is forced to be very small, i.e. $C \ll 1$. Note that while in the former case, setting $C=1$ merely means effectively we did not need the $C$ parameter, in the latter case $C \ll 1$ encodes the required $U(1)_{B-L}$ breaking. 
The new parameter $C \ll 1$ in the second scenario could be a ratio of mass scales or couplings within the new sector, or simply be controlled by a new $U(1)_{B-L}$-breaking interaction to the SM. We will refer to these models as scenarios with {\emph{small}} $U(1)_{B-L}$ breaking.

UV completions of the Weinberg operator have other important physical implications. The necessary source of $U(1)_{B-L}$ breaking indicates for example that the UV dynamics responsible for generating eq.~(\ref{eq:Weinberg_operator}) may also have the possibility to realize baryogenesis through leptogenesis~\cite{Fukugita:1986hr}. Furthermore, the parameters $m_{\rm NP},y$ control the possible collider signatures of the new particles involved, suggesting that models with $m_{\rm NP}\sim$ TeV and $y\sim1$ certainly represent the most promising ones experimentally.

Combining these considerations, we find that small neutrino masses may be obtained in three qualitatively different ways, depending on whether TeV/$m_{\rm NP}$ or $y$ or $C$ is the small parameter suppressing $m_\nu$:
\begin{itemize}
\item[(I)] high-scale scenarios TeV$/m_{\rm NP}\ll1$ in which $C$, $y$ are not necessarily small, such as the popular high scale seesaw model~\cite{original};
\item[(II)] scenarios with small couplings $y\ll1$ and unsuppressed $C$ and TeV/$m_{\rm NP}$, like in low scale seasaw models;
\item[(III)] scenarios with {\emph{small}} $U(1)_{B-L}$ breaking, $C \ll1$, 
where $y$ and TeV/$m_{\rm NP}$ may be unsuppressed. The inverse seesaw~\cite{inverse} or linear seesaw~\cite{Malinsky:2005bi} belong to this latter class.
\end{itemize}

In table~\ref{summary_table}, we summarize how the most common realizations of the above three classes of UV completions of eq.~(\ref{eq:Weinberg_operator}) compare with respect to the generation of small neutrino masses, the realization of successful leptogenesis, and the possibility of featuring interesting signatures at colliders. Needless to say, this table reflects our own perspective on the topic, as well as our biases as model-builders.
For example, in the table and the remainder of the paper we will often use the term {\emph{natural}}. To help the reader appreciate this terminology we hereby attempt to provide an operative definition of this concept, which we may call \emph{Dirac naturalness}: dimensionless couplings (like $y$ or the new physics and SM self-couplings) are of natural size if they are not too far from one, say of ${O}(10^{-2}) - {O}(1)$; mass scales ($m_{\rm NP}$) are of natural size if they are either generated via dimensional transmutation of natural couplings or are related to a more fundamental dynamical scale (for e.g. the TeV, the GUT or the Planck scales) by factors of order unity; in the absence of a symmetry reason, the differences among masses and among couplings should be of the same order as the respective masses and couplings themselves (i.e. \emph{anarchic} masses and couplings). Our naturalness criteria is more restrictive than t' Hooft's technical naturalness, which only calls for stability under quantum corrections.

We can now proceed to explain table \ref{summary_table}. In \emph{\bf High scale type-I seesaw} models the new physics is in the form of heavy Majorana right-handed neutrinos $N$ with coupling $y\sim10^{-2}-1$ to the SM leptons. Once both a mass $m_{\rm NP}\equiv M_N$ for $N$ and the coupling $y$ are turned on, $U(1)_{B-L}$ is broken collectively by $y^2/M_N$ and hence the model belongs to class (I). In these cases leptogenesis is realized naturally~\cite{Fukugita:1986hr,Davidson:2008bu}. Unfortunately, with a high scale $M_N\sim10^{10}-10^{14}$ GeV there are no detectable LHC or low-energy experimental signals. Small neutrino masses can be obtained quite elegantly. However, the required mass $m_{\rm NP}$ must be a few orders of magnitude smaller than the known fundamental scales (say the Planck or GUT scale of $\sim 10^{18}$ or $10^{16}$ GeV, respectively), 
at an intermediate value that is not fully understood. In view of our definitions above, we may view such a scale as \emph{almost} natural. 
%

%%%%%%%%%%%%%%%%%%%%%%%%%%%%%%%%%%%%%%%
% TABLE
%%%%%%%%%%%%%%%%%%%%%%%%%%%%%%%%%%%%%%%
\begin{table}
\begin{center}
{

	%\begin{adjustwidth}{0cm}{}
		%\centering
		%\scalebox{1.0}{
			%
			\begin{tabular}{ | c || c | c | c | c|}
				\hline 
				UV seesaw model & Natural $m_\nu$? & Signals? & Leptogenesis? \\
				%&   &  &  \tabularnewline
				\hline
				\hline
				High scale type I & Almost & No & Yes 
				\tabularnewline
				\hline
				TeV scale type I & Not really & Possible & Possible 
				\tabularnewline
				\hline
				TeV scale inverse/linear & Not really & Yes & Possible 
				\tabularnewline
				\hline
				\hline
				Hybrid (this work and~\cite{short}) & Possible & Yes & Yes 
				\tabularnewline
				\hline

			\end{tabular}
		} 
		\end{center}
		\caption{Our comparison of various UV completions of eq.~(\ref{eq:Weinberg_operator}): see explanation in text.}
		\label{summary_table}
%	\end{adjustwidth}
	
\end{table}
%%%%%%%%%%%%%%%%%%%%%%%%%%%%%%%%%%%%%%%
%%%%%%%%%%%%%%%%%%%%%%%%%%%%%%%%%%%%%%%

In \emph{\bf TeV scale type-I seesaw} the $U(1)_{B-L}$ symmetry is again maximally broken ($C \equiv 1$). However, here the small neutrino masses ($m_\nu = { O} (0.1)$ eV) are obtained for $m_{\rm NP}\equiv M_N \sim $ TeV with tiny couplings to the SM, $y \sim 10^{-6}$.\footnote{Unless we invoke some special textures \cite{Dev:2013oxa}, which we do not consider here.} This model thus belongs to our class (II). The smallness of $y$ is considered a tuning and hence neutrino mass is not natural according to our definition. The small $y$ also makes the direct production of the exotic $N$ unlikely. To make this scenario more visible one may consider extensions with additional gauge symmetries ($B-L$ or LR models) so that $N$ may be produced via the associated {\em gauge} couplings, which can be sizable, giving collider signals with {\em same}-sign dileptons due to Majorana nature of $N$ (see references in 
 \cite{Mohapatra:2016twe}). Overall, this model scores a ``possible'' in the experimental signals entry. Finally, leptogenesis is not natural unless one imposes quasi-degeneracy among singlets of different generations to resonantly enhance the CP violation~\cite{Pilaftsis:2009pk}. This cannot be achieved without additional ingredients for example in the form of flavor symmetry. Hence, leptogenesis scores a ``possible'' here as well.

\emph{\bf TeV scale inverse seesaw (ISS)~\cite{inverse} and linear seesaw (LSS)~\cite{Malinsky:2005bi}} have, besides the right-handed neutrinos (here denoted by $\Psi$) which couple to the SM lepton and Higgs with Yukawa $y$, additional fermion singlets with left-handed chirality $\Psi^c$. The latter are introduced such that the singlets can form a Dirac mass term $m_{\rm NP}\equiv m_\Psi \sim$ TeV preserving $U(1)_{B-L}$. Such a symmetry is broken by a small Majorana mass term for the singlets $\mu$ in the ISS, or by a small lepton number breaking Yukawa coupling $y'$ of $\Psi^c$ to the SM in the LSS. In the language introduced in eq.~(\ref{eq:Weinberg_operator}) this means 
%$$
\bea
C \sim\frac{\mu}{m_\Psi}\ll1~~~~~~~~~~~~C \sim \frac{y'}{y}\ll1
\eea
%$$ 
respectively, and these models belong to class (III). Since $\mu \ll $ TeV is not set by any fundamental scale, and similarly the coupling $y'$ must be very small, the observed neutrino mass is not obtained naturally according to our criteria. Yet, an attractive feature here is that experimental signals from singlets $\Psi, \Psi^c$ can arise from sizable Yukawa coupling $y$ --- including a contribution to the rare process $\mu \to e \gamma$ as well as direct production of singlets themselves (see for example \cite{Deppisch:2015qwa} and references therein). The model therefore scores a clear ``yes'' in the signal column.\footnote{Even though we get accessibility to the singlets, it is true that it is difficult to directly probe the very small $\mu$-term, i.e. lepton-number breaking. % cf.~same-sign dilepton signals which provide a ``smoking gun'' of this feature (albeit through new gauge couplings) in TeV scale type I seesaw. 
}

Unfortunately, as we have shown recently in ref.~\cite{short} (focusing on the case of strong washout) leptogenesis is not achieved naturally in the ISS (i.e. $\mu \neq 0$ and $y'= 0$),\footnote{Some of these results have been obtained by others (for example, in ref.~\cite{Deppisch:2010fr} and more recently, in ref.~\cite{Dolan:2018qpy}).} according to our naturalness criteria. {We will elaborate more on this in section \ref{inverse_lepto}. This includes the effects of quasi-degenerate mass among different generation singlets: while such possibility provides a sizable improvement of the final asymmetry, it can barely accommodate the observed value.  We also discuss the possibility of weak washout and demonstrate various subtleties, which have not been discussed in the previous literature---though it does not change qualitatively our earlier conclusions. We further extend our conclusions to LSS (i.e. $\mu = 0$ and $y' \neq 0$) and show that leptogenesis is not natural there either.} Another result of the present paper is that even turning on both $\mu, y'$ still requires very small couplings $y'\ll y \ll {\cal O}(10^{-2})$ to achieve a successful leptogenesis (see also for example refs.~\cite{Gu:2010xc,Dolan:2018qpy}). Our conclusion is that in this scenario leptogenesis scores a ``possible''. 

\begin{figure}[t]
\centering
\includegraphics[width=75mm]{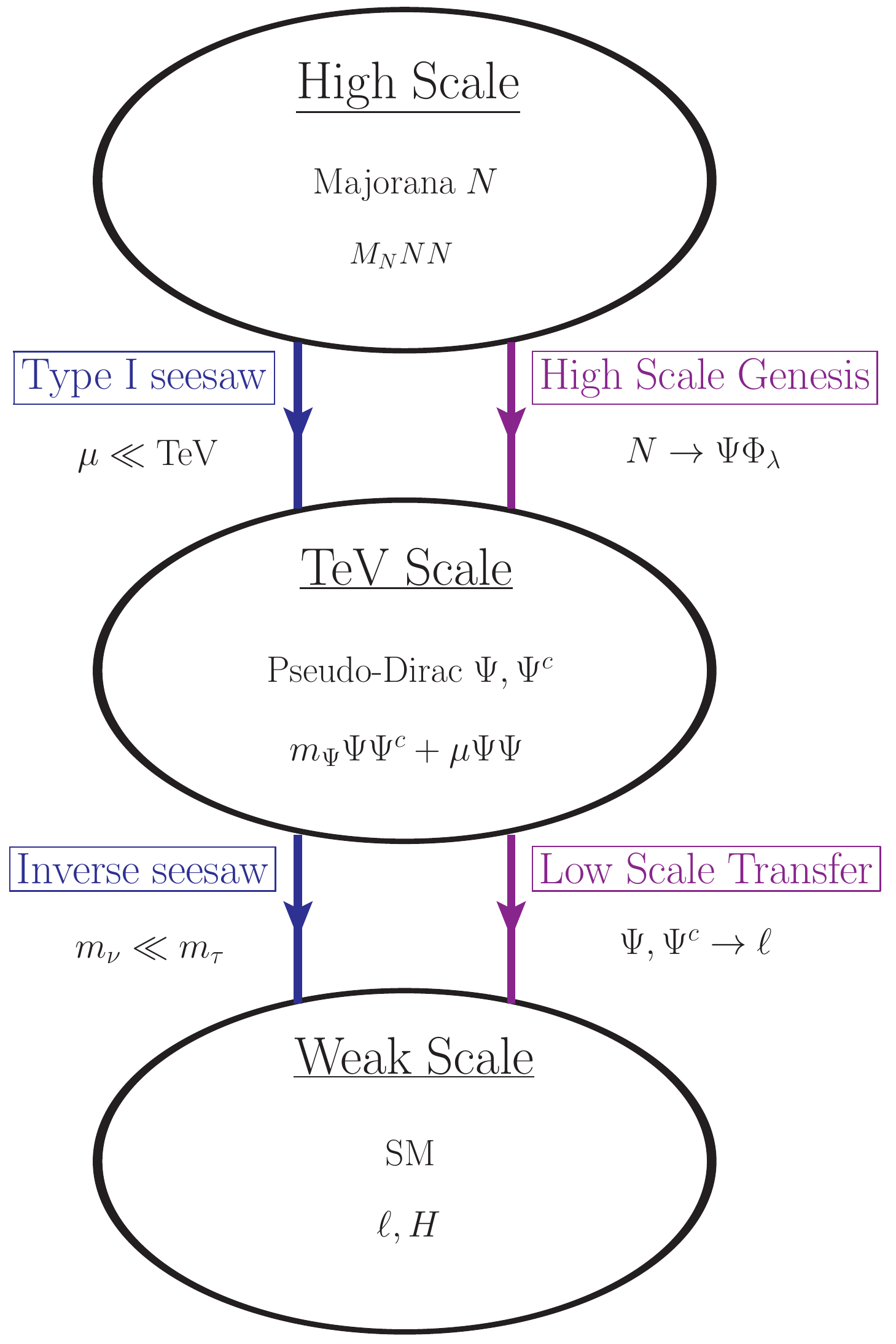} 
\caption{Schematic representation of physics (seesaw and genesis) of the hybrid model.}
\label{fig:scheme}
\end{figure}

Finally, table~\ref{summary_table} includes the \emph{\bf Hybrid seesaw} first presented in ref.~\cite{short}. This was designed to overcome \emph{simultaneously} the two limitations of the ISS: unnaturally small $\mu$ term and difficulty in leptogenesis. The essential idea of hybrid seesaw is to introduce, on top of the ISS module, namely a Dirac pair $\Psi,\Psi^c$ of fermions with $m_{\rm NP}\equiv m_\Psi\sim$ TeV and unsuppressed coupling to leptons $y\sim1$, a high scale type I seesaw module, namely heavy ($M_N\gg$ TeV) {\em Majorana} singlets $N$, see figure~\ref{fig:scheme}. The theory has no bare $\mu$-term, but the two modules suitably mix via a IR-scale mass term $m_{\rm IR}$ arising from a scalar vacuum expectation value. In this manner, integrating out the heavy Majorana singlet generates an effective $\mu\ll$ TeV for $\Psi$\footnote{The basic idea of this model is along the lines of ref.~\cite{Aoki:2015owa}, but those authors considered $M_N\sim$ TeV instead. For this reason Leptogenesis is not as successful as in our picture, and $\mu$ is not naturally small according to our criteria.}   
%%%
\bea
\mu \sim \frac{m_{\rm IR}^2}{M_N}.
\label{eq:eff_mu}
\eea
%%%
Taking now $m_{\rm IR}\sim$ TeV we can explain why $\mu/m_\Psi\ll1$, and therefore the smallness of neutrino masses.

%%%%%%%%%%%%%%%%%%%%%%%%%%%%%%%%%

%%%%%%%%%%%%%%%%%%%%%%%%%%%%%%%%%

The structure of hybrid seesaw, and in particular the characteristic mixing between the low and high scales modules, arises elegantly from warped extra-dimensions (dual to composite Higgs models)~\cite{Huber:2003sf} as shown by some of us previously~\cite{Agashe:2015izu}. In this sense, the hybrid seesaw could be taken as a ``toy'' version of the warped/composite one. Alternatively, the peculiar coupling structure in figure~\ref{fig:scheme} can be enforced in weakly-coupled 4D models via a gauge symmetry as we will see in appendix~\ref{app:gauge_model}. Because in the 5D completions $m_\Psi\sim m_{\rm IR}$ are both related to the same fundamental scale (the TeV), that arises dynamically, and simultaneously $M_N$ can be effectively reduced dynamically compared to the Planck scale~\cite{Agashe:2015izu}, then it is clear that neutrino masses can be fully natural in the hybrid picture {\emph{once UV-completed}} (i.e., strictly speaking, going beyond the hybrid model on which we will focus here). This explains the score ``possible'' in the appropriate entry in the table (i.e., why it's not quite an actual ``yes'').

How about leptogenesis in the hybrid seesaw model? We have just seen that neutrino masses are suppressed by $C \ll1$, suggesting that this is a scenario with small $U(1)_{B-L}$-breaking (class (III) above). However, this is not the complete story. As we will see in detail below (see also~\cite{short}), the high scale module violates {\emph{maximally}} a global $U(1)$ carried by $\Psi$. Because $U(1)$ number violation is large at scales $\sim M_N\gg m_\Psi$, leptogenesis can \emph{naturally} proceed through the decay of $N$ to $\Psi$ (analogously to type I seesaw in class (I)), followed by the asymmetry in $\Psi$ being transferred to the SM leptons. Hence, the hybrid model also turn out to score a ``yes'' in natural leptogenesis.

In particular, in ref.~\cite{short} we emphasized that high scale leptogenesis with anarchic couplings can be realized for $M_N \sim 10^{11} - 10^{16}$ GeV. In this paper, we will study this scenario in more detail and also explore the lower scale $M_N \gtrsim 10^6$ GeV where leptogenesis can be realized albeit with hierarchical Yukawa couplings (among different generations of $N$). Such a relaxation of the lower bound on the heavy singlet mass, compared to the ordinary type I seesaw, might be especially relevant for resolving the SUSY gravitino problem. Overall, due to the hybrid structure, the allowed mass window gets enlarged compared to the usual case, $10^9 - 10^{15}$ GeV.

Regarding possible experimental signals in the hybrid seesaw, besides signals associated with TeV scale fermions as in the conventional ISS (as mentioned above), the model generally predicts new TeV scale scalars potentially within the reach of present and future colliders as we have shown in ref.~\cite{short}. Certain realizations, like the gauge model presented in appendix~\ref{app:gauge_model}, also contain light states that may contribute to $\Delta N_{\rm eff}$ and might thus be probed by CMB-Stage-IV \cite{Abazajian:2013oma}.
Hence, we put a ``yes'' in the experimental signals. Remarkably, this model has the ability to realize the most attractive features of the high and low scale modules simultaneously.

\bigskip

Our paper is organized as follows. We begin in the next section with an overview of scenario with small lepton-number breaking ($C\ll1$), i.e. the ISS and LSS models, and discuss the constraint from the non-observation of $\mu \to e \gamma$. A thorough analysis of leptogenesis in these models is given in section \ref{inverse_lepto} (see also appendices \ref{app:sec3} and \ref{app:BE_ISS_LSS}).
Section \ref{sec:4_extension_big_picture} outlines our hybrid seesaw solution of the problems of the original ISS model. Explicit UV completions of the scenario are presented in appendices~\ref{app:gauge_model} (gauge model) and \ref{app:warped_seesaw} (warped/composite model). This is followed by a detailed discussion of leptogenesis in the hybrid model. In section \ref{sec:Formalism} (and appendices \ref{app:approximate_solutions} and \ref{app:Spectator-effects}) we will provide a \emph{systematic} derivation of the necessary analytic formalism, which we believe clarifies many of underlying physics. 
This formalism is then used in section \ref{results} to identify what parameter choices give the right baryon asymmetry, including some interesting benchmark points. 
We finally conclude in section \ref{conclude}, providing some directions for future work.

%%%%%%%%%%%%%%%%%
\section{Scenarios with small $U(1)_{B-L}$ breaking}
\label{review}

We begin with a review of what we will refer to as {\emph{small $U(1)_{B-L}$ breaking models}}, that according to our earlier definition have $C \ll1$ [see eq.~(\ref{eq:Weinberg_operator})]. These are characterized by an effective theory with exotic particles not far from the TeV scale and unsuppressed couplings to the SM, say of order $0.1 - 1$. This guarantees that these scenarios have testable consequences at colliders. Because all new degrees of freedom are heavy, the SM neutrinos are Majorana particles. To ensure that small neutrino masses are generated, these scenarios must possess an approximate lepton number broken by a small dimensionless parameter. The most minimal incarnations of this scenario has been called inverse seesaw and linear seesaw. We will focus on these mostly for simplicity sake. 

Let us add to the SM two Weyl fermions $\Psi$ and $\Psi^c$, singlet under the SM, carrying lepton number $L(\Psi)=+1$, $L(\Psi^c)=-1$ respectively. In principle {we can combine the pair of Weyl fermions into a Dirac fermion} with $\Psi$ (or $i\sigma^2{\Psi^c}^*$) playing the role of the left (right) chiralities, but we will not do it here for later convenience. The only $U(1)_{B-L}$ invariant couplings, besides the kinetic terms, are:
\bea\label{inverse1}
{\cal L}_{B-L}&\supset& m_{\Psi}\Psi \Psi^c + y \Psi^c H\ell + {  O}(1/\Lambda) + {\rm h.c.},
\eea
with $\ell,H$ the SM lepton and Higgs doublets, respectively. 
Gauge contractions are understood, and the flavor indices for $\ell$ 
(possibly carried by $\Psi,\Psi^c$ as well) are not displayed here for brevity. 
{We will include the flavor indices in later parts whenever they are relevant.} 
We will take $m_\Psi={  O}({\rm TeV})$ as a reference value. 
Possible higher dimensional operators (denoted by ${  O}(1/\Lambda)$ in eq.~(\ref{inverse1})) are assumed to be negligible because they are suppressed by a large scale $\Lambda$. We will assume $\Lambda$ is of the order of the Planck scale for definiteness.

In the theory eq.~(\ref{inverse1}) the active neutrinos remain exactly massless. In order to obtain a realistic theory with tiny neutrino masses without adding additional light degrees of freedom, we introduce small sources of $U(1)_{B-L}$ breaking.  At the renormalizable level there exist only three $(B-L)$-breaking couplings:\footnote{One may also add $Z\Psi^\dagger i\bar\sigma^\mu\partial_\mu\Psi^c+{\rm hc}$. However, after a field redefinition one realizes this is equivalent to a correction to the couplings we show.}
\bea\label{inverse2}
{\cal L}_{\cancel{ B-L}}&=& \frac{\mu}{2}\Psi\Psi+\frac{\mu'}{2}\Psi^c\Psi^c+y'\Psi H\ell+{ O}(1/\Lambda)+{\rm h.c.}.
\eea
The assumption that the $U(1)_{B-L}$ breaking terms are small reads $|\mu|,|\mu'|\ll |m_\Psi|$, $|y'|\ll|y|$. The terms $\mu,\mu'$ correspond to small Majorana masses for the fields $\Psi, {\Psi^c}$. Conventionally, the ISS model is defined by $y' = 0$ while the LSS model by $\mu = 0$. Generally, in both of these models, $\mu'$ is taken to be zero as well.

The new couplings appearing in eq.~(\ref{inverse2}) can all be assigned a spurionic lepton number, namely $L(\mu)=L(y')=-2$ and $L(\mu')=+2$. Because the accidental charges of $\mu, \mu'^*, y'$ are the same, in generic UV completions the new couplings in eq.~(\ref{inverse2}) may in fact arise from a unique fundamental coupling with $L=-2$. In that case, a natural consequence of naive dimensional analysis is that, at the order of magnitude level,
\bea\label{scaling}
\frac{y'}{y}\sim\frac{\mu}{m_\Psi}\sim\frac{\mu'^*}{m_\Psi}.
\eea
Of course it is possible to build a UV dynamics in such a way that this relation is violated. Yet, the scaling in eq.~(\ref{scaling}) is what one expects to emerge from truly generic UV theories. More generally, setting one of the couplings in eq.~(\ref{inverse2}) {to zero} is not always a radiatively stable assumption. For example, inspecting 1-loop diagrams we find that starting with a non-vanishing $y'$ one generates (from log-divergent piece)
\bea\label{loops}
y'\neq0~~~~\Longrightarrow~~~~\delta\mu\sim m_\Psi\frac{y^*y'^t}{16\pi^2},~~~~\delta\mu'\sim m_\Psi^t\frac{y'^*y^t}{16\pi^2}.
\eea
On the other hand, no renormalization effects are induced by $\mu,\mu'$ because these correspond to a soft-breaking of $U(1)_{B-L}$. That is, $\mu$ and $\mu'$ only self-renormalize and do not radiatively generate other terms.

Majorana masses $m_\nu$ for the active neutrinos, that have $L(m_\nu)=-2$, must be linear in the couplings of eq.~(\ref{inverse2}) to leading order in the small $(B-L)$-breaking. This can be readily verified by integrating out $\Psi,\Psi^c$ at tree-level to obtain, in the leading approximation:
\bea\label{MajoranaNu}
{\cal L}_{\rm EFT}&=&\frac{1}{2}(H\ell)^t\frac{m_\nu}{v^2}(H\ell)+{\rm h.c.}+{  O}(1/\Lambda).
\eea
where $v=174$ GeV and
\bea\label{neutrino_mass}
m_\nu&=&v^2\left[y^t\frac{1}{m_\Psi}\mu\frac{1}{m^t_\Psi}y-\left(y'^t\frac{1}{m^t_\Psi}y+y^t\frac{1}{m_\Psi}y'\right)\right].
\eea
Note that $\mu'$ does not enter because its $U(1)_{B-L}$ charge forces it to appear in front of $(H\ell)^t(H\ell)$ as complex conjugate, which is not possible at tree-level. With the relation eq.~(\ref{scaling}) the two contributions in eq.~(\ref{MajoranaNu}) are naturally of the same order. The parameter introduced in eq.~(\ref{eq:Weinberg_operator}) may now be identified as
\bea
C \equiv{\rm max}\left(\frac{y'}{y},\frac{\mu}{m_\Psi}\right).
\eea

Lacking a UV description of $U(1)_{B-L}$ breaking, it is fair to say that the smallness of $\mu,y'$ is merely an assumption in our effective field theory eqs.~(\ref{inverse1}) and (\ref{inverse2}). While this model does not truly explain the size of the SM neutrino masses, it provides an interesting laboratory to investigate the phenomenology of scenarios with small $U(1)_{B-L}$ breaking. A distinctive feature of these models is the presence of signatures in colliders (see for example \cite{Deppisch:2015qwa} and references therein). For $m_\Psi\lesssim1$ TeV and sizable $y$ it is in fact possible to produce the pseudo-Dirac fermions at the LHC via mixing with SM neutrinos and observe its subsequent resonant decay. Unfortunately, we will not be able to measure the tiny couplings $\propto\mu,\mu',y'$ and hence unambiguously connect the exotic particles to a mechanism for neutrino mass generation. The reason is that in the typical benchmark models from eq.~(\ref{MajoranaNu}) one derives from eq.~(\ref{neutrino_mass}) that $\mu/m_\Psi,y'/y\sim10^{-10}$, that is certainly out of reach of current and future colliders. 

Besides direct production of $\Psi^c$, there can be indirect signatures in rare processes, like $\mu\to e\gamma$ and the electron EDM. At leading order in $(yv)^2/m_\Psi^2$, the branching ratio can be written as \cite{Cheng:1980tp}
\bea\label{eq:mutoe}
\textrm{BR}_{\rm ISS}(\mu \to e \gamma)\simeq \frac{3\alpha_{\rm em}}{8\pi}
\left|\left(y^t\frac{v^2}{m^\dagger_\Psi m_\Psi}y^*\right)_{\mu e}\right|^2,
\eea
where $\alpha_{\rm em}\approx 1/137$ is the fine structure constant, $v\approx174$ GeV the SM Higgs VEV, 
and we have neglected corrections of order $m_W^2/m_\Psi^2$. The current experimental bound is $\textrm{BR}(\mu \to e \gamma)< 4\times 10^{-13}$ \cite{TheMEG:2016wtm}. For anarchic couplings and masses this translates into $y /m_{\Psi} \lesssim 2.7 \times 10^{-2} / {\rm TeV}$. However, the bound can be significantly relaxed by using flavor symmetries. One very efficient way to achieve this is to assume that the Lagrangian eq.~(\ref{inverse1}) has a global $U(1)_e\times U(1)_\mu\times U(1)_\tau$ symmetry under which the three generations of $\ell, e, \Psi, \Psi^c$ transform~\cite{Frigerio:2018uwx}.\footnote{One may use gauge symmetries to enforce this possibility.} This assumption forces $y, m_\Psi$, as well as the SM lepton Yukawa coupling, to be diagonal in flavor space and therefore $\mu\to e\gamma$ to vanish. The symmetry $U(1)_e\times U(1)_\mu\times U(1)_\tau$ is then weakly broken by the $(B-L)$-violating couplings in eq.~(\ref{inverse2}) to ensure large mixing angles in the PMNS matrix. As a result, we find a huge suppression ${  O}(C^4)$ with respect to the result eq.~(\ref{eq:mutoe}), i.e. $\textrm{BR}(\mu \to e \gamma)\sim\textrm{BR}_{\rm ISS}(\mu \to e \gamma)\, C^4$. Similarly, one can verify that all CP-odd phases can be removed from $y,m_\Psi$, and the first new physics contribution to the EDMs is suppressed by at least $C^4$. We thus see that the non-observation of rare processes does not represent a robust constraint on this scenario. The most model-independent constraints on $y,m_\Psi$ come from ElectroWeak (EW) precision tests and are of order $y/m_\Psi\lesssim0.1/$ TeV (see for example \cite{Deppisch:2015qwa} and references therein).

%%%%%%%%%%%%%%%%%%
\section{Leptogenesis with small $U(1)_{B-L}$ breaking}

\label{inverse_lepto}

In this section we present {\em analytic} estimations of the baryon asymmetry from thermal leptogenesis in TeV scale models with small $U(1)_{B-L}$ breaking.  We show the results for two specific models: 
the inverse seesaw and linear seesaw models, as well as combinations of the two. 
Our qualitative conclusions are however more general and may extend to a broader class of models with small lepton number violation. In section~\ref{subsec:lepto_inverse} we determine the size of the CP parameter, the washout factor and the final baryon asymmetry. Our results will demonstrate that TeV scale models with anarchic (i.e., roughly of same order but not degenerate) couplings and mass parameters tend to predict too small baryon asymmetry. An intuitive interpretation of the parametric dependence of these results is shown in section~\ref{subsec:NW} based on the (generalized) Nanopoulos-Weinberg theorem. Finally, in section~\ref{subsec:ways_out}, 
we identify a few possibilities that can give rise to successful (sub-)TeV scale leptogenesis with small lepton-number breaking. Conclusions similar to ours are obtained in the numerical analysis of ref.~\cite{Dolan:2018qpy}.

\subsection{Leptogenesis in TeV scale inverse and linear seesaw}\label{subsec:lepto_inverse}
In this section, instead of studying thermal leptogenesis in the most general model [eq.~(\ref{inverse1}) and eq.~(\ref{inverse2})], we illustrate the main results in two limiting cases, namely the ISS and LSS models. The Lagrangians we consider are 
\bea
-{\mathcal L}_{\rm ISS}&\supset& y_{a\al}\Psi_a^cH\ell_\al + m_{\Psi_a} \Psi_a\Psi^c_a+ \frac{\mu_{ab}}{2}\Psi_a\Psi_b+{\rm h.c.}, \label{eq:ISS}\\ 
-{\mathcal L}_{\rm LSS}&\supset& y_{a\al}\Psi_a^cH\ell_\al + m_{\Psi_a} \Psi_a\Psi^c_a+ y'_{a\al}\Psi_a H\ell_\al+{\rm h.c.},\label{eq:LSS}
\eea
where $\al = \{ e, \mu, \tau \}$ denotes SM lepton flavor index and $a,b$ are  the generation indices for $\Psi$. Without loss of generality, we work in the basis where $m_\Psi$ is diagonal and real.
For both models, we demand the singlet neutrinos come in two generations ($a,b=\{1,2\}$), which is the minimum number of generations required to achieve the realistic neutrino mass matrix. Qualitative results in such two-generation model will not differ much from three-generation one. In the rest of this section, we demand that $y\gtrsim 0.01$ and define\footnote{Since we mostly assume couplings and masses are anarchic in this section, we will simply use variables without generation or flavor indices to show the parametric dependence. }
\bea\label{defs}
\vep\equiv\mu/m_\Psi\ll1,~~~~~~\vep'\equiv y'/y\ll1.
\eea
These are the natural choice of parameters for both seesaw models to obtain the SM neutrino masses and testable collider signals. The smallness of neutrino mass is controlled by the smallness of $\vep$ or $\vep'$ [see eq.~(\ref{neutrino_mass})].

To be concrete here we will present the case of the ISS model. Similar conclusions can be drawn for the LSS model, as we emphasize at the end of section \ref{subsec:lepto_inverse} and a more quantitative analysis is shown in the appendix~\ref{app:sec3}. Starting from eq.~\eqref{eq:ISS}, we can write
\bea
\mu =
\begin{pmatrix}
\mu_1  & \bar\mu   \\
\bar\mu &\mu_2
\end{pmatrix},
\eea
where we define $\mu_a(\bar\mu)$ as the diagonal (off-diagonal) parts of $\mu$ matrix. In general, the $\mu$ matrix is complex. However, since we assume all the phases of each element are order one, and yet we will be doing order of magnitude parametric estimation, including those will make at most ${O} (1)$ changes, but will not modify the parametrics of our estimations. For the sake of simplicity, then we simply treat all elements as real numbers. Assuming $\mu_a \sim \bar\mu \ll m_{\Psi_a}, |m_{\Psi_2}-m_{\Psi_1}|$, we can diagonalize the $\Psi,\Psi^c$ mass matrix to first order in $\vep_a\equiv\mu_a/m_{\Psi_a}(a=1,2)$ and $\bar\mu/m_{\Psi_a} $. 
Defining four Majorana states ($\tilde{\Psi}_i,~i=1,2,3,4$) with real masses $(m_i, ~i=1,2,3,4)$ we have 
\bea\label{eq:diagonalization}
-\mathcal L_{\rm ISS}^{\rm mass} \supset h_{i\al}\tilde\Psi_{i}H\ell_\al + \frac{1}{2} m_ i \tilde\Psi_i\tilde\Psi_i+ \textrm{h.c.}.
\eea
To first order in $\vep_a$ and $\bar{\mu}/m_\Psi$, their masses and couplings $h_{i\al}$ are given as (ref.~\cite{Blanchet:2010kw}) 
\bea\label{eq:Mandh}
m_1\simeq m_{\Psi_1}\left(1-\frac{\vep_1}{2}\right)~&;&~h_{1\al}\simeq \frac{i }{\sqrt{2}}\left(y_{1\al}  +\frac{\vep_1}{4} y_{1\al}  +\bar\vep_1 y_{2\al}\right)\nn
m_2\simeq m_{\Psi_1}\left(1+\frac{\vep_1}{2}\right)~&;&~h_{2\al}\simeq  \frac{1}{\sqrt{2}}\left(y_{1\al} -\frac{\vep_1}{4} y_{1\al}  -\bar\vep_1y_{2\al}\right)\nn
m_3\simeq m_{\Psi_2}\left(1-\frac{\vep_2}{2}\right)~&;&~h_{3\al}\simeq \frac{i}{\sqrt{2}}\left(y_{2\al}+\frac{\vep_2}{4} y_{2\al} - \bar\vep_2y_{1\al}\right)\nn
m_4\simeq m_{\Psi_2}\left(1+\frac{\vep_2}{2}\right)~&;&~h_{4\al}\simeq \frac{1}{\sqrt{2}}\left(y_{2\al} -\frac{\vep_2}{4} y_{2\al} + \bar\vep_2y_{1\al}\right),
\eea
where 
\bea\label{eq:sina_sinb}
\bar \vep_1=\frac{\bar\mu m_{\Psi_2}}{m_{\Psi_2}^2-m_{\Psi_1}^2}~~,~~ \bar\vep_2= \frac{\bar\mu m_{\Psi_1}}{m_{\Psi_2}^2-m_{\Psi_1}^2}.
\eea 
From eq.~(\ref{eq:Mandh}), we see that $(\tilde \Psi_1,\tilde \Psi_2)$ and $(\tilde \Psi_3,\tilde \Psi_4)$ form pseudo-Dirac pairs with small Majorana mass splitting. The mass splitting between a pseudo-Dirac pair is only controlled by diagonal $\mu_a$ while both $\mu_a$ and $\bar\mu$ modify the Yukawa couplings. Taking the limit $\mu_a,\bar\mu \to 0$,  one can easily find that $m_1=m_2$, $m_3 = m_4$ and $h_{1\al}=ih_{2\al}$, $h_{3\al}=ih_{4\al}$, as expected for pure Dirac states.

\subsubsection{CP asymmetry}
Now we are ready to calculate the CP asymmetry from the decay of each Majorana state 
$\tilde \Psi_i \to \ell_\alpha H, (\ell_\alpha H)^*$. After summing over SM lepton flavor $\al$, we get:\footnote{Assuming anarchy of 
Yukawa couplings $h_{i\alpha}$, the lepton asymmetry produced will be distributed among all the lepton flavors in roughly equal proportion. 
For simplicity, we ignore the small differences in the various flavor asymmetries and sum over $\alpha$. When couplings are hierarchical flavor effects~\cite{Nardi:2006fx} could 
play a more relevant role, and we will briefly mention about it in section \ref{subsec:ways_out}.}
\bea\label{eq:epsilon}
\ep_{i} \equiv \frac{\sum_\al \left[\Gamma(\tilde \Psi_i \to \ell_\al H)-\Gamma(\tilde \Psi_i \to \overline \ell_\al H^*)\right]}
{\sum_\al \left[\Gamma(\tilde \Psi_i \to \ell_\al H)+\Gamma(\tilde \Psi_i \to \overline \ell_\al H^*)\right]}
=\frac{1}{8\pi}\sum_{j\neq i}\frac{\textrm{Im}[(h h^\dagger)^2_{ij}]}{(h h^\dagger)_{ii}}f_{ij},
\eea
where $f_{ij} \equiv f^{\rm v}_{ij}+f^{\rm self}_{ij}$ comprises a contribution from vertex corrections~\cite{Covi:1996wh}
\bea\label{eq:fv}
f^{\rm v}_{ij}=g\left(\frac{m^2_j}{m^2_i}\right)  ~~\textrm{;}~~g(x)=\sqrt{x}\left[1-(1+x)\ln(1+\frac{1}{x})\right],
\eea
as well as a self energy correction to the decay~\cite{Deppisch:2010fr}
\bea\label{eq:fself}
f^{\textrm{self}}_{ij}=\frac{(m^2_i-m^2_j)m_im_j}{(m^2_i-m^2_j)^2+m^2_i\Gamma^{2}_j}.
\eea
Here $\Gamma_j\equiv (h h^\dagger)_{jj} m_j/(8 \pi)$ is the decay width of $\tilde \Psi_j$.

Let's take a close look at $\ep_1$ and $\ep_2$ in eq.~(\ref{eq:epsilon}):
\bea\label{eq:ep1&ep2}
\ep_1&=&\frac{1}{8\pi (hh^\dagger)_{11}}\textrm{Im}[(h h^\dagger)^2_{12}f_{12}+(h h^\dagger)^2_{13}f_{13}+(h h^\dagger)^2_{14}f_{14}] , \nn
\ep_2&=&\frac{1}{8\pi (hh^\dagger)_{22}}\textrm{Im}[(h h^\dagger)^2_{21}f_{21}+(h h^\dagger)^2_{23}f_{23}+(h h^\dagger)^2_{24}f_{24}]  .
\eea
Given that the pseudo-Dirac pairs are almost degenerate in mass, the number density of two states are approximately the same. 
As a result (see appendix~\ref{app:BE_ISS_LSS}), it is appropriate to consider $\epsilon_1 + \epsilon_2$ and
$\epsilon_3 + \epsilon_4$ as the effective CP asymmetry for each generation.
Due to the pseudo-Dirac nature, one finds that
\bea
(h h^\dagger)^2_{13}\simeq-(h h^\dagger)^2_{23}\simeq -(h h^\dagger)^2_{14}\simeq (h h^\dagger)^2_{24}~;~f_{13}\simeq f_{14}\simeq f_{23}\simeq f_{24}\,.
\eea  
This means that when we consider the sum of $\ep_1$ and $\ep_2$,  parts involving $f_{13}$ and $f_{14}$ in $\ep_1$ 
will cancel against the corresponding parts with $f_{23}$ and $f_{24}$  in $\ep_2$ to first order. Also, if we consider the generic parameter region of the ISS, i.e., 
\bea\label{eq:generic_param}
\mu_a\sim \bar\mu\ll\Gamma_i \ll m_{\Psi_a} \sim |m_{\Psi_2}-m_{\Psi_1}| ,
\eea
and no hierarchies in mass or couplings among singlet generations and SM flavors,
we would get 
$\bar\vep_{1,2}\sim\bar\mu/m_{\Psi_{1,2}}$
and
\bea
&&-f^{\rm self}_{12}\simeq f^{\rm self}_{21}\sim \vep_1\left(\frac{m_{\Psi_1}}{\Gamma_2}\right)^2~,~f^{\rm v}_{12}- f^{\rm v}_{21}\sim \vep_1,\nn
&&(f_{13}-f_{14})\sim \vep_2 ~,~ (f_{13}-f_{14}-f_{23}+f_{24})\sim \vep_1\vep_2.
\eea
Therefore, the terms involving $f^{\rm self}_{12}$ and $ f^{\rm self}_{21}$ dominate in $\ep_1+\ep_2$, giving  [see eq.~(\ref{eq:ep1&ep2})] 
\bea\label{eq:ep1approx}
\ep \equiv\ep_1+\ep_2\sim \frac{\textrm{Im}[ (yy^\dagger)^2_{12}]}{ (yy^\dagger)^2_{11}}\bar\vep_1 \frac{\mu_1/m_{\Psi_1}}{(yy^\dagger)_{11}/(16\pi)} \sim \frac{\bar\mu}{m_{\Psi}}\frac{\mu}{\Gamma}~~~(\mu,\bar\mu\ll \Gamma),
\eea
where we have dropped the family indices for $\mu$ and $\Gamma$ to show only the parametric dependence.
Similarly, $\ep_3+\ep_4$ can be obtained by changing index $1\to 2$ and $2\to 1$ in eq.~(\ref{eq:ep1approx}), resulting in the same parametric dependence. 

For completeness, we also show the parametric dependence of $\ep_{1,2}$:
\bea
\ep_1\approx -\ep_2=O(\frac{\bar\mu}{m_{\Psi}}\frac{\mu}{\Gamma})+O(\frac{\mu}{m_{\Psi}}\frac{\Gamma}{m_\Psi}).
\eea
We only use $\ep_1+\ep_2$ instead of individual $\ep_1$ or $\ep_2$ in our study of leptogenesis. However, they are relevant for the argument in appendix~\ref{app:BE_ISS_LSS}.

If we assume $\mu\sim\bar\mu$ and enforce $m_\nu \sim 0.05$ eV via eq.~(\ref{neutrino_mass}), eq.~(\ref{eq:ep1approx}) becomes (see also ref.~\cite{Deppisch:2010fr})
\bea\label{eq:ep-result}
\ep \sim\frac{\mu}{m_\Psi}\frac{\mu}{\Gamma}\sim \frac{16\pi m_\nu^2 m_\Psi^2}{y^6v^4}\sim10^{-10}\left(\frac{m_\Psi}{\rm TeV}\right)^2 \left(\frac{10^{-2}}{y}\right)^{6} .
\eea
As we will see shortly [eq.~\eqref{eq:YB}], $|\ep |$ should be $\gtrsim 10^{-7}$ to generate the observed baryon asymmetry via leptogenesis 
and eq.~(\ref{eq:ep-result}) falls short by three orders of magnitude. 
From eq.~(\ref{eq:ep-result}), it seems that one can obtain a larger value by reducing Yukawa couplings $y$. 
However, this approach will not allow us to obtain sufficient baryon asymmetry once we, as required, include the washout effects. We will discuss this in the following section.

\subsubsection{Washout and baryon asymmetry}
The final baryon asymmetry through leptogenesis from decays of $\tilde \Psi_i \to \ell_\alpha H, (\ell_\alpha H)^*$ can be parametrized as follows 
\bea\label{eq:YB}
Y_{\Delta B}\equiv\frac{n_B-n_{\bar B}}{s}\sim 10^{-3}\ep\,\eta,
\eea
where $n_{B(\bar B)}$ is the number density of baryons (anti-baryons) and $s$ is the total entropy density of the thermal bath. 
The pre-factor $\sim 10^{-3}$ comes from relativistic number density of $\tilde \Psi_i$ normalized to the entropy density $s$. 
The efficiency factor $\eta$ is always less than unity and parametrizes the effect of washout processes. 
{It is obtained by solving the Boltzmann equations. The efficiency of leptogenesis can be parametrized by the so-called washout factor~\cite{Davidson:2008bu}} 
%%%
\bea
\label{eq:K}
K_i \equiv \frac{\Gamma_i}{H (T = m_i)}
\eea
%%%
where $H(T) \sim \sqrt{g_*}\, T^2/M_{\rm Pl}$ is the Hubble rate 
with $T$ being the thermal bath (photon) temperature, $g_*$ the number of relativistic degrees of freedom 
and $M_{\rm Pl} = 1.22\times 10^{19}$ GeV the Planck mass. 
In the ISS scenario, due to the approximate lepton number conservation, the washout from inverse decay is actually controlled by~\cite{Blanchet:2009kk}\footnote{The appearance of $\delta^2$ may be understood as follows. In the limit $\mu \to 0$, since lepton number is preserved, no process can washout (or produce) the asymmetry. Therefore, the effective washout factor must vanish as $\mu \to 0$. Another (more technical) way to see this is to recall that the washout from the inverse decay can be obtained by the on-shell part of $\Delta L =2$ $H\ell \leftrightarrow (H\ell)^*$ scattering. Due to the near degeneracy, this scattering gets contribution from both s-channel $\tilde{\Psi}_1$ and $\tilde{\Psi}_2$ and importantly, most of their contributions cancel. The surviving piece comes from interference of the two and is proportional to $\delta^2$.}
\bea\label{eq:Keff}
K^{ \rm eff} \sim K \delta^2 ,
\eea
%%%
where 
{
$\delta\equiv |\Delta m|/\Gamma \simeq\mu /\Gamma$ with $\Delta m = m_2 - m_1$ or $m_4 - m_3$. Also, we dropped generation index for simplicity of notation and we will do so below when there is no chance of confusion. Consistently, this quantity vanishes in 
the lepton number conserving limit. 
Notice that we can express eq.~\eqref{eq:ep1approx} as
%%%
\bea 
\ep \sim \frac{\Gamma}{m_\Psi} \delta^2,
\label{eq:ep_delta}
\eea
%%%
where we have taken $\bar\mu \sim \mu$.
}

If $K^{ \rm eff}> $ a few, the washout from inverse decay ($H \ell_\alpha, \left( H \ell_\alpha \right)^* \to \tilde{\Psi}_i$) is efficient (strong washout regime) and $\eta\sim 1/K^{\rm eff}$ 
(see appendix~\ref{app:strong_washout}).
In this regime, substituting eqs.~(\ref{eq:ep_delta}) and (\ref{eq:Keff}) into eq.~(\ref{eq:YB}), the baryon asymmetry is estimated to be
%%%
\bea\label{eq:general-YB}\label{eq:etaB_general}
Y_{\Delta B}\sim10^{-3}  \sqrt{g_*}\frac{m_{\Psi}}{M_{\rm Pl}}\sim 10^{-18}\left(\frac{m_{\Psi}}{1\textrm{TeV}}\right),
\eea
%%%
where we have taken $\sqrt{g_*}\sim 10$. This analytic estimation was first obtained in our earlier paper~\cite{short}. Clearly, a TeV scale $m_\Psi$ will result in a too small asymmetry compared 
to the observed value $Y_{\Delta B}^{\rm obs}\approx9\times10^{-11}$~\cite{Ade:2015xua}. 
Remarkably, in the strong washout regime, the final baryon asymmetry ($Y_{\Delta B}$) for the ISS model with anarchic couplings and masses 
reduces to the simple formula [eq.~(\ref{eq:general-YB})] which does not depend on $\mu$ and $y$.

To complete our discussion, we also need to consider the weak washout regime, where $K^{\rm eff}<1$. 
ISS model has a peculiar feature that the production of singlets is controlled by $K$ [eq.~\eqref{eq:K}], 
whereas the washout is controlled by $K^{\rm eff}$ [eq.~\eqref{eq:Keff}]. 
Assuming no initial abundance of $\tilde \Psi_i$,  there are two cases in weak washout region 
and the corresponding efficiency factors $\eta$ are 
%%%
\bea\label{eq:eff_weak_new}
\eta\sim\left\{
\begin{array}{l}
  K^{\rm eff} ~~~~~~~~~~~( K^{\rm eff}<1~ \textrm{and}~K>1~\textrm{with no initial $\tilde \Psi_i$})\\
  K\times K^{\rm eff} ~~~~( K^{\rm eff}<1~ \textrm{and}~K<1~\textrm{with no initial $\tilde \Psi_i$}),
 \end{array}
 \right. 
\eea
%%%
as derived in appendix \ref{app:weak_washout}.
We emphasize that such parametric dependence of $\eta$ is qualitatively different from that of usual type I seesaw (i.e., $\eta\sim K^2$). To the best of our knowledge, this analytic result, especially which of $K^{\rm eff}$, $K$ should appear in $\eta$, has not been discussed in the literature. 
If we, on the other hand, assume $\tilde \Psi_i$ has been kept in thermal equilibrium with SM particles by interactions other than those due to 
Yukawa coupling $y$,\footnote{For instance, if $\tilde \Psi_i$ is charged under new gauge symmetries (e.g. $U(1)_{B-L}$), they can acquire an initial thermal abundance.}
the efficiency factor is of the order
%%%
\bea
\eta\sim O(1)~~~~( K^{\rm eff}<1~\textrm{with thermal initial $\tilde \Psi_i$}).
\eea
%%%

Putting everything together, in the weak washout regime, we have
\bea\label{3cases}
Y_{\Delta B}\sim 10^{-3} \ep \,\eta \sim 
\left\{
\begin{array}{l}
 10^{-3}  \sqrt{g_*}\frac{m_\Psi}{M_{\rm pl}} (K^{\rm eff})~~~~~~~~~~~( K^{\rm eff}<1~\textrm{with thermal initial $\tilde \Psi_i$})\\
 10^{-3}  \sqrt{g_*}\frac{m_\Psi}{M_{\rm pl}} (K^{\rm eff})^2~~~~~~~~~~( K^{\rm eff}<1~ \textrm{and}~K>1~\textrm{with no initial $\tilde \Psi_i$})\\
   10^{-3} \sqrt{g_*}\frac{m_\Psi}{M_{\rm pl}}(K^{\rm eff})^2 K~~~~~~~( K^{\rm eff}<1~ \textrm{and}~K<1~\textrm{with no initial $\tilde \Psi_i$}).
    \end{array}
 \right. 
\eea
{We see that in all cases the final baryon asymmetry in the weak washout regime $K^{\rm eff}<1$ 
is smaller compared to that of strong washout in eq.~(\ref{eq:general-YB}). Therefore, the TeV scale ISS model with anarchic mass and coupling cannot provide successful leptogenesis.

An analogous calculation for the LSS model is shown in appendix \ref{app:sec3} 
and the parametric dependences of the final baryon asymmetry of the two seesaw models 
are in fact the same, as we summarize in table~\ref{tab:results}.
Therefore, we conclude that TeV scale ISS and LSS model with anarchic parameters $(y, m_\Psi,\mu\,\textrm{or}\,y')$ and sizable $y$
cannot give rise to successful leptogenesis.}

\renewcommand{\arraystretch}{2}
\begin{table}[t]
\centering
\begin{tabular}{|c|c|c|c|}
\hline
 Model & CP asymmetry ($\ep$) & Efficiency ($\eta$) & Baryon asymmetry ($Y_{\Delta B}$) \\
 \hline \hline
Inverse seesaw &\large $ \frac{\Gamma}{m_\Psi} \delta^2 $ & $\lesssim \left(\frac{\Gamma}{H} \delta^2\right)^{-1}$& $\lesssim10^{-3} \sqrt{g_*}\frac{m_\Psi}{M_{\rm pl}}\sim10^{-18}\left(\frac{m_{\Psi}}{1\textrm{TeV}}\right)$\\
\hline
Linear seesaw & $\frac{\Gamma}{m_\Psi} \vep^{\prime 2}$ & $\lesssim \left(\frac{\Gamma}{H} \vep^{\prime 2}\right)^{-1}$& $\lesssim10^{-3}  \sqrt{g_*}\frac{m_\Psi}{M_{\rm pl}}\sim10^{-18}\left(\frac{m_{\Psi}}{1\textrm{TeV}}\right)$ \\
 \hline
\end{tabular}
\caption{Summary of the parametric dependence of CP asymmetry, washout, baryon asymmetry in inverse [see eq.~(\ref{eq:ISS})] and linear seesaw
[see eq.~(\ref{eq:LSS})]. The parameters $\varepsilon,\varepsilon'$ are defined in eq.~(\ref{defs}), wheres $\delta$ below eq.~(\ref{eq:Keff}). 
\label{tab:results} }
\end{table}

\renewcommand{\arraystretch}{1}

\subsection{Nanopoulos-Weinberg theorem}\label{subsec:NW}

As discussed in the previous section, the CP asymmetries in TeV scale ISS and LSS models are small because they are respectively $\ep \propto \delta^2 $\footnote{One might wonder why the lepton-number violation is captured by $\vep \sim \mu/m_{\Psi}$ in the case of the neutrino mass, and by $\delta \sim \mu / \Gamma$ in the case of CP-violation (and leptogenesis). This may follow from the fact that while the generation of $m_\nu$ is off-shell phenomenon (i.e. simply integrate out $\tilde{\Psi}$'s), that of CP-violation and related asymmetry generation occurs near on-shell. Especially, when the genesis goes through the resonance-enhancement, on top of parametric lepton-number violation $\vep$, it acquires extra kinematic (resonance-)enhancement $\sim m_{\Psi} / \Gamma$, yielding the associated net breaking parameter $\sim \left(\frac{\mu}{m_{\Psi}}\right) \left( \frac{m_{\Psi}}{\Gamma} \right) \sim \delta$.   }
and $\vep^{\prime 2}$, where $\delta,\vep'$ are the tiny parameters characterizing the small lepton number violation. We will argue in this section that this feature can indeed be anticipated due to Nanopoulos-Weinberg (NW) theorem (ref.~\cite{Nanopoulos:1979gx}) and similar conclusions can be drawn in some variations of ISS or LSS models, or combination of both. 
  
The NW theorem states that, in the CP-violating decay process, if the particle can decay only through baryon (lepton) number violating parameters (e.g.~Type-I),
a nonzero CP asymmetry can be generated starting at third order in baryon (lepton) number violating parameters. In addition, the generalized version of the NW theorem (ref.~\cite{Adhikari:2001yr}) says that, if the decaying particle, on the other hand, can decay through both baryon (lepton) number violating and conserving couplings, the CP asymmetry may be generated at second order in baryon (lepton) number violating parameters. 

Now we apply both theorems to check our results for the ISS and LSS models. The CP asymmetry in decay width is given in eq.~(\ref{eq:ep1approx}) for ISS and eq.~(\ref{eq:ep'}) for LSS:
\bea\label{NW1}
\sum_{i}\sum_f|\Gamma(\tilde \Psi_i \to f)-\Gamma(\tilde \Psi_i \to \bar f)|\propto\left\{
\begin{array}{ll}
\textrm{Im}[(yy^\dagger)_{12}^2]\delta^2 ~~~~~~~(\textrm{ISS})\\
~\\
\textrm{Im}[(yy^\dagger)_{12}(y'y^{\prime \dagger})_{12}]~(\textrm{LSS}) ,
\end{array}
\right. 
\eea
where we sum over almost degenerate $\tilde \Psi_i$ states and all final states $f$. 

For the ISS, if we assign the lepton number charges $L(\ell)=L(\Psi)=-L(\Psi^c)=1$, the Yukawa coupling $y$ is lepton number conserving and $\mu$ is the only lepton number violating parameter. Then $\Psi,\Psi^c$ can decay also via number-conserving interactions and, following the extended version of the NW theorem, the CP asymmetry should be ${O}(\mu^2)$. The CP asymmetry in eq.~(\ref{NW1}) indeed contains $\delta^2$, hence $\propto\mu^2$. 

Similarly for the LSS, we can always assign lepton number such that only one of $y$ or $y'$ violates lepton number. Since $\Psi,\Psi^c$ can decay through either $y$ or $y'$,  it always follows the extended NW theorem. Therefore, we expect the CP asymmetry is proportional to two powers of $y$ and two powers of $y'$, which matches the result in eq.~(\ref{NW1}).

In general, NW theorem forces the CP asymmetry from singlets decay to be ${O}(\delta^2)$ or ${O}(\vep^{\prime 2})$, which is suppressed in models with small lepton number breaking. Adding further lepton number conserving decay channels or new generations of leptons would not alter this result.

\subsection{Possible variations to achieve successful leptogenesis}\label{subsec:ways_out}

Our discussion so far assumed anarchic couplings and masses and considered either a small $\mu$ or a small $y'$, separately. In this subsection we relax these assumptions with the aim of looking for models with small $U(1)_{B-L}$ violation that can result in larger final asymmetry compared to eq.~(\ref{eq:etaB_general}).
%%%%%%%%
\subsubsection{Inverse seesaw with degeneracy among different generations}
%%%%%%%%

We first consider the possibility that the singlet masses are quasi-degenerate among different \emph{generations}:
%%%
\bea
\Delta m_\Psi\equiv |m_{\Psi_2} - m_{\Psi_1}|,
\eea
with $\mu\ll\Delta m_\Psi \ll  m_{\Psi_1}, m_{\Psi_2}$ so that our previous formulae in section \ref{subsec:lepto_inverse} still apply.
%%%
{Although quasi-degeneracy in mass within a pseudo-Dirac pair is naturally obtained due to approximate lepton number, 
to realize quasi-degeneracy in mass among singlets of different generations in a natural way, 
an approximate family symmetry is necessary as was done, for example, in the resonant leptogenesis scenario~\cite{Pilaftsis:2003gt}.} In scenarios with minimal flavor violation, even if $\Delta m_\Psi$ is set to zero at the tree level, generally Yukawa couplings might break the family symmetry, generating $\Delta m_\Psi$ at loop level of the size
\bea\label{natDelta}
\frac{\Delta m_\Psi}{m_\Psi}\gtrsim \frac{y^2}{16\pi^2}.
\eea

\noindent In this case, the $\bar\vep_{1,2}$ which parametrically is given by (see eq.~(\ref{eq:sina_sinb})) 
\bea\label{eq:sina_deg}
\bar\vep_1\sim\bar\vep_2\sim \frac{\bar\mu}{\Delta m_\Psi},
\eea
can be enhanced.
Substituting eq.~(\ref{eq:sina_deg}) into eq.~(\ref{eq:ep1approx}), one has
\bea\label{eq:ep-deg-off}
\ep \sim \frac{\mu}{m_\Psi}\frac{\mu}{\Gamma} \frac{m_{\Psi}}{\Delta m_\Psi}.
\eea
When two generations are nearly degenerate, thus, the CP asymmetry is enhanced compared 
to eq.~(\ref{eq:ep1approx}) by a factor of $\frac{m_{\Psi}}{\Delta m_\Psi}$. 
The washouts are nevertheless unchanged.\footnote{The contribution to $K^{\rm eff}$ has two pieces in the non-degenerate case: $K^{\rm eff}=K \left[O(\frac{\mu^2}{\Gamma^2})+O(\frac{\mu^2}{m_\Psi^2})\right]$. The second term is suppressed compared to the first one and thus we only keep the first term in the previous estimation. In the case we discussed here, where there is degeneracy among different generations, the first term is still unchanged. This is because the first term is controlled by the mass splitting within each generation, which will not be modified by the degeneracy among different generations. The second term, however, is enhanced by $\frac{m_\Psi^2}{\Delta m_\Psi^2}$: $K \frac{\mu^2}{m_\Psi^2}\frac{m_\Psi^2}{\Delta m_\Psi^2}$. Now these two terms are comparable due to the assumption in eq.~(\ref{natDelta}) and the parametric dependence of $K^{\rm eff}$ remain the same as in eq.~(\ref{eq:Keff}). }
So the final result scales as
\bea
Y_{\Delta B} \sim 10^{-3} \sqrt{g_*} \frac{m_{\Psi_1}}{M_{\rm pl}}\frac{m_{\Psi}}{\Delta m_\Psi}.
\eea
The right size of $Y_{\Delta B}$ may be obtained by choosing the right size for $\frac{m_{\Psi}}{\Delta m_\Psi}$. However, we are not completely free to choose its value here. In particular, our analysis is done under the assumption that $\Delta m_\Psi \gg \mu$\footnote{Implicitly, we also assumed $\Gamma \gg \mu$ to get a concrete expression. However, a straightforward check can confirm that while CP and washout factor will change (basically replacing $\mu/\Gamma$ with $\Gamma/\mu$), the final asymmetry will be the same as the one we show above.} and (technical) naturalness indicates that $\Delta m_\Psi / m_\Psi \gtrsim y^2 / 16\pi^2$. Combining these two with the constraint from the neutrino mass, i.e. $m_\nu \sim y^2 v^2 \mu /m_\Psi^2$, gives rise to an upper bound on the enhancement factor $m_\Psi / \Delta m_\Psi \ll 10^7$. Therefore, we conclude that while degeneracy among different singlet generation can induce a significant enhancement in the final asymmetry, whether or not the actual observed quantity can be accounted requires a careful numerical study. We find it quite likely that the observed asymmetry may be explained by this effect, but only in a small corner of the parameter space with $y\sim 10^{-3}$ for $m_\Psi\sim$ TeV.

%%%%%%%
\subsubsection{Inverse seesaw $+$ linear seesaw}\label{sec:ISS+LSS}
%%%%%%%

ISS and LSS models were treated separately in the previous discussions, see eqs.~(\ref{eq:ISS}) and (\ref{eq:LSS}). Now we consider scenarios in which both $\mu,y'$ are non-vanishing:\footnote{In principle there could also be $\mu'\Psi^c\Psi^c$ term, see eq.~(\ref{inverse2}). However, such a term does not enter neutrino mass formula and has similar effects as $\mu$ in leptogenesis. Therefore we neglect it in this study. }
\bea
-{\mathcal L}_{\rm ISS+LSS} \supset y_{a\al}\Psi_a^c H\ell_\al + (m_\Psi)_a \Psi_a\Psi^c_a 
+ \frac{(\mu)_{ab}}{2}\Psi_a\Psi_b + y'_{a\al}\Psi_a H\ell_\al+ {\rm h.c.}.
\eea
For this model we will only consider one generation of singlets. Nothing qualitatively new happens when more generations are included (unless they are nearly degenerate, in which case one can use the results of the previous subsection). As previously shown in ref.~\cite{Blanchet:2009kk}, the CP asymmetry is parametrized as
\bea\label{eq:ep_ISS+LSS}
\ep\sim\frac{\Gamma}{m_\Psi}\left(O(\delta^2)+O(\delta\vep'\frac{m_\Psi}{\Gamma})+O(\vep^{\prime2})\right).
\eea
Following the analysis in ref.~\cite{Blanchet:2009kk}, the washout can be worked out as
\bea\label{eq:K_ISS+LSS}
K^{\rm eff}\sim \frac{\Gamma}{H}\left(O(\delta^2)+O(\vep^{\prime2})+O(\vep\vep^{\prime})\right).
\eea
Based on eqs.~(\ref{eq:ep_ISS+LSS}) and (\ref{eq:K_ISS+LSS}), it is obvious that in the limit where $y'\to 0$ or $\mu\to 0$, 
we recover the results for the ISS or LSS (at tree level) models (see table~\ref{tab:results}).

Now we would like to find if there exists a range of parameters where the asymmetry is larger than eq.~(\ref{eq:etaB_general}).
To do this, let us first focus on the strong washout regime for definiteness. Under this hypothesis the final asymmetry is given by $\ep/K^{\rm eff}$ and
\bea\label{sb}
Y_{\Delta B}\sim10^{-3} \sqrt{g_*} \frac{m_{\Psi}}{M_{\rm pl}}\times\left[\frac{1+{\cal O}\left(\frac{\varepsilon'^2}{\delta^2}\right)+{\cal O}\left(\frac{\varepsilon'm_\Psi}{\delta\Gamma}\right)}{1+{\cal O}\left(\frac{\varepsilon'^2}{\delta^2}\right)+{\cal O}\left(\frac{\varepsilon'\Gamma}{\delta m_\Psi}\right)}\right].
\eea
For $\varepsilon'/\delta\ll \Gamma/m_\Psi$ or $\varepsilon'/\delta\gg m_\Psi/\Gamma$ one can readily see that the square bracket in eq.~(\ref{sb}) becomes of order unity. In these limits, one can check that terms only involving $\mu$ (when $\varepsilon'/\delta\ll \Gamma/m_\Psi$) or $y'$ (when $\varepsilon'/\delta\gg m_\Psi/\Gamma$) will be dominant in both neutrino mass formula [eq.~(\ref{MajoranaNu})] and leptogenesis [see eqs.~(\ref{eq:ep_ISS+LSS}) and (\ref{eq:K_ISS+LSS})]. Clearly, these limits correspond to the cases studied above, namely the ISS and LSS respectively. 

The only unexplored region of parameter space is $\Gamma/m_\Psi\ll\varepsilon'/\delta\ll m_\Psi/\Gamma$, where we have 
\bea\label{eq:[]}
[\cdots]\sim\frac{m_\Psi}{\Gamma}\frac{{\cal O}\left(\frac{\varepsilon'}{\delta}\right)}{1+{\cal O}\left(\frac{\varepsilon'^2}{\delta^2}\right)}.
\eea
Here $[\cdots]$ refers to the expression inside the square bracket in eq.~(\ref{sb}) and is maximized at $\varepsilon'\sim\delta$. Interestingly, in this regime, the neutrino mass formula is dominated by terms containing $y'$, whereas both $\mu$ and $y'$ have a significant impact on the asymmetry. Because both $y',\mu$ are necessary here, this case does not correspond to any model we discussed before. The final asymmetry is given by  $Y_{\Delta B}\sim10^{-3} \sqrt{g_*} \frac{m_{\Psi}}{M_{\rm pl}}\frac{m_{\Psi}}{\Gamma}$. This result is enhanced by a factor $m_\Psi/\Gamma$ compared to the typical value in eq.~(\ref{eq:etaB_general}).

\begin{figure}[t]
\centering
\includegraphics[width=120mm,height=60mm]{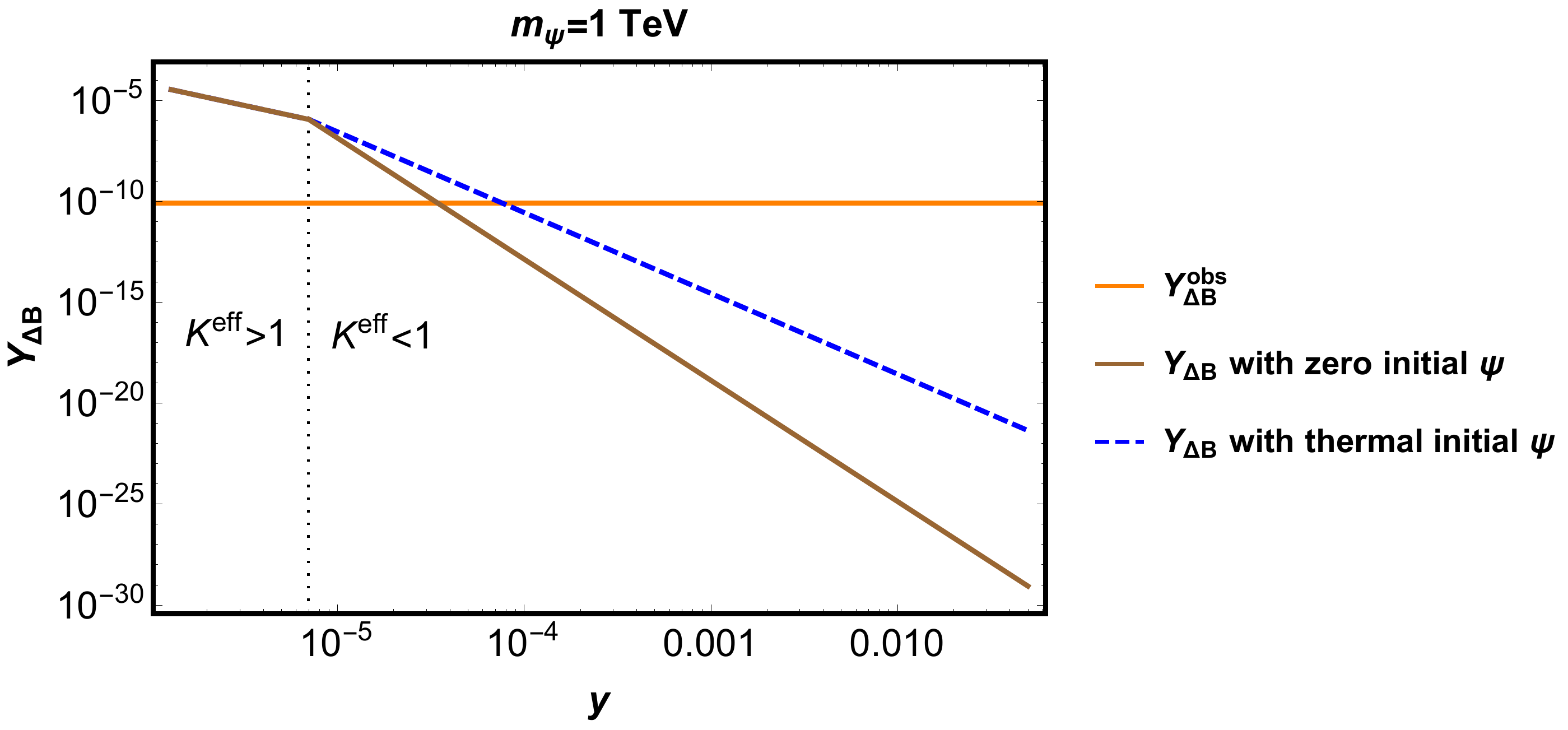} 
\caption{The baryon asymmetry $Y_{\Delta B}$ as a function of $y$ in the case where $\delta\sim \vep'$ in ISS+LSS models. Here $m_\Psi$ is fixed to be 1 TeV. The blue dashed line shows result with initial thermal abundance of $\tilde \Psi_i$ while the solid brown line shows the results with no initial $\tilde \Psi_i$ abundance. The vertical dotted lines indicate the border between the strong and weak washout regions. The orange line shows where the observed baryon asymmetry is obtained. Here we only plot the region where $y>y'$, meaning $y>\sqrt{yy'}\sim\sqrt{m_\nu m_\Psi/v^2}\approx 10^{-6}$. When $y<y'$, the results can be simply obtained by the exchange the role of $y$ and $y'$. }
\label{fig:ISS+LSS}
\end{figure}

Let us therefore consider $\delta\sim\varepsilon'$. First of all, such condition might be realized quite naturally starting with a LSS framework and generating a $\mu$ term from radiative corrections. This way one expects [see eq.~(\ref{loops})]
$$
\delta=\frac{\mu}{\Gamma}\sim\frac{yy'm_\Psi}{16\pi^2}\times\frac{{16\pi}}{y^2m_\Psi}\sim\frac{1}{\pi}\varepsilon',
$$
which is not far from the required relation. Then we can relax the assumption of strong washout ($K^{\rm eff}>1$) and estimate the final baryon asymmetry more generally. In this regime $K^{\rm eff}\propto y^{\prime 2}$, $K\propto y^2$ and the SM neutrino mass $m_\nu\propto yy'$. This implies that in the weak washout region (i.e., $K^{\rm eff}<1$) we always have $K > 1$ for $m_\Psi\sim$ TeV. We thus have only two of the options previously considered in eq.~(\ref{3cases}). Finally, the baryon asymmetry scales as
\bea\label{eq:YB_ISS+LSS}
&~&\ep\sim \vep^{\prime2},~~~~~K^{\rm eff}\sim \frac{\Gamma}{H} \vep^{\prime2},\nn
&\Rightarrow&Y_{\Delta B}\sim\left\{
\begin{array}{ll}
 10^{-3}\frac{\ep}{K^{\rm eff}}\sim 10^{-3} \sqrt{g_*} \frac{m_{\Psi}}{M_{\rm pl}}\frac{m_{\Psi}}{\Gamma}~~~~~~~~~~~(K^{\rm eff}>1)\\
 10^{-3} \ep K^{\rm eff }~~~~~~~~~~~~~~~~~~~~~~~~~~~~~~~~~~(K^{\rm eff}<1~\textrm{with no initial $\tilde \Psi_i$})\\
  10^{-3} \ep ~~~~~~~~~~~~~~~~~~~~~~~~~~~~~~~~~~~~~~~(K^{\rm eff}<1~\textrm{with thermal initial $\tilde \Psi_i$}).
 \end{array}
 \right.
\eea
These values are shown in figure~\ref{fig:ISS+LSS} as a function of $y$ with $m_\Psi=1$ TeV. Figure~\ref{fig:ISS+LSS} indicates the observed baryon asymmetry can be obtained if $y=O(10^{-5}-10^{-4})$.  

To conclude, we found that successful leptogenesis is achievable in scenarios of ISS + LSS with $\delta\sim\varepsilon'$, provided the Yukawa couplings are small enough.  
The Yukawa coupling needed for leptogenesis, $y=O(10^{-5}-10^{-4})$, clearly lies outside of the window of our naturalness criteria and is also too small to provide signals at colliders.
Therefore, we will not consider this option any further.

%%%%%%%%%%%%%%%%%%%%%%%%%%%%%%%%%

%%%%%%%%%%%%%%%%%%%%%%%%%%%%%%%%%

%%%%%%
\subsubsection{Other mechanisms}\label{subsec:others}
%%%%%%

There are several alternative options that may allow us to achieve a successful TeV scale (or lower) leptogenesis in scenarios with small $U(1)_{B-L}$ breaking. We here mention a few that were originally realized in the context of type I seesaw model with singlet fermions. We will however not discuss them in detail because they all require unnatural couplings or flavor symmetries.

As a first option, if there is certain hierarchical structure in the Yukawa coupling $y_{a \alpha}$ (i.e. deviations from anarchical as well as natural values), 
lepton flavor effects can play an important role~\cite{Nardi:2006fx} in enhancing the efficiency since 
an optimal regime can be realized by having the lepton asymmetry stored in the lepton flavors that suffer the least washout. As we have touched upon earlier, a second option is allowing quasi-degeneracy in singlet mass of different generations---as in resonant leptogenesis~\cite{Pilaftsis:2003gt}. This can be realized by imposing approximate family symmetry.
In ref.~\cite{GonzalezGarcia:2009qd}, while total lepton number violation can be very small (or even conserved), 
both lepton flavor effects and quasi-degenaracy among mass of singlets have been utilized to achieve leptogenesis 
at around TeV scale.

Finally we should mention an alternative mechanism for leptogenesis. While the present paper focuses on leptogenesis from decays of singlets, the lepton asymmetry can also be realized via flavor oscillation among singlets, as first pointed out by Akhmedov, Rubakov and Smirnov (ARS)~\cite{Akhmedov:1998qx}. 
One distinguishing feature of the ARS mechanism is that leptogenesis must occur at a scale higher than the singlet mass, $T> m_{\rm NP}$, when oscillations among sterile neutrinos can be important. Although the total lepton number is approximately conserved, flavor oscillation among singlets can create an asymmetry in some singlet flavor. The singlets that are in thermal equilibrium can subsequently transfer their asymmetry to the SM lepton doublets and finally, via the EW sphalerons, to the baryon sector. Requiring the generation of lepton asymmetry takes place while the EW sphalerons are still active ($T \gtrsim 100$ GeV), implies the mass of new singlets involved in ARS leptogenesis must be well below the weak scale, $m_{\rm NP}<100$ GeV. They may hence be probed in neutrinoless double beta decay experiments and high intensity beam experiments. In this scenario, a hierarchy in $y_{a \alpha}$ is needed such that at least one of the singlets does not reach thermal equilibrium until after EW sphaleron processes freeze out to prevent the washout of the asymmetry. According to our earlier definition, a certain amount of unnaturalness is thus required to realize this mechanism as well. In the context of ISS model, ARS leptogenesis with GeV scale singlets has been studied in refs.~\cite{Abada:2015rta}.

%%%%%%%%%%%%%%%
\section{General idea of the hybrid seesaw}
\label{sec:4_extension_big_picture}

Low-scale seesaw models with small lepton-number violation are confronted with several issues that make them not fully satisfactory. In particular, the
required smallness of lepton-number breaking terms, the central ingredient for the seesaw mechanism, is often left unexplained. Even though the requirement $C\ll1$ is consistent with the criteria of technical naturalness, {one finds it not fully convincing because it has no clear origin within that description. In this sense the smallness of neutrino masses is not truly explained.} The second major issue was discussed in section~\ref{inverse_lepto} and corresponds to the difficulty with regard to the question of explaining the observed baryon-anti-baryon asymmetry of the Universe via leptogenesis.

In this paper, we will show that there exists a rather simple and motivated extension that addresses {\em both} issues. Before we get to more technical discussions, however, in this section we present a qualitative description of our model. We hope this makes the big picture and expected outcomes more transparent, which often could be obscured by otherwise essential details. For concreteness of discussion, in the rest of the paper, we will focus on an extension of the inverse seesaw model.

\subsection{Natural $\mu$ term and successful leptogenesis}

Our hybrid seesaw model is based on the following Lagrangian:
\bea
-{\cal L} & \supset & y_{a\alpha}\Psi_{a}^{c}H\ell_{\alpha}+\kappa_{a}\Psi_{a}^{c}\Phi_{\kappa}\Psi_{a}+\lambda_{ia}N_{i}\Phi_{\lambda}\Psi_{a}+\frac{1}{2}M_{N_{i}}N_{i}N_{i}+{\rm H.c.},\label{eq:hybrid_model}
\eea  
where $H$ and $\ell_{\alpha}$ ($\alpha=e,\mu,\tau$) are the SM Higgs and lepton doublets and the $SU(2)_{L}$ contraction is left implicit. The new fermions ${N_{i}}$, $\left(\Psi_{a}^{c},\Psi_{a}\right)$ and the two complex scalars $\Phi_{\kappa}$ and $\Phi_{\lambda}$ are SM gauge singlets. We assume $\Phi_{\kappa}$ and $\Phi_{\lambda}$ have masses and develop vacuum expectation values of order the TeV scale. As a result $\left(\Psi_{a}^{c},\Psi_{a}\right)$ get a mass of that order as well. On the other hand we will take $M_{N_{i}}\gg$ TeV. Our model is therefore the marriage of a TeV scale module ($\Psi,\Psi^c,\Phi_\lambda,\Phi_\kappa$) and a heavy module ($N_i$).

The model in eq.~(\ref{eq:hybrid_model}) represents a very special combination of the standard type I and the inverse seesaw. Specifically, the fermions $N_{i}$ are analogous to those of the type I seesaw, whereas $\left(\Psi_{a}^{c},\Psi_{a}\right)$ play the role of the pseudo-Dirac fermions present in the usual inverse seesaw model of section~\ref{review}. However, to realize our hybrid version of the seesaw it is crucial that there is {\em no} direct coupling between the heavy module $N$ and the SM, i.e. $\ell H$. The heavy sector interacts with the SM only via the TeV module, see figure~\ref{fig:scheme}. This ensures that the virtual exchange of $N_i$ does not generate neutrino masses, but rather a small Majorana mass term for $\Psi_{a}$ (after the scalar $\Phi_{\lambda}$ acquires VEV).

To be more specific, the connection with eqs.~(\ref{inverse1}) and (\ref{inverse2}) in section~\ref{review} or eq.~(\ref{eq:ISS}) in section~\ref{inverse_lepto} can be made clear by noting that integrating out the heavy Majorana singlet $N$, we get 
\bea
\mu_{ab} & = & \sum_{i}\frac{\lambda_{ia}\lambda_{ib}\left\langle \Phi_{\lambda}\right\rangle^{2}}{M_{N_{i}}} \label{eq:mu_hybrid} \\
m_{\Psi_a} & = & \kappa_a \left\langle \Phi_\kappa \right\rangle. 
\eea

Using this in eq.~(\ref{neutrino_mass}), the SM neutrino mass is found to be (dropping flavor indices for simplicity)
\bea\label{eq:neutrino_mass_in_hybrid_model}
m_\nu  \sim  \left[ \frac{(yv)^2}{M_N} \right] \left( \frac{\lambda \langle \Phi_\lambda \rangle}{\kappa \langle \Phi_\kappa \rangle} \right)^2. 
\eea
The first factor in eq.~(\ref{eq:neutrino_mass_in_hybrid_model}) is the usual neutrino mass formula in high-scale type-I seesaw, {i.e., the one we would have obtained {\em had} $N$ directly coupled to $\ell H$.} Instead, here the TeV-scale particles $\Psi,\Psi^c,\Phi_\lambda,\Phi_\kappa$ mediate lepton number violation from the heavy singlet $N$ to the SM sector.
This is the origin of the second factor in the SM neutrino mass formula above, which may thus 
be viewed as a ``modulation'' by TeV-scale physics.

Moving on to the leptogenesis, we have seen in section~\ref{inverse_lepto} that generically models with $C\ll1$ and natural couplings and masses fail to produce the observed baryon asymmetry. Hence, the inverse seesaw should be equipped with a primordial baryon asymmetry. In 
the hybrid model of
eq.~(\ref{eq:hybrid_model}) the latter in fact originates from the decays
\bea
N_i & \rightarrow & \Phi_\lambda \Psi_a.
\eea
These do not induce an asymmetry in the SM $\ell$ directly, but first in $\Psi$. The asymmetry in $\Psi$ is then distributed to $\Psi^c$ via sizable $\kappa$ and then eventually into SM lepton number asymmetry via a large Yukawa $y$ in eq.~(\ref{eq:hybrid_model}). 
Again, just like in the case of neutrino mass generation, we see that the TeV-scale particles ($\Psi, \Psi^c, \Phi_\kappa$ and $\Phi_\lambda$) acts as a mediator of lepton number violation from $N_i$ to the SM, see figure~\ref{fig:scheme}. 
In addition, 
decays of $\Psi, \Psi^c$ can lead to washout of the UV asymmetry.
Thus, this process is completed through an interesting and subtle interplay between physics at UV and IR
(details of which are discussed in the following two sections).
Remarkably, with a single move, we have cured the two most important hurdles of the inverse seesaw model. 
Namely,
the structure of the hybrid seesaw model is such that the small neutrino masses are controlled by the small Majorana mass of $\Psi$ as in the usual inverse seesaw model. The twist here is that the smallness of this Majorana mass is 
 explained by a version of high-scale type I seesaw
 and baryogenesis is then primarily achieved by the decay of the associated heavy fermions (as in the standard type I high-scale seesaw).

There are, however, several aspects that tell us eq.~(\ref{eq:hybrid_model}) is incomplete. (a) We introduced new scalar fields that undergo symmetry breaking phase transition\footnote{As we will discuss in detail later, \emph{dynamical} scalars, as opposed to their VEVs, are required in order to be able to set thermal equilibrium between the SM and the singlet sectors in the early Universe and enable leptogenesis within the singlet sector.} and, given that phenomenologically the size of their VEV needs to be ${O}$(TeV), those scalars suffer from a naturalness problem. Unless we can explain why the new scalars are at the TeV scale, we have provided no convincing explanation of why the neutrinos are light. To achieve this the ultimate theory must therefore be able to solve the hierarchy problem. (b) As we emphasized above, the Lagrangian eq.~(\ref{eq:hybrid_model}) is not the most general one involving the SM and the new fields introduced here. Symmetries or new mechanisms must be invoked in order to avoid other interaction terms ({for example, a direct coupling of $N$ to SM lepton and Higgs or bare Majorana mass terms for $\Psi$, $\Psi^c$}) that would otherwise completely spoil our conclusions. This problem is much easier to solve than the previous one, since the required global symmetries may for example emerge as accidental symmetries of an underlying UV completion of 
eq.~(\ref{eq:hybrid_model}) with gauge symmetries (as demonstrated in appendix \ref{app:gauge_model}).

Interestingly, as shown by some of us in \cite{Agashe:2015izu}, a straightforward attempt at an implementation of {\em high}-scale seesaw mechanism in the framework of warped extra-dimensions/composite Higgs \cite{Huber:2003sf} actually realizes the above-discussed hybrid marriage: remarkably, in that UV completion of eq.~(\ref{eq:hybrid_model}) both issues (a) and (b) are naturally addressed. Warped extra dimensions are known to address the hierarchy problem, thus it is not a big surprise to find that issue (a) can be overcome within that framework. 
Issue (b) can also be nicely addressed due to 5D locality.
In appendix~\ref{app:warped_seesaw} we provide a brief review of warped seesaw and discuss how the two issues (a) and (b) of the 4D hybrid seesaw model eq.~(\ref{eq:hybrid_model}) are elegantly evaded. There we will also see that the TeV-modulation factor in eq.~(\ref{eq:neutrino_mass_in_hybrid_model}) is provided in warped/composite scenarios by the renormalization group evolution and can easily be much larger or smaller than unity even if the underlying theory has no large nor small parameters. What this means is that the bare mass $M_N$ in eq.~(\ref{eq:neutrino_mass_in_hybrid_model}) might even be taken to be comparable to the Planck scale. In that case the warped version of eq.~(\ref{eq:hybrid_model}) is ultimately defined by a unique mass scale of order the Planck scale and couplings of order unity. The ratio TeV$/$Planck and all other small parameters are generated via renormalization group effects.

Rather than focusing on a specific solution of the above issues (a) and (b), in this paper we will take a more model-independent approach and analyze in detail the physics of the low energy picture 
eq.~(\ref{eq:hybrid_model}). In our minds eq.~(\ref{eq:hybrid_model}) should be interpreted as a toy model capturing the main qualitative features of the warped realization or any other UV completion of the hybrid framework.

Before moving on to study the details of our toy model, we mention that the same Lagrangian eq.~(\ref{eq:hybrid_model}) was considered previously in \cite{Aoki:2015owa} with $M_{N_i} = {O} ({\rm TeV})$. From our results, however, $M_{N_i} \gg {O} ({\rm TeV})$ turns out to be a necessary condition to achieve a successful baryogenesis. We will hence not consider that possibility.

\section{Formalism for the hybrid genesis} \label{sec:Formalism}

Our hybrid seesaw model is defined by the Lagrangian in eq.~(\ref{eq:hybrid_model}). Without loss of generality, we work in the basis where $\kappa$ and $M_{N}$ are real positive and diagonal. For definiteness, we have chosen a minimal model $a=1,2,3$ and $i=1,2$ required to explain
two light neutrino mass differences and leptogenesis.\footnote{
In section~\ref{sec:LowMN}, when we consider certain hierarchy in $y$, 
three generation of $N_i$ is required to obtain a realistic neutrino mass matrix.}
The hybrid seesaw model consists of states at high scale, $\sim M_N$, and states at $\sim$ TeV scale. As we will discuss in detail, the entire process of genesis is comprised of two steps: high scale leptogenesis (both generation and washout) at $T\sim M_N$ and low scale washout at $T\sim m_\Psi$. In particular, one important result that we show in section~\ref{subsubsec:Survival of the asymmetry at intermediate temperatures} and section~\ref{subsec:Low-scale-washout} is that seemingly complicated physics at intermediate scales does not induce additional washout of the asymmetry generated at $T \sim M_N$, and hence establishing a clean two-step structure of hybrid genesis.

We begin with a general discussion of generation of asymmetries in particle number based on symmetry argument in section~\ref{subsec:generalities}, which is valid for {\em any} model. In section~\ref{subsec:hybrid-model}, we provide a qualitative assessment of leptogenesis specific to our hybrid seesaw model, and then present a quantitative study in section~\ref{subsec:quantitative}. The formalism developed in this section will be used in section~\ref{results} to map out in detail the parameter space which works for leptogenesis in our hybrid model.

\subsection{Generalities\label{subsec:generalities}}

Before delving into the specifics of our hybrid seesaw model, we provide a brief discussion on generic aspects of the generation of particle number asymmetries viewed from symmetries of the underlying physics. 
Once the Lagrangian is given, all the symmetries (and corresponding charges of fields) of the theory can be analyzed. In particular, all the $U(1)$ symmetries can be identified and as we will demonstrate below, they will play an important role in understanding particle asymmetry generation. 

Viewing each parameter in the Lagrangian as symmetry breaking parameter, by comparing the rates of processes to the Hubble rate, one realizes that the notion of symmetry can be more general than the symmetry of the Lagrangian. Namely, some of $U(1)$ symmetries unseen in the Lagrangian may arise when processes mediated by couplings that break those symmetries are slow compared to the Hubble rate. In this sense, the notion of symmetry in the history of the Universe, now including those already seen from the Lagrangian, are to be understood as temperature dependent concept. In particular, they would be broken or restored, depending on the temperature $T$. Let us take the case of the EW sphaleron processes as an example to illustrate this idea. Due to the mixed $SU(2)_L$ anomaly, the EW sphaleron configuration breaks $B+L$~\cite{tHooft:1976rip} while preserving $B-L$. At high temperatures its rate is given approximately by $\Gamma_{\cancel{B+L}} \sim 250 \alpha_W^5 T$. At temperature $T \gtrsim 10^{12}$ GeV, the EW sphaleron processes are slower than the Hubble rate and hence inactive. In that regime, $U(1)_B$ and $U(1)_L$ are separately good symmetries. At intermediate temperatures, $100$ GeV $\lesssim T \lesssim 10^{12}$ GeV, on the other hand, the EW sphaleron processes are fast and only the $B-L$ remains as a good symmetry. At even lower temperature, $T \lesssim 100$ GeV, the process gets Boltzmann suppressed, $\Gamma_{\rm sph} \propto e^{-E_{\rm sph}/T}$ where $E_{\rm sph} \sim m_W/\alpha_W$, again making both $U(1)_B$ and $U(1)_L$ good symmetries.

To be more specific, now we will define \emph{exact}, \emph{effective} and \emph{approximate} symmetries as follows. 
Let us take $\Gamma_{\cancel{x}}$ to be the rate of a process that violates a specific $U(1)_x$. For example,  the EW sphaleron processes contribute to $\Gamma_{\cancel{B+L}}$. 
At a given temperature $T^*$, $\Gamma_{\cancel x}(T^*)$ falls into one of the following three possibilities: (i) $\Gamma_{\cancel x}(T^*) \gg H(T^*)$; (ii) $\Gamma_{\cancel x}(T^*)  \ll H(T^*)$; and (iii) $\Gamma_{\cancel x}(T^*) \sim H(T^*)$. For case (i), the $x$-violating process is fast enough and thus the corresponding $U(1)_x$ is broken at $T^*$. In case (ii), although the symmetry-violating process exists, it is very slow compared to the Hubble rate. To a good approximation, the corresponding $U(1)_x$ is a good symmetry at $T^*$ and therefore we call it an \emph{effective} symmetry. 
We emphasize that there is a special case in (ii) where $\Gamma_{\cancel x}(T^*) = 0$, meaning there is no $x$-violating process for such $U(1)_x$. Typically, gauged $U(1)$ symmetries like $U(1)_Y$ of the SM will have this property.
For an obvious reason, we call such symmetry as \emph{exact} symmetry. It is crucial to identify exact/effective symmetries because they act as conservation laws at the temperature of interest and determine the spectator effects. Finally, processes of type (iii) have rates of the order of the Hubble rate and are to be described by non-equilibrium dynamics using the Boltzmann equations (BEs). The associated symmetry is special in that it is neither a perfectly good effective symmetry nor gets completely violated.\footnote{For these processes, two out of three Sakharov conditions i.e. the out-of-equilibrium and $U(1)_x$ violation conditions, are met. If the last ingredient i.e. $C$ and $CP$ violation is also met, a nonzero $U(1)_x$ asymmetry can develop.}  Therefore, in the rest of the discussion, we will refer to it as an \emph{approximate} symmetry.

For a particle $i$, we can describe asymmetry in its \emph{number} density as $n_{\Delta i} \equiv n_i - n_{i^*}$ with $n_{i}$ and $n_{i^*}$ respectively the number density of itself and its antiparticle. If they carry a charge $q_i^x$ under a $U(1)_x$, they will contribute to the corresponding \emph{charge} asymmetry:
\beq
n_{\Delta x} = \sum_i q_i^x n_{\Delta i}.
\label{eq:charge_asymmetry}
\eeq
It is shown in refs.~\cite{Antaramian:1993nt,Fong:2015vna} that one can invert the relation above to express $n_{\Delta i}$ in term of $n_{\Delta x}$ as follows\footnote{A pedagogical derivation of this result is given in ref.~\cite{Fong:2015vna}. Explicitly, we have $R_{ix} = g_i \xi_i \sum_{y} q_i^y (J^{-1})_{yx}$ where $J_{yx} = \sum_i g_i \xi_i q_i^x q_i^y$ with $g_i$ the gauge/family degrees of freedom of particle $i$. Also, $\xi_i$ is the statistical factor which goes to 1 (2) for relativistic fermion (boson) and becomes exponentially suppressed for non-relativistic particle.}
\beq
n_{\Delta i} = \sum_x R_{ix} n_{\Delta x},
\label{eq:particle_asymmetry}
\eeq
where $R_{ix}$ can be constructed from charges carried by all the particles under all $U(1)_x$ of a model. Hence the analysis of asymmetry generation which involves various particles and interactions in the thermal bath at certain range of temperature $T^*$ boils down to identifying  \emph{exact}/\emph{effective} or \emph{approximate} $U(1)$ symmetries as discussed above.  
From eq.~\eqref{eq:particle_asymmetry}, it is clear that in the absence of $U(1)$ or only with exact/effective $U(1)$ such that $n_{\Delta x} = 0$ for all $x$, all particle asymmetries vanish. Hence, for successful genesis, it is necessary to have at least one approximate $U(1)$ which allows for $n_{\Delta x} \neq 0$, although existence of such $U(1)$ alone is not a sufficient condition. The actual size of final asymmetry requires further quantitative study of non-equilibrium physics via BEs. In the next section, we will discuss the viability of hybrid-genesis by identifying the \emph{exact/effective} symmetries as well as \emph{approximate} symmetries. The latter allows the development of nonzero asymmetries at a specific temperature regime.

\subsection{Hybrid genesis: qualitative description\label{subsec:hybrid-model}}

We now move on to our hybrid model:
we start by discussing a crucial ingredient which is common to all leptogenesis models, namely, the EW sphaleron processes which communicate the asymmetry in the lepton sector to the baryon sector. In particular, they are active in the temperature range $T_{{\rm EWSp}}^{+}>T>T_{{\rm EWSp}}^{-}$.
The upper bound is estimated to be $T_{{\rm EWSp}}^{+}\sim10^{12}$
GeV \cite{Khlebnikov:1988sr} while the lower bound is determined
from lattice simulation to be $T_{{\rm EWSp}}^{-}=132$ GeV and occurs
after EW phase transition at $T=159$ GeV \cite{DOnofrio:2014rug}.
Generically, the genesis will occur through one of the following two scenarios:
%%%
\begin{enumerate}
\item[(A)]  If high scale genesis takes place and completes at $T_g > T_{{\rm EWSp}}^{+}$, since baryon number $B$ remains to be a good symmetry, 
genesis occurs through generation of an asymmetry in the approximate symmetry $U(1)_L$ (lepton number). We denote $Y_{\Delta L} \equiv n_{\Delta L}/s$
where $n_{\Delta L}$ is lepton charge asymmetry defined as in eq.~\eqref{eq:charge_asymmetry} normalized by entropic density $s=\frac{2\pi^{2}}{45}g_{\star}T^{3}$. Here, $g_{\star}$ is the number of relativistic degrees of freedom of the Universe at temperature $T$.\footnote{
	As we will see later, in our model, we will have to extend the definition of lepton number to include particles beyond the SM. Here and for the rest of the work, we assume all lepton flavors $L_{\alpha}$ are not conserved. This is in accordance with our consideration where all the dimensionless couplings are taken to have `natural' values $\gtrsim{\cal O}\left(10^{-2}\right)$. For instance, taking $\left|y_{a\alpha}\right|\sim 0.05-0.5$,
	lepton flavors are not conserved for $T\lesssim10^{13}-10^{15}$ GeV (for the estimation, one can use the rate calculated in for e.g. ref.
	\cite{Garbrecht:2013urw}). This allows us to assume that the asymmetry is equally distributed 
	among the three lepton flavors, simplifying the analysis.} 

When the temperature drops below
$T_{{\rm EWSp}}^{+}$, while both $L$ and $B$ are no longer conserved
by the EW sphaleron processes, $B-L$ remains conserved
and the asymmetry in this conserved charge is related to the generated lepton asymmetry 
as $Y_{\Delta (B-L)}\left(T<T_{{\rm EWSp}}^{+}\right)=-Y_{\Delta L}\left(T_{g}\right)$. 

\item[(B)] On the other hand, if leptogenesis takes place and completes at $T_{{\rm EWSp}}^{-}<T_{g}<T_{{\rm EWSp}}^{+}$,
instead of $L$, the generation of asymmetry is described directly in terms of $Y_{\Delta (B-L)}\left(T_{g}\right)$.
\end{enumerate}
%%%

Barring the low scale washout that we will discuss later in section
\ref{subsec:Low-scale-washout}, the baryon asymmetry will be frozen at $T_{{\rm EWSp}}^{-}$ and we have\footnote{$Y_{\Delta B}$ includes the contributions of the quarks which are in chemical equilibrium. For instance, if all quarks are in chemical equilibrium, we simply have $Y_{\Delta B} = \sum_a \left(Y_{\Delta Q_a} + Y_{\Delta u_a} + Y_{\Delta d_a} \right)$.}
%%%
\beq
Y_{\Delta B}\left(T_{{\rm EWSp}}^{-}\right) = d\,Y_{\Delta (B-L)},
\label{eq:B-L_to_B}
\eeq
%%%
where $d$ is an order one number which depends on number of relativistic degrees of freedom at $T_{{\rm EWSp}}^{-}$. Assuming only the SM number of relativistic degrees of freedom (excluding the top quark)
at $T_{{\rm EWSp}}^{-}$ after EW symmetry breaking, we have $d=\frac{30}{97}$ 
\cite{Harvey:1990qw} which is the value we will use in this work.

Having understood these two cases separately, it is useful to introduce a new symbol $\Delta$ to denote asymmetry in both cases as follows:
\bea
\Delta = \left\lbrace
\begin{array}{ll}
- \Delta L, \hspace{1.3cm} {\rm scenario \; (A)}, \\
\Delta (B-L), \hspace{0.5cm}  {\rm scenario \; (B)}.
\end{array} \right.
\label{eq:Delta_general}
\eea

In principle, leptogenesis can happen across $T_{{\rm EWSp}}^{+}$. However, such possibility may correspond to a very small portion of the parameter space, and for simplicity, we will not consider this possibility further.
In practice, for $M_{N_{1}}>10^{12}$
GeV, we will assume scenario (A) while for $M_{N_{1}}<10^{12}$
GeV, we will consider scenario (B).

\subsubsection{Symmetries of the hybrid model}\label{subsubsec:Symmetry-hybrid-model}

We now identify the exact/effective $U(1)$ symmetries as well as approximate ones of the hybrid model. 
From the Lagrangian eq.~\eqref{eq:hybrid_model}, we have seven types of fields $\{ H, \Phi_\lambda,$ $\Phi_\kappa, \ell_\alpha, \Psi^c_a, \Psi_a, N_i $\}. Let us first identify exact symmetries of the theory. For this, we note that the Majorana mass of $N_i$ implies that they cannot carry any conserved charge. Together with hypercharge conservation and three interaction terms in eq.~\eqref{eq:hybrid_model}, we get five constraints and have $7 - 1 - 4 = 2$ \emph{exact} $U(1)$ symmetries, provided the scalar potential does not break them (tree-level breaking) and they can be made gauge-anomaly free (loop-level breaking).
These two symmetries are chosen to be $U(1)_{B-L}$ and $U(1)_{\lambda-B}$ with particle charge assignments shown in table \ref{tab:global_charges}. We denote the first one as $U(1)_{B-L}$ since, although it is not exactly the same as $(B-L)$ (accidental) symmetry of the SM, as far as charges of SM particles are concern, it coincides with the baryon minus lepton number of the SM. Notice that $SU(2)_L-SU(2)_L-U(1)_{B-L}$ mixed anomaly vanishes. For this reason, in scenario (B) when EW sphaleron processes are in thermal equilibrium, $U(1)_{B-L}$ remains conserved. Under $U(1)_{\lambda-B}$, the rest of SM particles carry the charges same as $U(1)_{L-B}$. One readily see that $U(1)_{\lambda - B}$ is also free from $SU(2)_L-SU(2)_L-U(1)_{\lambda-B}$ anomaly and hence is also preserved by the EW sphaleron processes.

For later purpose, we will call {\bf fully-symmetric} those realizations in which both $U(1)$ symmetries are preserved. The model based on gauge symmetries presented in appendix \ref{app:gauge_model} is such an example: in that case $U(1)_{B-L}$ is a gauge symmetry while $U(1)_{\lambda-B}$ arises as an accidental global symmetry. 
On the other hand, in the scenario where eq.~\eqref{eq:hybrid_model} originates from warped extra dimension, the two global symmetries are absent since $\Phi_{\kappa,\lambda}$ are identified with a single \emph{real} field --- the dilaton. As we indicated in previous sections, in this case we view eq.~\eqref{eq:hybrid_model} as a good proxy or toy version of would-be effective theory coming from warped extra dimensional theory. Scenarios like these, in which the scalars $\Phi_{\kappa,\lambda}$ are real, will be called {\bf non-symmetric} models. We expect that a study of this case may capture main features of physics of genesis in warped extra dimensional theory.
As we will see in the next section, while detailed dynamics can differ, the difference in the final asymmetry between the fully-symmetric and non-symmetric scenarios is just order one. Finally, in section~\ref{subsubsec:Survival of the asymmetry at intermediate temperatures} we will briefly comment on the case where only one combination of $U(1)_{B-L}$, $U(1)_{\lambda-B}$ is preserved by the scalar potential.

%%%%%%%%%%%%%%%%%%%%%
\begin{table}[t]
	\begin{center}
		\begin{tabular}{c|cc} 
			\rule{0pt}{1.2em}%
			& $U(1)_{B-L}$ & $U(1)_{\lambda-B}$ \\
			\hline
			$\ell_\alpha$ &  $-1$ & $1$   \\
			$\Psi_a$ &  $0$ & $1$   \\
			$\Psi^c_a$ &  $+1$ & $-1$   \\
			$N$ &  $0$ & $0$ \\
			$\Phi_\kappa$ & $-1$ & $0$ \\
			$\Phi_\lambda$ & $0$ & $-1$ 
		\end{tabular}
\caption{Charge assignments of $\ell_\alpha, \Psi_a, \Psi^c_a, \Phi_\kappa, \Phi_\lambda$ under the two global symmetries of the \textbf{fully-symmetric model}. The former coincides with the baryon minus lepton number of the SM particles. Besides the lepton doublet $\ell_\alpha$, we do not show the charges of the rest of SM particles. In the gauge model presented in Appendix \ref{app:gauge_model}, $U(1)_{B-L}$ is a gauge symmetry while $U(1)_{\lambda-B}$ remains an accidental global symmetry. Under $U(1)_{\lambda-B}$, the rest of SM particles carry the charges same as $U(1)_{L-B}$. On the other hand, the two symmetries are absent in the \textbf{non-symmetric model} originated from warped extra dimension since $\Phi_{\kappa,\lambda}$, which are identified with the dilaton, are real.\label{tab:global_charges}}
	\end{center}
\end{table}  
%%%%%%%%%%%%%%%%%%%%%%%

Having identified the exact symmetries, we now move on to finding the approximate ones. Recall that when we counted the number of constraints to figure out the exact $U(1)$'s, we used the fact that $M_N$ disallows charges for $N_i$'s. In the limit $\lambda_{ia} \to 0$ or $M_{N_i} \to 0$, however, a new $U(1)$ emerges. This approximate lepton number is broken by the coexistence of $\lambda_{ia}$ and $M_{N_i}$ and, as we discuss below, it is in this charge that the asymmetry gets generated via high scale genesis. We define such approximate symmetry as the \emph{extended} lepton number $L'$ and the associated $Y_{\Delta L'}$ (hence its $U(1)$ charges) is given as\footnote{The contribution from right-handed charged leptons $e_\alpha$ in eq.~\eqref{eq:YLprime} will be absent if the corresponding charge lepton Yukawa interactions are out of thermal equilibrium for e.g. $T \gtrsim 10^{12}$ GeV~\cite{Nardi:2006fx}.} 
\beq
Y_{\Delta L'} = \sum_\alpha Y_{\Delta \ell_\alpha} + \sum_\alpha Y_{\Delta e_\alpha} 
+ \sum_a Y_{\Delta \Psi_a} - \sum_a Y_{\Delta \Psi_a^c},
\label{eq:YLprime} 
\eeq
where $e_\alpha$ denotes the SM right-handed lepton for a given flavor $\alpha$.
It may be worth mentioning that the above extended lepton number ($L'$) is to be distinguished from the lepton number ($L$) of the SM. Notice, however, that when all heavy states ($\Psi_a$, $\Psi_a^c$ and also $\Phi_\kappa$) eventually disappear from the thermal bath, the two coincide.
Similarly to eq.~(\ref{eq:Delta_general}) defined for general case, we define $\Delta$ for the hybrid model. Since there is little chance for confusion and we will consider only hybrid model from now on, we decided to use the same symbol. 
\bea
\Delta = \left\lbrace
\begin{array}{ll}
- \Delta L', \hspace{1.3cm} {\rm scenario \; (A)} \\
\Delta (B-L'), \hspace{0.5cm}  {\rm scenario \; (B)}.
\end{array} \right.
\label{eq:Delta_hybrid_model}
\eea
The breaking of this approximate symmetry at high temperatures is captured by the following processes: decays and inverse decays $N_i \leftrightarrow \Phi_\lambda \Psi_a, N_i \leftrightarrow (\Phi_\lambda \Psi_a)^*$ and scatterings $\Phi_\lambda \Psi_a \leftrightarrow (\Phi_\lambda \Psi_b)^*$, $\Psi_a \Psi_b \leftrightarrow (\Phi_\lambda \Phi_\lambda)^*$. Below, we discuss how these processes can be studied to understand the generation and washout of the asymmetry in the $\Delta$ charge.

\subsubsection{{High scale leptogenesis ($T\sim M_N$)}\label{subsec:High-scale-leptogenesis}}

The dynamics of genesis at high scale $\sim M_N$ is essentially the same as that of the usual type-I seesaw model: the high scale leptogenesis proceeds via out of equilibrium decay of heavy $N_i$ and if involved couplings provide needed CP-violation, non-zero asymmetry may be generated in approximate $U(1)_{L'}$ charge.

Starting with the generation, asymmetry is created via out-of-equilibrium decay of $N_i$: $N_i \rightarrow \Phi_\lambda \Psi_a, N_i \rightarrow (\Phi_\lambda \Psi_a)^*$. Concretely, $N_i$ decays more often to $\Phi_\lambda \Psi_a$ than to $(\Phi_\lambda \Psi_a)^*$ if these processes occur with CP-violation. A non-zero CP-violation arises through the interference of tree and one-loop diagrams. When this happens, the number density of $\Psi_a$ may be larger than that of $\Psi_a^*$, i.e. non-zero asymmetry in $Y_{\Delta \Psi_a}$ is created.

However, this immediately raises the question of erasing the asymmetry via the inverse decay: $\Phi_\lambda \Psi_a \rightarrow N_i, (\Phi_\lambda \Psi_a)^* \rightarrow  N_i$. Intuitively, if the number density of $\Psi$ is larger than that of $\Psi^*$, the corresponding inverse decay will tend to occur more rapidly than the other, thus coverting more $\Psi$ (and $\Phi_\lambda$) into $N_i$ than $\Psi^*$. This, combined with the above story of decay, then leads to null net asymmetry. Indeed, this reasoning can be shown to be correct if everything happens in equilibrium environment. Namely, it is non-equilibrium condition that enables actual creation of net asymmetry. This condition is met by virtue of the expansion of the Universe. That is, as the temperature cools down below the mass of $N_i$, unlike the decay, the inverse decay becomes Boltzmann suppressed: thermal energy that $(\Psi \Phi_\lambda)$ or $(\Psi \Phi_\lambda)^*$ carries becomes insufficient to create $N_i$ with mass $M_N > T$. When the washout process due to the inverse decay becomes effectively inactive, a net asymmetry can eventually be generated.

However, the inverse decay is not the only washout process to consider. The scattering processes, $\Phi_\lambda \Psi_a \leftrightarrow (\Phi_\lambda \Psi_b)^*$, $\Psi_a \Psi_b \leftrightarrow (\Phi_\lambda \Phi_\lambda)^*$, violate $\Delta$ by two units and can erase the $\Delta$ asymmetry. As is well-known, by unitarity, the on-shell contribution to these scattering amplitude is the same as the inverse decay. For this reason, in order to avoid double-counting, in writing down the BEs in section~\ref{subsec:quantitative} we will treat the inverse decay and \emph{off-shell} part of $\Delta=2$ scattering as separate source of washout.

\subsubsection{Survival of the asymmetry at $T \lesssim M_N$} 
\label{subsubsec:Survival of the asymmetry at intermediate temperatures}

Having discussed generation and standard mechanisms for washout of asymmetry at high scale ($T \gtrsim M_N$), we now move on to the consideration of physics at intermediate scales, $\langle \Phi_\lambda \rangle < T < M_N$, as well as other potentially dangerous washout processes. In principle, these dynamics can erase previously created asymmetry and hence successful genesis necessitates any such washout processes, including those at intermediate scales, to be under control.

%%%%%%%%%%%%%%%%%%%%%%%%%%%%%%%%%%%%%%%%%%
% washout from \Phi_\lambda decay
% when M_{N_1} < m_\Phi (T)
%%%%%%%%%%%%%%%%%%%%%%%%%%%%%%%%%%%%%%%%%%

We first discuss a subtle washout effect that can potentially appear in the {\bf fully-symmetric model}. As we mentioned earlier, there are two exact/effective symmetries $U(1)_{B-L}$ and $U(1)_{\lambda-B}$ in this model, which impose conservation laws for the global charges of $U(1)_{B-L}$ and $U(1)_{\lambda-B}$ in table \ref{tab:global_charges}. In terms of $\Delta$ that we defined in eq.~(\ref{eq:Delta_hybrid_model}), the conservation laws can be expressed as 
%%%
\bea
Y_{\Delta} + \sum_a Y_{\Delta\Psi_a} -Y_{\Delta\Phi_\kappa} &=& 0, \label{eq:U1_B-L} \\
Y_{\Delta} + Y_{\Delta\Phi_\lambda}  &=&0 .  \label{eq:U1_lambda}  
\eea
%%%
If there exists a process depleting $\Phi_\lambda$, e.g. $\Phi_\lambda$ decaying into particles in the model, $Y_{\Delta\Phi_\lambda}$ would vanish and this will result in $Y_{\Delta}\to 0$ due to the conservation of $U(1)_{\lambda-B}$ [see eq.~\eqref{eq:U1_lambda}].
For a concrete example, let's imagine $M_{N_1}\ll M_{N_2}$ and consider that the high scale genesis is mostly done by the decay of $N_2$ while the generation of asymmetry and washout from $N_1$ are negligible. At temperature $T\sim M_{N_2}$, $\Phi_\lambda$ could get a thermal mass $m_{\Phi_\lambda}(T)\sim T\gg M_{N_1}$ and thus the decay $\Phi_\lambda\to N_1\Psi^*$ could be kinematically allowed. In that case, this decay is the dominant depletion process for $\Phi_\lambda$ at temperatures $T \gg \langle \Phi_\lambda \rangle$. 
If such decay is fast compared to Hubble, all asymmetry generated by $N_2$ will then be washed out and hence leptogenesis fails. 

It is also clear from the above example that such washout process can be forbidden assuming $M_{N_i}$ are of the same order and the high scale asymmetry is primarily generated from $N_1$ decay.
In this case when $T\lesssim M_{N_1}$, $\Phi_\lambda$ cannot decay and the high scale asymmetry survives. 
Given that such choice of $M_{N_i}$'s, i.e. no hierarchy, is more natural according to our naturalness criteria, we only consider this for the rest of our discussion.

There are further washout processes that should be taken into account after $\Phi_\lambda$ has acquired a VEV at much lower temperatures of the order $T\sim \langle{\Phi_\lambda}\rangle$. We will discuss these ``\emph{low-scale washout}'' processes partly below and the rest in the next section.

For {\bf non-symmetric model}, both $U(1)_{B-L}$ and $U(1)_{\lambda-B}$ are absent, and $\Phi_\lambda$ cannot carry an asymmetry.  
Thus, once the asymmetry is generated at high scale, in contrast to the \textbf{fully-symmetric model}, one does not have to worry about the washout from depletion of $\Phi_\lambda$ discussed above.

%%%%%%%%%%%%%%%%%%%%%%%%%%%%%%%%%%%%%%%%%%
% breaking of U(1) by scalar potential
% and resulting washout
%%%%%%%%%%%%%%%%%%%%%%%%%%%%%%%%%%%%%%%%%%

Next, we briefly comment on the possibility of a scenario where only one linear combination of $U(1)_{B-L}$ and $U(1)_{\lambda-B}$ survives due to the scalar potential. Such scenario is different from both the fully-symmetric and non-symmetric models and requires a separate consideration. When one combination of two $U(1)$'s is broken, it is possible for $\Phi_\lambda$ to decay and thus erase the primordial asymmetry. One such an example is obtained starting from the two global symmetries in table \ref{tab:global_charges} 
with eq.~\eqref{eq:hybrid_model} but allowing their breaking in the scalar potential. 
For instance, let us consider the potential interaction $\Phi_\lambda \Phi_\kappa^* |\Phi_\kappa|^2$ 
which preserves only $U(1)_{(B-L) + (\lambda -B)}$. In such a case, we have only one conservation law for unbroken $U(1)$, not two, and the statement that the existence of process leading to $Y_{\Delta \Phi_\lambda} \to 0$ implies $Y_{\Delta} \to 0$ is not generally true anymore. Instead, the fate of the asymmetry depends more on details of the dynamics. Still, it is possible to argue that since more breakings tend to enable more asymmetry transferring channel, if dynamics caused by those breakings transmit all the asymmetries eventually into the SM sector before EW sphaleron processes are turned off, genesis fails, assuming zero net primordial asymmetry. This is simply because the sum of net asymmetry in the SM sector plus net asymmetry of the singlet sector is zero by initial condition and asymmetry transmitting dynamics moved all the asymmetries to the SM sector. Importantly, the above statement is regardless of the details of asymmetry transferring physics. As long as they are efficient enough and completed above $T^-_{\rm EWSp}$, it is a correct statement. To illustrate the idea, let us take the example above with $\Phi_\lambda \Phi_\kappa^* |\Phi_\kappa|^2$-term in the scalar potential.
If thermal masses for the scalars satisfy $m_{\Phi_\lambda}(T) > 3 m_{\Phi_\kappa}(T)$, the decay $\Phi_\lambda \to \Phi_\kappa \Phi_\kappa \Phi_\kappa^*$ is allowed. Since $\Phi_\kappa$ couples to $\Psi$ and $\Psi^c$ with unsuppressed coupling $\kappa$, it may decay/scatter into those. Finally, asymmetry stored in $\Psi$ and $\Psi^c$ may get processed to the SM via Yukawa coupling either by decay or scattering process. If all this is done at temperatures above $T^-_{\rm EWSp}$, as per the argument above, we get zero net asymmetry. We will not consider these scenarios anymore, and next go back to the discussion of fully-symmetric and non-symmetric models.

%%%%%%%%%%%%%%%%%%%%%%%%%%%%%%%%%%%%%%%%%%
% washout at intermediate scale
%%%%%%%%%%%%%%%%%%%%%%%%%%%%%%%%%%%%%%%%%%

We finally discuss washouts at scales below $M_{N_i}$. In particular, we will argue that provided above mentioned two subtle (and easy to avoid) washouts are absent and if washout from off-shell $\Delta =2$ scattering is small, then there is no additional washout effects at intermediate scales. Further washouts we need to consider is, therefore, those occurring at TeV scale after scalars get VEVs. Notice that this is quite a remarkable fact in that although physics happens in the entire energy range, the study of genesis can be structured in clean two steps: high scale genesis and low scale washout. 

Integrating out heavy singlets $N_{i}$, the effective Lagrangian at $\langle \Phi_\lambda \rangle < T < M_N$ is given by
%
%
%%%%%%%%%%%%%%%%%%%%%
\begin{eqnarray}
-{\cal L} & \supset & y_{a\alpha}\Psi_{a}^{c}H\ell_{\alpha}+\kappa_{a}\Psi_{a}^{c}\Phi_{\kappa}\Psi_{a}+\sum_{i}\frac{\lambda_{ia}\lambda_{ib}}{M_{N_{i}}}\Phi_{\lambda}\Psi_{a}\Phi_{\lambda}\Psi_{b}+{\rm H.c.}.\label{eq:hybrid_model_eff}
\end{eqnarray}
The dimension five operator violates $L'$ by two units which could contribute to the washout of $Y_\Delta$. The corresponding process at high $T \sim M_N$ is that of $\Delta =2$ scattering mediated by off-shell $N$ and will be taken into account whenever they are relevant (see section~\ref{subsubsec:therm_rate} for details). Since the rate for this process $\Gamma_{\Delta =2}\propto T^3 $, which drops faster than Hubble rate, $\Gamma_{\Delta =2} < H \propto T^2$ is always true at lower temperatures if it is enforced at a high temperature. Namely, requiring $\Gamma_{\Delta =2} < H$ at high temperature guarantees that washout from the above dimension five operator is under control at \emph{all} intermediate temperatures.

%%%%%%%%%%%%%%%%%%%%%%%%%%%%%%%%%%%%%%%%%%
% summary
%%%%%%%%%%%%%%%%%%%%%%%%%%%%%%%%%%%%%%%%%%

To summarize, assuming no depletion of the asymmetry from $\Phi_\lambda$ decay (both kinds discussed above), once all washout processes in high scale ($T \gtrsim M_N$) involving $N$-exchange are taken under control, the preservation of the asymmetry is a robust feature of our model, at least for temperatures at which $\langle\Phi_{\kappa,\lambda}\rangle=0$ (what happens after the scalars have acquired a VEV will be discussed in the next subsection). In the model which arises from gauge symmetry we considered in appendix \ref{app:gauge_model} (fully-symmetric model), this is ensured by assuming high scale asymmetry is generated from the lightest $N$ decay. This is because in the gauge model, at the renormalizable level, no symmetry breaking terms in the scalar potential is allowed by gauge invariance and possible symmetry breaking higher dimensional terms are highly-Planck-suppressed. See appendix \ref{app:gauge_model} for more detail. No new source of asymmetry violation at intermediate temperatures is possible in non-symmetric models. For scenarios with a remnant global symmetry, i.e. ``intermediate models'', the conclusion is however model-dependent.

\subsubsection{Low scale washout ($T\sim m_\Psi$)\label{subsec:Low-scale-washout}}

We define the temperature region $T\lesssim \langle \Phi_{\kappa,\lambda}\rangle$ as low scale or TeV scale.
In principle, a large entropy production during thermal phase transition(s) of $\Phi_\lambda$ and/or $\Phi_\kappa$ can result in undesired dilution of asymmetry generated from high scale. In order to avoid this, throughout the discussion we assume that phase transition is smooth and hence no large entropy production occurs. This ensures no significant dilution of the asymmetry from phase transitions and the only washout out effects we need to consider at low scale are the dynamical processes involving relevant particles discussed below.

Once scalars get VEVs, there can be new kinds of processes generated by the higher dimensional operator in \eqref{eq:hybrid_model_eff} with some or all of scalars set to their VEVs. Washouts mediated by these processes may be significant even after suppressing those with all physical $\Phi_\lambda$ (as we did in section~\ref{subsubsec:Survival of the asymmetry at intermediate temperatures}). Therefore, they need to be treated separately and we will call them as ``low scale washout''.

We first consider an operator with \emph{one} of $\Phi_\lambda$ set to its VEV: $\sim \lambda^2 \langle \Phi_{ \lambda } \rangle { \Phi }_{ \lambda } \Psi_a \Psi_b / M_N$. This operator can generate several kinds of washout dynamics. As we now show, however, each of those new effects are automatically suppressed assuming a large separation of two physical scales: $\langle \Phi_\lambda \rangle \ll M_N$ and $m_\Psi \sim m_{\Phi_{\lambda,\kappa}} \sim \langle \Phi_{\lambda,\kappa} \rangle$. In order to see this more explicitly, we first note that the condition that $\Delta =2$ washout from off-shell scattering at high scale module ($T \sim M_{N_1}$) is small can be expressed schematically as 
\bea
\frac{\lambda_i^4}{16\pi^3} \left. \frac{T^3}{M_{N_i}^2} \right\vert_{T=M_{N_1}} < \sqrt{g_\star} \left. \frac{T^2}{M_{\rm Pl}} \right\vert_{T=M_{N_1}} \;\; \Rightarrow \;\; \frac{\lambda_i^4}{16\pi^3} \left( \frac{M_{N_1}}{M_{N_i}} \right)^2 < \sqrt{g_\star} \frac{M_{N_1}}{M_{\rm Pl}}.
\label{eq:UV_off_sell_scatt_condition}
\eea 
where $i=1,2$ denotes singlet generation. As we will discuss more later, the dominant $\Delta =2$ washout scattering in the UV module comes from off-shell exchange of $N_2$. Above, however, we show the condition for both $N_1$ and $N_2$ by keeping the index $i$ general. We do this because at scales $T \ll M_N$ the local higher dimensional operator $\lambda_i^2 \langle \Phi_\lambda \rangle^2 \Psi^2 / M_i$ will be generated as a result of integrating out both $N_1$ and $N_2$, and yet the effects of the two will appear as a single operator. Assuming no degeneracy of $M_{N_1}$ and $M_{N_2}$, on the other hand, we can safely drop the interference effects and the matching of effects may be done for each rate.

Next, we discuss four leading washout processes that above mentioned dimension 5 operator generates and argue that all of them are rather generically suppressed.
\begin{itemize}
\item[(1)] The inverse decay $\Psi \Psi \to \Phi_\lambda$:
The condition that this process is slower than Hubble rate at $T \sim m_{\Phi_\lambda}$ can be written as
\bea
\frac{\lambda_i^4}{16\pi^3} \left( \frac{M_{N_1}}{M_{N_i}} \right)^2 \left[ \pi^2 \frac{\langle \Phi_\lambda \rangle}{M_{N_1}} \right] < \sqrt{g_\star} \frac{M_{N_1}}{M_{\rm Pl}}.
\eea
where we used $m_{\Phi_\lambda} \sim \langle \Phi_\lambda \rangle$.
Comparing this to eq.~(\ref{eq:UV_off_sell_scatt_condition}), we see that the washout from this inverse decay is a small effect if the quantity in the square bracket is less than one:
\bea
\pi^2 \frac{\langle \Phi_\lambda \rangle}{M_{N_1}} < 1
\label{eq:low_scale_washout_1_inverse_decay}
\eea
and it is clear that with assumed gap $M_{N_1} \gg \langle \Phi_\lambda \rangle$ this condition is easily met.

\item[(2)] $\Psi \Phi_\lambda \to \Psi^c \Phi_\kappa$ and its associated $t$-channel $\Delta =2$ scattering:
Such process may be generated by usage of one factor of $\lambda_i^2 \langle \Phi_\lambda \rangle / M_{N_i}$ from dimension 5 operator above and one factor of $\kappa$. Following similar steps, this washout can be small if
\bea
\kappa^2 \frac{\langle \Phi_\lambda \rangle}{M_{N_1}} \frac{\langle \Phi_\lambda \rangle}{m_\Psi} <1
\label{eq:low_scale_washout_2_Psi_Phi_lambda_to_Psic_Phi_kappa}
\eea
Again, with $M_{N_1} \gg \langle \Phi_\lambda \rangle$, $m_\Psi \sim \langle \Phi_\lambda \rangle$, and $\kappa \sim \mathcal{O} (1)$ that we are assuming, the above condition is easily satisfied.

\item[(3)] $\Psi \Phi_\lambda \to \left( \Psi \Phi_\lambda \right)^*$ and its associated $t$-channel $\Delta =2$ scattering: There are two contributions to be added at the amplitude level. One is from local vertex of $\lambda_i^2 \Phi_\lambda^2 \Psi^2 / M_i$ and we already argued in section~\ref{subsubsec:Survival of the asymmetry at intermediate temperatures} that it is suppressed. The other diagram can be constructed by two factors of $\lambda_i^2 \langle \Phi_\lambda \rangle / M_{N_i}$ and one insertion of $\mu \sim \lambda^2 \langle \Phi_\lambda \rangle^2 / M_{N_1}$. Note, howerver, that the second is much more suppressed compared to the first: in the UV, it corresponds to eight-point correlator with four $\Phi_\lambda$'s set to VEV. 
Therefore, neglecting subdominant latter contribution, it is a robust fact that washout from $\Psi \Phi_\lambda \to \left( \Psi \Phi_\lambda \right)^*$ scattering is suppressed once the corresponding process at UV scale is small.

\item[(4)] scattering $\Psi \Psi \leftrightarrow (\Psi \Psi)^*$ mediated by off-shell $\Phi_{ \lambda }$: Noting that on-shell part of such scattering is the inverse decay $\Psi \Psi \to \Phi_\lambda$ and that off-shell contribution is sub-dominant to the on-shell contribution, it can be safely dropped once the inverse decay is suppressed via eq.~(\ref{eq:low_scale_washout_1_inverse_decay}).

\end{itemize}

With the above discussion, now the only remaining washouts to discuss are when both $\Phi_\lambda$ in \eqref{eq:hybrid_model_eff} have a VEV. Because we will always take $\langle\Phi_{\kappa}\rangle\leq\langle\Phi_{\lambda}\rangle$, we can limit our discussion to temperatures in which both scalars have acquired VEVs. In this regime $\Psi_{a}$ and $\Psi_{a}^{c}$
form three pairs of pseudo-Dirac fermions, $\tilde{\Psi}_{i}\, (i=1,...,6)$ with masses $m_i$. Their mass splitting as well as strength of washout are controlled by eq.~(\ref{eq:mu_hybrid}).
%
%%%
Notice that in this temperature range we can match our hybrid seesaw at low scale to the ISS model eq.~\eqref{eq:ISS}. The scatterings controlled by $\mu$ violate $L'$ and could erase exponentially the asymmetry $Y_{\Delta}$ generated at high scale. The formulas for low scale washout will be presented in section \ref{subsec:quantitative}.

Finally, there are also new washout processes pertaining to the gauge model of appendix \ref{app:gauge_model}.  
Gauge bosons associated to $U(1)_{B-L}$ could mediate new washout processes like $\tilde{\Psi}_i \tilde{\Psi}_i \to f \bar f$ where $f$ is any fermion charged under $U(1)_{B-L}$. 
 However, these processes are suppressed as $\mu^2/T^2$ at $T>m_i$ and as $\mu^2/m_i^2$ for $T$ below the critical temperature at which $\Phi_\kappa$ gets a VEV. Unless we consider highly non-generic models in which the gauge boson mass is $\sim 2 m_i$ these processes are not resonantly enhanced, and therefore do not induce significant washout.

%%%%%%%%%%%%%%%%%%%%%

\subsubsection{{Initial conditions and assumptions}\label{subsec:Initial-conditions}}

In the standard cosmological model, it is assumed that after inflation, 
inflatons decay populate the Universe with particles which 
thermalize among themselves to a so-called reheating temperature.

For $M_N \lesssim 10^{15}$ GeV, the genesis occurs at temperatures where the SM particles could be thermalized by the SM interactions as well as new interactions in our model. There may be a few options for the reheating. When inflatons decay only to the SM particles, SM partcles thermalize themselves through gauge and Yukawa interactions. Then, singlet sector states, $\Psi_a^c$, $\Psi_a$, and $\Phi_\kappa$, can be populated via interactions $y$ and $\kappa$. The singlet scalars $\Phi_\kappa$ and $\Phi_\lambda$ can also be populated through scalar interactions like $|H|^2 |\Phi_{\kappa,\lambda}|^2$. If, on the other hand, inflatons only decay to singlet sector particles, $H$ and $\ell_\alpha$ can be produced from the aforementioned interactions and then through the SM gauge and Yukawa interactions, the rest of the SM particles can be populated. When inflatons reheat both sectors simultaneously, then thermalization happens naturally through various interactions metioned so far. Therefore, we see that regardless of the assumption about the reheating, both sectors will be thermalized and we will consider the contribution to the total energy density of the Universe from  both the SM particles and singlet sector particles: $g_\star = 121.25$.

If genesis takes place at $M_{N_i} \gtrsim 10^{15}$ GeV, on the other hand, the SM particles 
might not be thermalized by the SM interactions~\cite{Enqvist:1993fm}. 
If inflatons decay dominantly to the SM particles, 
we cannot describe this scenario within a standard radiation-dominated thermal bath. 
Since a separate treatment is called for, we will not consider this scenario further.
Instead, we will consider the situation where inflatons decay only to the singlet 
sector particles and they are thermalized through interactions in our model eq.~\eqref{eq:hybrid_model} 
as well as interactions in scalar potential. For instance, interactions with a large $\kappa_a$ 
can thermalize $\Psi_a$ and $\Psi_a^c$ while the scalars $\Phi_\lambda$ and
$\Phi_\kappa$ can be thermalized through interaction like $\left|\Phi_{\kappa}\Phi_{\lambda}\right|^{2}$. When $\Psi_a, \Psi_a^c, \Phi_\lambda$ and $\Phi_\kappa$ are all thermalized 
the total relativistic degrees of freedom is $g_\star = 14.5$\footnote{It is reasonable to assume that the decays of inflatons 
to heavy $N_i$ are kinematically forbidden. Furthermore, since they are not relativistic, 
they do not contribution significantly to the energy density of the Universe.} and we use this number to calculate the high scale genesis. When temperature cools down, interactions involving $y$ and SM interactions will eventually be in equilibrium and thus SM particles are thermalized through coupling to singlets.

Starting from zero initial $N_i$, a thermalized $\Psi_a$ and $\Phi_\lambda$ 
can generate $N_i$ through inverse decays. In the gauge model of appendix \ref{app:gauge_model}, $\Psi_{a}^{c}$ and
$\Phi_{\kappa}$ can also be thermalized by $U(1)_{B-L}$ gauge interaction.
After $U(1)_X$ symmetry breaking at around genesis scale, if
$U(1)_X$ gauge boson is not much heavier than the reheating temperature,
it could also thermalize $N_{i}$, $\Psi_{a}$, $\Phi_{\kappa}$ and
$\Phi_{\lambda}$. Motivated by the above considerations, we will consider two possible
initial conditions for $N_i$ abundance: zero $N_{i}$ abundance and thermalized $N_{i}$
abundance.

%%%%%%%%%%%%%%%%%%%%%

\subsection{Hybrid genesis: quantitative description}\label{subsec:quantitative}
In this section, we will discuss CP violation in section~\ref{subsubsec:CP_violation}, washout processes from inverse decay and off-shell scatterings in section~\ref{subsubsec:therm_rate} and finally in section~\ref{subsubsec:Boltzmann-equations}, we write down the BEs of hybrid genesis. Under reasonable assumptions, the formal solution to the BEs can be written down including both the inverse decay and off-shell $\Delta=2$ scattering. It will be in the form of integral, which can readily be evaluated numerically. On the other hand, keeping only the inverse decay term allows us to derive approximate analytical solutions in appendix.~\ref{app:approx_sols}. In this way, our strategy will be to use these analytical solutions when the washout from off-shell $\Delta =2$ scattering is negligible while evaluate numerically when it is relevant.

%%%%%%%%%%%%%%%%%%%%%%%%%%%%%%%%%%%%%%%%%
% CP-violation
%%%%%%%%%%%%%%%%%%%%%%%%%%%%%%%%%%%%%%%%%

\subsubsection{CP violation}\label{subsubsec:CP_violation}

To quantify the CP violation in the decays of $N_{i}\to \Phi_{\lambda}\Psi_{a}$ and
$N_{i}\to (\Phi_{\lambda}\Psi_{a})^*$, we define the CP parameter as follows [see eq.~(\ref{eq:CP_parameter0})]
%%%
\begin{eqnarray}
\epsilon_{ia} & \equiv & \frac{\Gamma\left(N_{i}\to\Phi_{\lambda}\Psi_{a}\right)-\Gamma\left(N_{i}\to\Phi_{\lambda}^{*} {\Psi}_{a}^*\right)}{\Gamma_{N_{i}}},
\label{eq:CP_parameter}
\end{eqnarray}
%%%
where $\Gamma(P)$ is the partial decay width for process $P$ and the total decay width of $N_{i}$ (at tree-level) is
%%%
\begin{eqnarray}
\Gamma_{N_{i}} & \equiv & \sum_{a}\left[\Gamma\left(N_{i}\to\Phi_{\lambda}\Psi_{a}\right)
+\Gamma\left(N_{i}\to\Phi_{\lambda}^{*}{\Psi}_{a}^*\right)\right]=\frac{\left(\lambda\lambda^{\dagger}\right)_{ii}M_{N_{i}}}{16\pi}.
\label{eq:Gamma_N_1}
\end{eqnarray}
%%%
The leading CP violation in the decays comes from the interference between tree-level and one-loop diagrams and eq.~(\ref{eq:CP_parameter}) can be written down as \cite{Covi:1996wh}
%%%%%%%%%%
\begin{eqnarray}
\epsilon_{ia} & = & \frac{1}{8\pi}\frac{1}{\left(\lambda\lambda^{\dagger}\right)_{ii}}\sum_{j\neq i}{\rm Im}\left[\left(\lambda\lambda^{\dagger}\right)_{ij}\lambda_{ia}\lambda_{ja}^{*}\right]g\left(\frac{M_{N_{j}}^{2}}{M_{N_{i}}^{2}}\right)\nonumber \\
&  & +\frac{1}{16\pi}\frac{1}{\left(\lambda\lambda^{\dagger}\right)_{ii}}\sum_{j\neq i}{\rm Im}\left[\left(\lambda\lambda^{\dagger}\right)_{ji}\lambda_{ia}\lambda_{ja}^{*}\right]\frac{M_{N_{i}}^{2}}{M_{N_{i}}^{2}-M_{N_{j}}^{2}},\label{eq:CP_parameter_explicit}
\end{eqnarray}
where the loop function is given by\footnote{This includes both self-energy and vertex corrections with the first term in the square bracket for the former while the rest of the terms for the latter. A factor of $\frac{1}{2}$ in the self energy term compared to the standard leptogenesis case is due to the fact that $\Phi_{\lambda}$ and $\Psi_{a}$ are singlets instead of doublets under $SU(2)_{L}$. For the same reason, the second term in eq.~(\ref{eq:CP_parameter}) coming from self-energy diagrams also has a factor of $\frac{1}{2}$. Such term becomes CP-invariant once summed over flavor $a$.}
%%%%%%%%%%%%%%%%%
\begin{eqnarray}
g\left(x\right) & = & \sqrt{x}\left[\frac{1}{2}\frac{1}{1-x}+1-\left(1+x\right)\ln\frac{1+x}{x}\right].
\end{eqnarray}
%was %%%%%%%%
Assuming a modest hierarchy, $M_{N_{2}} / M_{N_{1}} \sim$ a few and that the main contribution to asymmetry generation to come from the decays of $N_{1}$, we will expand $g\left(M_{N_{1}}/M_{N_{2}}\right)$ at leading order in $M_{N_{1}}/M_{N_{2}}$. Furthermore we will ignore the $\Psi_{a}$ flavor effect by summing over $a$.\footnote{This is justified assuming a large $\kappa$ in eq.~(\ref{eq:hybrid_model}), which results in fast flavor equilibrating scatterings $\Psi_{a}^{c}\Psi_{a}\leftrightarrow\Psi_{b}^{c}\Psi_{b}$. See section \ref{subsubsec:Boltzmann-equations} for further discussion.\label{fn:flavor}} Doing so, we have
\begin{eqnarray}
\epsilon_{1} & \equiv & 
\sum_a \epsilon_{1a} 
\approx -\frac{1}{8\pi}\frac{1}{\left(\lambda\lambda^{\dagger}\right)_{11}}
\left|\left(\lambda\lambda^{\dagger}\right)_{12}\right|^{2}\sin\left(\phi_{12}\right)\frac{M_{N_{1}}}{M_{N_{2}}} \no \\ 
& \approx & -\frac{1}{8\pi}\left(\lambda\lambda^{\dagger}\right)_{22}\sin\left(\phi_{12}\right)\frac{M_{N_{1}}}{M_{N_{2}}},
\label{eq:CP_approx}
\end{eqnarray}
%%%
where we have defined $\phi_{12}\equiv\arg\left[\left(\lambda\lambda^{\dagger}\right)_{12}^{2}\right]$. In the last approximation above (and the rest of the work), we assume
%%%
\begin{eqnarray}
\lambda_{ia}\sim\lambda_{ib},
\label{eq:anarc}
\end{eqnarray}
%%%
for any $a,b$ and $i$. However, we will allow $\lambda_{ia}/\lambda_{ja}$ with $i\neq j$ to vary within a few orders of magnitude.

%%%%%%%%%%%%%%%%%%%%%%%%%%%%%%%%%%%%%%%%%
% basics of BE and 
% thermal averaged rate
%%%%%%%%%%%%%%%%%%%%%%%%%%%%%%%%%%%%%%%%%

\subsubsection{Thermal averaged reaction densities}\label{subsubsec:therm_rate}

Here we will describe the thermal averaged reaction densities [defined in eq.~(\ref{eq:thermal_reaction_densities})] which appear in the BEs to describe the decay and scattering processes (see appendix~\ref{app:BE_gen} for details).

In the following, we assume $N_i$ to be massive while all other particles to be massless. Firstly, we will consider the (inverse) decay $N_1\leftrightarrow \Phi_{\lambda}\Psi_{a}$ (and the corresponding CP conjugate processes).\footnote{This is equivalent to the on-shell part of $\Delta =2$ scattering $\Phi_{\lambda}\Psi_{a}\leftrightarrow (\Phi_{\lambda} {\Psi}_{a})^*$ mediated by $N_1$ where the subdominant off-shell contribution can be ignored.} 
From eqs.~(\ref{eq:decay_rec_den}) and (\ref{eq:total_decay_rec_den}), the total decay reaction density is
%%%
\bea
\gamma_{N_1}
= n_{N_1}^{\rm eq} \Gamma(N_1 \to \Phi_{\lambda}\Psi_{a})
\frac{{\cal K}_1(z)}{{\cal K}_2 (z)},
\label{eq:gamma_N1}
\eea
%%%
where $z=M_{N_1}/T$ and ${\cal K}_n(z)$ is the modified Bessel function of the second kind of order $n$. The decay term is proportional to $\frac{n_{N_1}}{n_{N_1}^{\rm eq}} \gamma_{N_1}$ which is not Boltzmann suppressed. On the other hand, inverse decay is simply proportional to $ \gamma_{N_1}$ and will be Boltzmann suppressed for $T < M_{N_1}$. If $\Gamma_{N_1} \ll H(T=M_{N_1})$, the decay happens at $\Gamma_{N_1} \sim H(T \ll M_{N_1})$ when the inverse decay is Boltzmann suppressed. In this case, the asymmetry is efficiently generated while washout due to inverse decay is suppressed. The degree of out-of-equilibrium decay for $N_i \to \Phi_{\lambda}\Psi_{a}$ is usually quantified by the washout factor already introduced in eq.~(\ref{eq:K})
\begin{eqnarray}\label{eq:K_i}
K_i & \equiv & \frac{\Gamma_{N_{i}}}{H (T=M_{N_i})}. \label{eq:washout_factor}
\end{eqnarray}
The case of $K_{i}<1$ is known as the weak washout regime (washout of the asymmetry from the inverse decay is not effective) while $K_{i}>1$ as the strong washout regime (washout of the asymmetry from the inverse decay becomes relevant).

Next, as we maximize the CP parameter in eq.~\eqref{eq:CP_approx} by increasing $\lambda_{2a}$, 
the $\Delta = 2$ scatterings $\Phi_{\lambda}\Psi_{a}\leftrightarrow (\Phi_{\lambda} {\Psi}_{b})^*$ and $\Psi_{a}\Psi_{b}\leftrightarrow (\Phi_{\lambda} \Phi_{\lambda})^*$ from off-shell exchange of heavier $N_2$ can become relevant. We estimate the scattering rate for the processes above for $T\sim M_{N_{1}}\ll M_{N_{2}}$ as follows [see eqs.~(\ref{eq:scatt_rec_den}) and (\ref{eq:scatt_rate_gen})]
%%%
\bea
\Gamma_{\rm scatt}^{ab} 
= 
\frac{\gamma_{\rm scatt}^{ab}}{n^{\rm eq}} 
\approx
\frac{1}{16 \pi^{3}}
\frac{\left|\lambda_{2a}\right|^{2}\left|\lambda_{2b}\right|^{2}}{M_{N_2}^{2}}T^{3},
\label{eq:scatt_rate_N2}
\eea
%%%
where $n^{\rm eq} = \frac{T^3}{\pi^2}$. With the assumption eq.~(\ref{eq:anarc}), we can relate the scattering rate above to the CP parameter eq.~\eqref{eq:CP_approx} as follows
%%%
\bea
\Gamma_{\rm scatt}^{ab} &\approx& \frac{4}{\pi}\frac{\epsilon_{1}^{2}}{\sin^{2}\left(\phi_{12}\right)}
\frac{M_{N_1}}{z^3}.
\label{eq:scatt_rate}
\eea
%%%
From the above, it becomes clear that as one increases the CP parameter $\epsilon_1$, the scattering washout rate will increase and vice versa. As we will see in section~\ref{sec:LowMN}, requiring this washout scattering to be under control in general implies a lower bound on $M_{N_1}$~\cite{Racker:2013lua}.

%%%%%%%%%%%%%%%%%%%%%%%%%%%%%%%%%%%%%%

\subsubsection{Boltzmann equations}\label{subsubsec:Boltzmann-equations}

We now study in detail the generation/washout of the asymmetry at $T\sim M_{N_1}$ and the low scale washout at $T\sim m_{\Phi_{\kappa,\lambda}}$.
The high scale genesis can be described by the following BEs:
\begin{eqnarray}
sH z_{\scalebox{0.5}{UV}} \frac{dY_{N_{1}}}{d z_{\scalebox{0.5}{UV}}} & = & -\gamma_{N_{1}}\left(\frac{Y_{N_{1}}}{Y_{N_{1}}^{{\rm eq}}}-1\right), \label{eq:BE_UV_1}\\
sH z_{\scalebox{0.5}{UV}} \frac{dY_{\Delta}}{d z_{\scalebox{0.5}{UV}}} & = & - \epsilon_{1}\gamma_{N_{1}}\left(\frac{Y_{N_{1}}}{Y_{N_{1}}^{{\rm eq}}}-1\right) + \frac{1}{2}\sum_{a}P_{a}\gamma_{N_{1}}\left(\frac{Y_{\Delta\Psi_{a}}}{Y^{{\rm eq}}} + \frac{1}{2} \frac{Y_{\Delta\Phi_{\lambda}}}{Y^{{\rm eq}}}\right)\nonumber \\
 &  & + \sum_{a}\gamma_{{\rm scatt}}^{aa}\left(\frac{Y_{\Delta\Psi_{a}}}{Y^{{\rm eq}}}
+ \frac{1}{2} \frac{Y_{\Delta\Phi_{\lambda}}}{Y^{{\rm eq}}}\right) \nonumber \\
& & + \sum_{a}\sum_{b\neq a}\gamma_{{\rm scatt}}^{ab} \left( \frac{1}{2} \frac{Y_{\Delta\Psi_{a}}}{Y^{{\rm eq}}}
+ \frac{1}{2} \frac{Y_{\Delta\Psi_{b}}}{Y^{{\rm eq}}}+ \frac{1}{2} \frac{Y_{\Delta\Phi_{\lambda}}}{Y^{{\rm eq}}} \right), \label{eq:BE_UV_2}
\end{eqnarray}
where $z_{\scalebox{0.5}{UV}} \equiv M_{N_{1}}/T$, $Y_{\Delta i}\equiv Y_{i}-Y_{i^*}$,  $P_{a}\equiv\frac{\lambda_{1a}\lambda_{1a}^{*}}{\left(\lambda\lambda^{\dagger}\right)_{11}}$ and $Y^{\rm eq}$ is defined in eq.~(\ref{eq:Y_eq}). The rates $\gamma_{N_1}$ and $\gamma^{ab}_{\rm scatt}$ are defined in eq.~(\ref{eq:gamma_N1}) and eq.~(\ref{eq:scatt_rate_N2}), respectively. 
We also used the fact that in our case, both $\Psi_{a}$ and
$\Phi_{\lambda}$ are relativistic $\zeta_{\Psi_{a}}=1$, $\zeta_{\Phi_{\lambda}}=2$
and have one gauge degree of freedom $g_{\Psi_{a}}=g_{\Phi_{\lambda}}=1$. 
Here we take $g_{\star}=121.25$ since $\Psi_{a}$, $\Psi_{a}^{c}$,
$\Phi_{\kappa}$ and $\Phi_{\lambda}$ all contribute to number of relativistic degrees of freedom.

We now discuss each terms in eq.~(\ref{eq:BE_UV_1}) and eq.~(\ref{eq:BE_UV_2}). In eq.~(\ref{eq:BE_UV_1}), which describes the evolution of the number density of $N_1$, the first term on the right hand side is the reduction of number density by decay and the second term is the production of $N_1$ via inverse decay. In principle, several scattering terms that produce/remove $N_1$ appear on the right hand side of this equation and we ignore these subleading terms. Moving onto eq.~(\ref{eq:BE_UV_2}), this equation determines evolution of asymmetry $\Delta$. The first term on the right hand side proportional to the CP parameter describes the production of asymmetry via out-of-equilibrium decay of $N_1$. The remaining terms are for washout processes: the second, third, and fourth term respectively denoting washout from inverse decay, $s$-channel $\Delta =2$ scattering $\Psi \Phi_\lambda \to \left( \Psi \Phi_\lambda \right)^*$ and its related $t$-channel process with \emph{same} flavor $\Psi$, and the same scattering but with \emph{different} $\Psi$ flavors $a$ and $b \neq a$.

Assuming that all $\Psi_{a}^{c}\Psi_{a}\leftrightarrow\Psi_{b}^{c}\Psi_{b}$
are in thermal equilibrium due to large $\kappa$, the asymmetry will
be equally distributed among all generations of the $\Psi_{a}$. In this case we have $Y_{\Delta\Psi_{1}}=Y_{\Delta\Psi_{2}}=Y_{\Delta\Psi_{3}}\equiv\frac{1}{3}Y_{\Delta\Psi}$
where $Y_{\Delta\Psi}=Y_{\Delta\Psi_{1}}+Y_{\Delta\Psi_{2}}+Y_{\Delta\Psi_{3}}$.
With this assumption, we can sum over flavor in the scattering rate $\gamma_{{\rm scatt}}\equiv\sum_{a,b}\gamma_{{\rm scatt}}^{ab}$ and eq.~(\ref{eq:BE_UV_2}) can be simplified to 
\begin{eqnarray}\label{1}
sH z_{\scalebox{0.5}{UV}} \frac{dY_{\Delta}}{d z_{\scalebox{0.5}{UV}}} & = & - \epsilon_{1}\gamma_{N_{1}}\left(\frac{Y_{N_{1}}}{Y_{N_{1}}^{{\rm eq}}}-1\right) 
+ \frac{1}{2}\gamma_{N_{1}}\left(\frac{Y_{\Delta\Psi}}{3Y^{{\rm eq}}} + \frac{Y_{\Delta\Phi_{\lambda}}}{2 Y^{{\rm eq}}}\right)\nonumber \\
 &  & + \gamma_{{\rm scatt}}\left(\frac{Y_{\Delta\Psi}}{3Y^{{\rm eq}}} + \frac{Y_{\Delta\Phi_{\lambda}}}{2 Y^{{\rm eq}}}\right),
\end{eqnarray}
where we have made use of $\sum_{a}P_{a}=1$.  

In order to solve the equations above in closed form, we need to express
$Y_{\Delta\Psi}$ and $Y_{\Delta\Phi_{\lambda}}$ in term of $Y_{\Delta}$. According to symmetry consideration we presented in section~\ref{subsec:generalities}, all particle asymmetries can be relate to the charge $Y_{\Delta}$ [see eq.~\eqref{eq:particle_asymmetry}]. So, we write $Y_{\Delta \Psi} = - c_\Psi Y_\Delta$ and $Y_{\Delta \Phi_\lambda} = - c_{\Phi_\lambda} Y_\Delta$, with $c_\Phi, c_{\Phi_\lambda} > 0$. In terms of these, we can rewrite eq.~\eqref{1} as 
%%%%%%%%%%%%
\begin{eqnarray}
sH z_{\scalebox{0.5}{UV}} \frac{dY_{\Delta}}{d z_{\scalebox{0.5}{UV}}} & = & - \epsilon_{1}\gamma_{N_{1}}\left(\frac{Y_{N_{1}}}{Y_{N_{1}}^{{\rm eq}}}-1\right) - \left(\frac{1}{2}\gamma_{N_{1}}+\gamma_{{\rm scatt}}\right)\left(c_{\Psi}\frac{Y_{\Delta}}{3Y^{{\rm eq}}}+c_{\Phi_{\lambda}}\frac{Y_{\Delta}}{2 Y^{{\rm eq}}}\right)\nonumber \\
 & \equiv & - \epsilon_{1}\gamma_{N_{1}}\left(\frac{Y_{N_{1}}}{Y_{N_{1}}^{{\rm eq}}}-1\right) - \left(\frac{1}{2}\gamma_{N_{1}}+\gamma_{{\rm scatt}}\right)c_{W1}\frac{Y_{\Delta}}{Y^{{\rm eq}}},\label{eq:BE_Yq}
\end{eqnarray}
where we have defined
%%%
\begin{eqnarray}\label{eq:defc}
c_{W1} & \equiv & \frac{c_{\Psi}}{3}+\frac{c_{\Phi_{\lambda}}}{2}. 
\end{eqnarray}
%%%
The coefficient $c_{W1}$ will be determined by the symmetries of the model and by the particle contents in the thermal bath
%also the species of particles which are in thermal equilibrium 
(see Appendix \ref{app:Spectator-effects} for details). 
Formally, the solution to eq.~\eqref{eq:BE_Yq} is given by 
%%%
\begin{eqnarray}
Y_{\Delta} \left( z_{\scalebox{0.5}{UV}} \right) 
& = & Y_{\Delta}\left( z_{\scalebox{0.5}{UV},i} \right)e^{-\frac{c_{W_1}}{Y^{{\rm eq}}}\int_{z_{\scalebox{0.3}{UV},i}}^{z_{\scalebox{0.3}{UV}}}dz'W\left(z'\right)}
+ \epsilon_1 \int_{z_{\scalebox{0.3}{UV},i}}^{z_{\scalebox{0.3}{UV}}}dz'\frac{dY_{N_1}}{dz'}e^{-\frac{c_{W_1}}{Y^{{\rm eq}}}\int_{z'}^{z_{\scalebox{0.3}{UV}}}dz''W\left(z''\right)},
\label{eq:formal_solution}
\end{eqnarray}
%%%
where $z_{\scalebox{0.5}{UV},i}$ is the initial temperature, $W(z) \equiv \frac{1}{sHz}\left(\frac{1}{2}\gamma_{N_{1}}+\gamma_{{\rm scatt}}\right)$ is the total washout factor
and $Y_{\Delta}(z_{\scalebox{0.5}{UV},i})$ is a preexisting asymmetry. 
The approximate solution including only decays and inverse decays are presented in appendix \ref{app:approximate_solutions} and can be summarized as 
%%%
\bea\label{eq:Y_q_zf}
Y_{\Delta} (z_{\scalebox{0.5}{UV}} \to \infty)= \epsilon_1 \eta_{N_1} Y_{N_1}^{\textrm{eq}} (0)
\eea
%%%
where $\eta_{N_1}$ is the efficiency factor : 
%%%
\bea\label{eq:eta_N_1}
\eta_{N_1}\sim 
\left\{
\begin{array}{l}
1/(K_1\ln K_1)  ~~ {\rm for} ~ K_1 \gg 1 ~ [\textrm{eq.~\eqref{eq:eff_strong}}]\\
K^2_1 ~~~~~~~~~~~~~~~ {\rm for} ~ K_1 \ll 1 ~ \textrm{with zero initial $Y_{N_1}$ ~[eq.~\eqref{eq:eff_weak_a}}]\\
O(1)~~~~~~~~~~~~~ {\rm for} ~ K_1 \ll 1 ~\textrm{with thermal initial $Y_{N_1}$ ~[eq.~\eqref{eq:eff_thermal}}]\\
 \end{array}
 \right. 
\eea
%%%
and $K_1$ is defined in eq.~(\ref{eq:K_i}). When off-shell $\Delta = 2$ scatterings are relevant, we took into account their effects numerically.

Next we will study the effect of low scale washout. For simplicity, we assume that all three pseudo-Dirac pairs $\tilde{\Psi}_i$ have comparable masses and denote the common mass scale as $m_{\Psi}$. With this assumption BE's for all three pairs of $\tilde{\Psi}_{i}$ can be written with a common $z_{\scalebox{0.5}{IR}} \equiv\frac{m_{\Psi}}{T}$. We will work up to leading order in the mass difference. At low scale $T\sim m_{\Psi}$,
the washout is described by (here we ignore the asymmetry generation from $\Psi$ decay as it is shown to be negligible in section~\ref{inverse_lepto})
%%%%%%%%%%%%%%%
\begin{eqnarray}
sH z_{\scalebox{0.5}{IR}} \frac{dY_{\Delta}}{d z_{\scalebox{0.5}{IR}}} & = & \frac{1}{2}\gamma_{\Psi}\left(\frac{Y_{\Delta\ell}}{6Y^{{\rm eq}}}+\frac{Y_{\Delta H}}{4 Y^{{\rm eq}}}\right)\nonumber \\
 & = & -\frac{1}{2}\gamma_{\Psi}\left(c_{\ell}\frac{Y_{\Delta}}{6 Y^{{\rm eq}}}
+c_{H}\frac{Y_{\Delta}}{4 Y^{{\rm eq}}}\right)\equiv-\frac{1}{2}c_{W2}\gamma_{\Psi}\frac{Y_{\Delta}}{Y^{{\rm eq}}},\label{eq:BE_washout}
\end{eqnarray}
where $\gamma_{\Psi}$ describes the washout from $\ell H\leftrightarrow\bar{\ell}H^{*}$
with on-shell pseudo-Dirac fermions $\tilde{\Psi}_i$ and we used $g_{\ell}=g_{H}=2$ and $\zeta_{\ell}=1$, $\zeta_{H}=2$. Also, as we already did above, we wrote $Y_{\Delta \ell} = - c_\ell Y_\Delta$ and $Y_{\Delta H} = - c_H Y_\Delta$, with $c_\ell, c_H \geq 0$. We have also assumed lepton flavors to equilibrate such that $Y_{\Delta\ell_{e}}=Y_{\Delta\ell_{\mu}}=Y_{\Delta\ell_{\tau}}\equiv\frac{1}{3}Y_{\Delta\ell}$ where $Y_{\Delta\ell}=Y_{\Delta\ell_{e}}+Y_{\Delta\ell_{\mu}}+Y_{\Delta\ell_{\tau}}$. We have defined $c_{W2}\equiv$ $\frac{c_{\ell}}{6}+\frac{c_{H}}{4}$ and here we take $g_{\star}=106.75$ since $\Psi_{a}$, $\Psi_{a}^{c}$, $\Phi_{\kappa}$ and $\Phi_{\lambda}$ no longer contribute to number of relativistic degrees of freedom. For $T_{{\rm EWsp}}^{-}<T\lesssim10^{4}$ GeV, we have $c_{\ell}=\frac{42}{79}$, $c_{H}=\frac{16}{79}$ and $c_{W2}=\frac{11}{79}$. Since the eq.~(\ref{eq:BE_washout}) is homogeneous, the solution can be obtained straightforwardly (or equivalently by keeping the first term of eq.~\eqref{eq:formal_solution} and setting the second term to zero)
%%%%%%%%%%%%%%%
\begin{eqnarray}
Y_{\Delta}\left( z_{\scalebox{0.5}{IR},f} \right) & = & Y_{\Delta}\left( z_{\scalebox{0.5}{IR},i} \right)e^{-\frac{6}{\pi^{2}}c_{W2}K_{\Psi}^{{\rm eff}}f( z_{\scalebox{0.3}{IR},i} , z_{\scalebox{0.3}{IR},f} )},
\label{eq:YDelta_lowscale}
\end{eqnarray}
%%%%%%%%%%%%%%%%%
where $z_{\scalebox{0.5}{IR},i}$ and $z_{\scalebox{0.5}{IR},f}$ denote respectively the initial and final
temperatures within the IR physics for which eq.~(\ref{eq:BE_washout}) is solved. The function $f(z_{\scalebox{0.5}{IR},i},z_{\scalebox{0.5}{IR},f})$ is defined in eq.~(\ref{eq:f_function}).
In addition, the low scale effective washout factor is defined as
[c.f. $T \ll M_{N_1}$ eq.~(\ref{eq:Keff})]
%%%%%%%%%%%%%%%%%
\begin{eqnarray}
K_{\Psi}^{{\rm eff}} & \equiv & \left.\sum_{a}\frac{\Gamma_{\Psi_{a}}}{H}\left(\frac{\left|\mu_{aa}\right|}{\Gamma_{\Psi_{a}}}\right)^{2}\right|_{T=m_{\Psi_a}}.
\label{eq:Keff2}
\end{eqnarray}
where $\Gamma_{\Psi_{a}}\equiv \frac{1}{16\pi}(yy^\dagger)_{aa} m_{\Psi_{a}}$.
%%%%%%%%%%%%%%%%%
Since for our case $M_{N_{1}}\gg m_{\Psi}$, we can take $z_{\scalebox{0.5}{IR},i} \to 0$. The initial abundance of $Y_{\Delta} (z_{\scalebox{0.5}{IR},i})$ of the IR solution is obtained from the final asymmetry $Y_{\Delta} (z_{\scalebox{0.5}{UV},f} \to \infty)$ of the UV genesis. Namely,
\bea
Y_{\Delta}\left( z_{\scalebox{0.5}{IR},i} \right)= Y_{\Delta} \left( z_{\scalebox{0.5}{UV},f} \right).
\eea
As for $z_{\scalebox{0.5}{IR},f}$, it is bounded by $T_{{\rm EWsp}}^{-}=132$ GeV where
EW sphaleron processes cease to be effective. If $m_{\Psi}\gg T_{{\rm EWsp}}^{-}$,
we can take $z_{\scalebox{0.5}{IR},f} \to\infty$ and use $f(0,\infty)=\frac{3\pi}{2}$. 
Already for $m_{\Psi}=1$ TeV, we have $f(0,\frac{1000}{132})=4.564$
which is only about 3 \% different from $\frac{3\pi}{2}$.
On the other hand, taking $m_{\Psi}=500$ GeV, we have $f(0,\frac{500}{132})=3.058$.
After the low scale washout, when the EW sphaleron processes get out
of equilibrium at $T_{{\rm EWSp}}^{-}=132$ GeV, 
the baryon asymmetry is frozen as given in eq.~\eqref{eq:B-L_to_B}.
To summarize, the complete formula for the baryon asymmetry generated in our hybrid seesaw model is given as
\bea
Y_{\Delta B} =d\times \epsilon_1 \eta_{N_1} Y_{N_1}^{\textrm{eq}} (0) \times e^{-\frac{6}{\pi^{2}}c_{W2}K_{\Psi}^{{\rm eff}}f(z_{\scalebox{0.3}{IR},i},z_{\scalebox{0.3}{IR},f})}.
\eea
This equation clearly demonstrates the interplay of the high scale and low scale physics in hybrid-genesis. The part $ \epsilon_1 \eta_{N_1} Y_{N_1}^{\textrm{eq}} (0)$ shows the generation of the asymmetry from the high scale $N_1$ decay [eq.~\eqref{eq:Y_q_zf}], while the exponential factor encodes the washout effect from the low scale [eq.~\eqref{eq:YDelta_lowscale}]. The coefficient $d$ is the factor related to the EW sphaleron processes shown in eq.~\eqref{eq:B-L_to_B}.

%%%%%%%%%%%%
\section{Results}

\label{results}

In this section we will use the formalism developed in section~\ref{sec:Formalism} to identify the region of parameter space of our model that accounts for the observed baryon asymmetry. Most of the results of this section are based on {\em analytic} expressions derived in section~\ref{sec:Formalism} and appendix \ref{app:approximate_solutions}; however, when the washout effects from off-shell scattering become significant we used numerical methods. Fortunately, most of the plots given below can be understood analytically. For the readers' convenience, we provide a list of formulae relevant to leptogenesis in table~\ref{tab:summary_formulae}. In particular, we show the parametric dependence of each quantity on the two main parameters of the effective theory in the IR, i.e., $m_\Psi$ and $y$, and present most of the plots in the plane $y-m_\Psi$.

We discuss the results only for our \textbf{fully-symmetric model} for concreteness. Noticing that the main difference between \textbf{fully-symmetric} and \textbf{non-symmetric} models are the existence of exact $U(1)$ symmetries and that this will mainly lead to a difference in spectator effects, we conclude based on the argument in appendix~\ref{app:Spectator-effects} that their final asymmetry will differ only up to an order one factor. 

%%%%%%%%%%%%%%%%%%%%%%%%%%

\afterpage{
%%%%%%%%%%%%%%%%%%%%%%
\begin{landscape}
\begin{table}[t]
\begin{adjustwidth}{0cm}{}
\centering
\renewcommand{\arraystretch}{1.6}
\setlength{\arrayrulewidth}{0.3 mm}
\begin{tabular}{|c|c|c|c|c|}
\hline 
\multicolumn{5}{|c|}{SM neutrino mass in hybrid seesaw} \\
\hline
No. & Quantity & Expression & Dependence on model parameters & Ref. \\
\hline
\hypertarget{eq:T1}{T1}  & Neutrino mass & $m_\nu \sim \frac{y^2 v^2}{m_{\Psi}^2} \mu$  &  $\mu \sim \frac{m_\nu}{v^2} \left( \frac{m_{\Psi}^2}{y^2} \right)$ & eq.~(\ref{MajoranaNu}) \\
\hypertarget{eq:T2}{T2} & $\mu$ & $\mu \sim \frac{\lambda_2^2 \langle \Phi_\lambda \rangle^2}{M_N}$ & $\lambda_2^2 \sim \frac{m_\nu M_N}{v^2 \langle \Phi_\lambda \rangle^2} \left( \frac{m_{\Psi}^2}{y^2} \right)$ & eq.~(\ref{eq:mu_hybrid}) \\
\hline \hline
\multicolumn{5}{|c|}{Leptogenesis in high scale module} \\
\hline
No. & Quantity & Expression & Dependence on model parameters & Ref. \\
\hline
\hypertarget{eq:T3}{T3}  & CP parameter & $\epsilon_{1} 
\sim \frac{\lambda_2^2}{8 \pi} $
& $\epsilon_1 \sim \frac{m_\nu M_N}{8 \pi v^2 \langle \Phi_\lambda \rangle^2} \left( \frac{m_{\Psi}^2}{y^2} \right)$ & eq.~(\ref{eq:CP_approx}) \\
\hypertarget{eq:T4}{T4} & Washout factor & $K_{1} \equiv \frac{\Gamma_{N_1}}{H(T=M_{N_1})}\sim \frac{\lambda_1^2}{16 \pi \sqrt{g_\star}} \frac{M_{\rm pl}}{M_N} $ & $K_1 \sim \frac{r^2 M_{\rm pl} m_\nu}{16 \pi \sqrt{g_\star} v^2 \langle \Phi_\lambda \rangle^2} \left( \frac{m_\Psi^2}{y^2} \right)$ & eq.~(\ref{eq:washout_factor}) \\
\hypertarget{eq:T5}{T5}& Asymmetry in strong washout & $Y^s_{\Delta}(z_{\scalebox{0.5}{UV},f})  \sim 10^{-3} \frac{\epsilon_{1}} {K_1 \ln K_1}$ & $Y^s_\Delta (z_{\scalebox{0.5}{UV},f}) \sim \frac{\sqrt{g_\star}}{r^2} \frac{M_N}{M_{\rm pl}} /\ln K_1$ & eq.~(\ref{eq:eff_strong}) \\
\hypertarget{eq:T6}{T6}& \multirow{2}{*}{Asymmetry in weak washout} 
 & $Y^w_{\Delta} (z_{\scalebox{0.5}{UV},f}) \sim 10^{-3} \epsilon_{1} K_1^2$ ($Y_{N_1} (z_{\scalebox{0.5}{UV},i})=0$) &$Y^w_\Delta (z_{\scalebox{0.5}{UV},f}) \sim \frac{r^4 M_{\rm pl}^2 m_\nu^3 M_N}{g_\star v^6 \langle \Phi_\lambda \rangle^6} \left( \frac{m_\Psi^6}{y^6} \right)$ & eq.~(\ref{eq:eff_weak_a}) \\
\hypertarget{eq:T7}{T7}&   &$Y_{\Delta}^{\rm th, w} (z_{\scalebox{0.5}{UV},f})\sim 10^{-3} \epsilon_{1} $ ~(thermal $Y_{N_1} (z_{\scalebox{0.5}{UV},i})$)& $Y^{\rm th, w}_\Delta (z_{\scalebox{0.5}{UV},f}) \sim \frac{m_\nu M_N}{8 \pi v^2 \langle \Phi_\lambda \rangle^2} \left( \frac{m_{\Psi}^2}{y^2} \right)$  & eq.~(\ref{eq:eff_thermal}) \\
\hypertarget{eq:T8}{T8}& Off-shell $N_2$ scattering & $K_{N_2}^{\rm scatt} =  \frac{\Gamma_{N_2}^{\rm scatt}}{H(T=M_{N_1})}\sim \frac{\lambda_2^4}{16 \pi^3 \sqrt{g_\star}} \frac{M_{\rm pl}}{M_N}$ & $K_{N_2}^{\rm scatt} \sim \frac{1}{\sqrt{g_\star}} \frac{M_{\rm pl} m_\nu^2 M_{N}}{v^4 \langle \Phi_\lambda \rangle^4} \left( \frac{m_\Psi^4}{y^4} \right)$ & eq.~(\ref{eq:scatt_rate}) \\
\hline \hline
\multicolumn{5}{|c|}{The washout in TeV scale module} \\
\hline
No. & Quantity & Expression & Dependence on model parameters & Ref. \\
\hline
\hypertarget{eq:T9}{T9}& Effective washout factor & $K^{\rm eff}_\Psi \sim \frac{\Gamma_\Psi}{H({T=m_\Psi})}  \left( \frac{\mu}{\Gamma_\Psi} \right)^2$ &$K^{\rm eff}_\Psi \sim \frac{16\pi}{\sqrt{g_\star}} \frac{M_{\rm pl} m_\nu^2}{v^4} \left( \frac{m_\Psi}{y^6} \right)$ & eq.~(\ref{eq:Keff2}) \\
\hypertarget{eq:T10}{T10}& Low scale washout & $Y_{\Delta}\left(z_{\scalebox{0.5}{IR},f}\right) \sim Y_{\Delta} (z_{\scalebox{0.5}{IR},i}) e^{-K^{\rm eff}_\Psi  }$ &  $Y_{\Delta}\left(z_{\scalebox{0.5}{IR},f}\right) \sim Y_{\Delta} (z_{\scalebox{0.5}{IR},i}) e^{-\frac{16\pi}{\sqrt{g_*}} \frac{M_{\rm pl} m_\nu^2}{v^4} \left( \frac{m_{\Psi}}{y^6} \right) }$  & eq.~(\ref{eq:YDelta_lowscale}) \\
\hline \hline 
\multicolumn{5}{|c|}{Constraints} \\
\hline
No. & Quantity & Expression & Dependence on model parameters & Ref. \\
\hline
\hypertarget{eq:T11}{T11}& $\mu \to e \gamma$ & $\textrm{BR}(\mu \to e \gamma)\simeq \frac{3\alpha_{\rm em}}{8\pi}
\left|\left(y^t\frac{v^2}{m^\dagger_\Psi m_\Psi}y^*\right)_{\mu e}\right|^2$  & $\textrm{BR}(\mu \to e \gamma) \sim \frac{3 \alpha_{\rm em}}{8\pi} v^4 \left( \frac{y^4}{m_\Psi^4} \right)$ & eq.~(\ref{eq:mutoe}) \\
\hline
\end{tabular}
\caption{Summary of formulae for hybrid-leptogenesis with $r \equiv \lambda_1/\lambda_2$. High scale and TeV scale modules are defined in eq.~(\ref{eq:two_modules}). Generally we assume that $M_N\sim M_{N_1}\sim M_{N_2}$ and $r\leq 1$
\label{tab:summary_formulae} }
\end{adjustwidth}
\end{table}
\end{landscape}
}
%%%%%%%%%%%%%%%%%%%%%%

\subsection{General parameter space}

As discussed in section \ref{sec:Formalism}, there are two scales relevant for leptogenesis in our model. The asymmetry is first generated at high temperatures through decays of the Majorana singlet $N_i$. This primordial asymmetry is then prone to further washout at the TeV scale. In other words, the asymmetry that is generated and survives the high temperature washout effects can be taken as an initial condition for the TeV-scale leptogensis. As was shown in section \ref{inverse_lepto}, the asymmetry generated at the TeV scale by
itself is generically negligible, so we only take into account the washout processes at this scale. Furthermore, as we discuss in detail in section \ref{subsec:benchmark_points}, the allowed range for $M_N$ in our hybrid model is roughly of order $10^6 -10^{16}$ GeV. This corresponds to relaxed bounds on both (upper and lower) sides compared to the standard Type I scenario, and in what follows, we will study this full mass range.

In order to avoid technical details obscuring the main physics, we assume two generations of $N_i ~(i = 1,2)$ and their Yukawa couplings to be anarchic, except for the specific cases described below. 
In particular, $\Psi_a$ flavor effects can be ignored due to the assumption in eq.~(\ref{eq:anarc}) as well as fast flavor equilibrating scatterings mentioned in footnote \ref{fn:flavor}.
The same assumption as in eq.~(\ref{eq:anarc}) allows us to simply denote their Yukawa couplings as $\lambda _i $ ($ \lambda _i \sim \lambda_{ia}$ for all $a$).  
As we discuss in section \ref{sec:LowMN}, in order to lower the scale of high scale leptogenesis down to $10^{6}$ GeV, hierarchies in the couplings $\lambda_{i}$ are required and the third generation of $N$ is also needed to fit neutrino observables. For this reason it is useful to define the ratio $r \equiv \frac{\lambda_1}{\lambda_2}$. The mass of $N_i$ is denoted as $M_{N_i}$ or simply $M_N$ when all $M_{N_i} $ are of the same order but not degenerate.
 For the other couplings $ y$ and $\kappa$, we assume anarchical structure and for simplicity, 
we treat them as numbers rather than matrices.

Under the reasonable assumptions mentioned above, our hybrid model can be parametrized by six parameters: 
\bea\label{eq:par}
y,\, m_\Psi \, \lambda_2,\, r,\, M_N, \langle \Phi_\lambda \rangle. 
\eea
Since hybrid-genesis intrinsically features two scales, it is convenient to classify these parameters in two modules :
\bea\label{eq:two_modules}
\textrm{High scale module}&:& \lambda_i N_{i}\Phi_{\lambda}\Psi+\frac{1}{2}M_{N_{i}}N_{i}N_{i}+{\rm h.c.},\nn
\textrm{TeV scale module}&: &y\Psi^{c}H\ell+m_\Psi\Psi^{c}\Psi+\frac{1}{2}\mu \Psi\Psi+{\rm h.c.}
\eea
Here the high scale module is the part of the full Lagrangian in eq.~\eqref{eq:hybrid_model} which only contains particles relevant to high scale leptogenesis and has three parameters: $\lambda_2, r , M_N $. (Note that $\langle \Phi_\lambda \rangle$ does not affect high scale genesis, so it is not included here.) 
 The TeV scale module is basically the ISS model (eq.~\eqref{eq:ISS}) with $m_{\Psi}=\kappa \langle \Phi_\kappa \rangle$ and $\mu$ determined by high scale parameters as in \hyperlink{eq:T2}{T2} in table~\ref{tab:summary_formulae}. Because $\langle \Phi_\lambda \rangle$ is a more fundamental quantity than $\mu$, we interpret the former as independent.

There are two constraints on six parameters in eq.~\eqref{eq:par}: one is the observed neutrino mass $m_\nu$ and the other the observed baryon asymmetry. This leaves us with four independent parameters. The above simplifications allow us to write down simple relations as in Table \ref{tab:summary_formulae}.

\subsubsection{TeV scale module }
\label{sec:param_TeV}

\begin{figure}[t!]
\centering
\includegraphics[width=140mm]{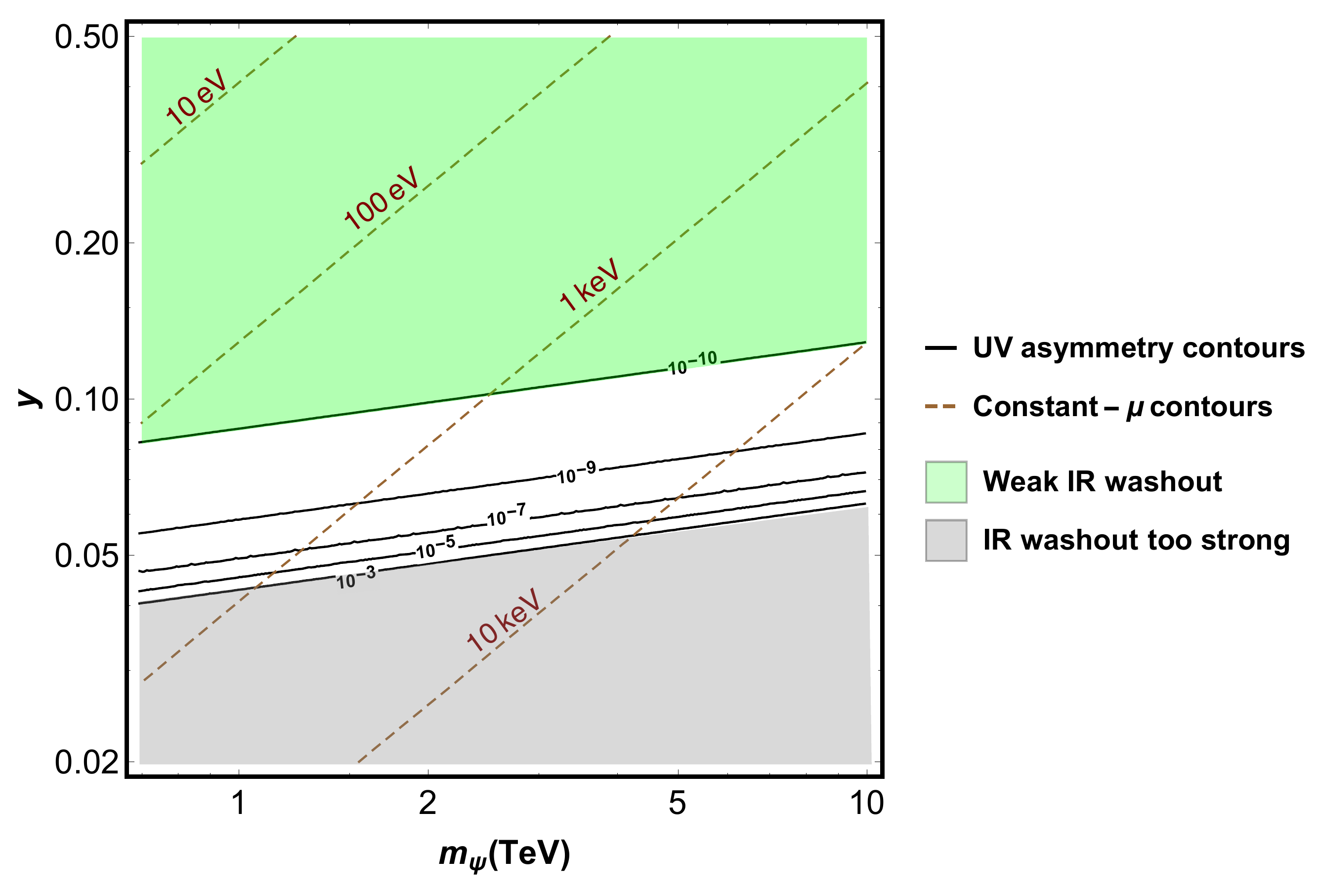}
\caption{Contours of needed UV asymmetry (solid black lines) and $\mu$ (dashed brown lines) in the $(m_{\Psi},y)$ plane to obtain the observed baryon asymmetry $Y_{\Delta B} \sim 10^{-10}$ and $m_\nu = 0.05 $ eV.  
In the green shaded region, washout at the TeV scale is negligible so that the UV asymmetry needs to be the observed baryon asymmetry $\sim 10^{-10}$. On the other hand, for smaller values of $y$, due to exponential sensitivity in $y$ and $m_\Psi$, UV asymmetry lines get closer to each other. In the gray shaded region, TeV scale washout becomes so large that even saturating maximal allowed $Y_{\Delta} (z_{\,_{UV},f}) \sim 10^{-3}$  results in too small final asymmetry to explain the observation. \label{fig:TeVPlot} }
\end{figure}

We first study the impact of our parameters in the TeV scale module in eq.~(\ref{eq:two_modules}). Consider figure \ref{fig:TeVPlot}, where $y$ and $m_\Psi$ are treated as independent. For a chosen $m_{\Psi}$ and $y$, the quantity $\mu$ will be fixed by the SM neutrino mass $m_\nu$ via \hyperlink{eq:T1}{T1} in table~\ref{tab:summary_formulae}.
This is presented in figure \ref{fig:TeVPlot}, where the dashed lines on the plot are contours of constant 
$\mu$ and we have fixed $m_\nu = 0.05 $ eV. From table~\ref{tab:summary_formulae}, we see that $\mu \sim m_\Psi^2 / y^2$.

Once $m_{\Psi}$, $y$ and $m_\nu$ are fixed, the low scale effective washout factor (\hyperlink{eq:T9}{T9} in table~\ref{tab:summary_formulae}) is fixed as well. Therefore, we can determine the required amount of asymmetry generated at the high scale $Y_{\Delta} (z_{\scalebox{0.5}{UV},f})$ (using \hyperlink{eq:T9}{T9} in table~\ref{tab:summary_formulae}) 
in order to match the observed value 
$Y_{\Delta B} \sim 10^{-10}$~\cite{Ade:2015xua} through eq.~\eqref{eq:B-L_to_B}.
Contours of the needed UV asymmetry are shown as solid lines in figure \ref{fig:TeVPlot}. 
As can be seen from table~\ref{tab:summary_formulae}, the required $Y_{\Delta}\left(z_{i}\right)$ depends on $(m_\Psi, y)$ via $K^{\rm eff}_\Psi \sim \frac{16\pi}{\sqrt{g_\star}} \frac{M_{\rm pl} m_\nu^2}{v^4} \left( \frac{m_\Psi}{y^6} \right)$. Therefore, the required UV asymmetry lines will simply be parallel to constant $\sim \frac{m_\Psi}{y^6}$. 
In the green shaded region of the plot, the washout effect at the TeV scale is negligible and hence the UV asymmetry will need to be of order the observed size. 
For smaller $y$ values, on the other hand, the washout from the TeV scale is exponentially strong so the final asymmetry becomes sensitive to TeV scale parameters:
this is reflected in the UV asymmetry contours getting closer and closer to each other as $y$ gets smaller.
For the gray shaded region, $y\lesssim 0.04$, the washout is so strong that even the maximal $Y_\Delta(z_{\scalebox{0.5}{UV},f})\sim 10^{-3}$ would not be enough. For this reason no UV asymmetry contours are present in that region.

%%%%%%%%%%%%%%%%%%%%%%%%

\subsubsection{High scale module}
\label{sec:param_UV}

\begin{figure}[t]
\centering
\includegraphics[width=150mm, height=100mm]{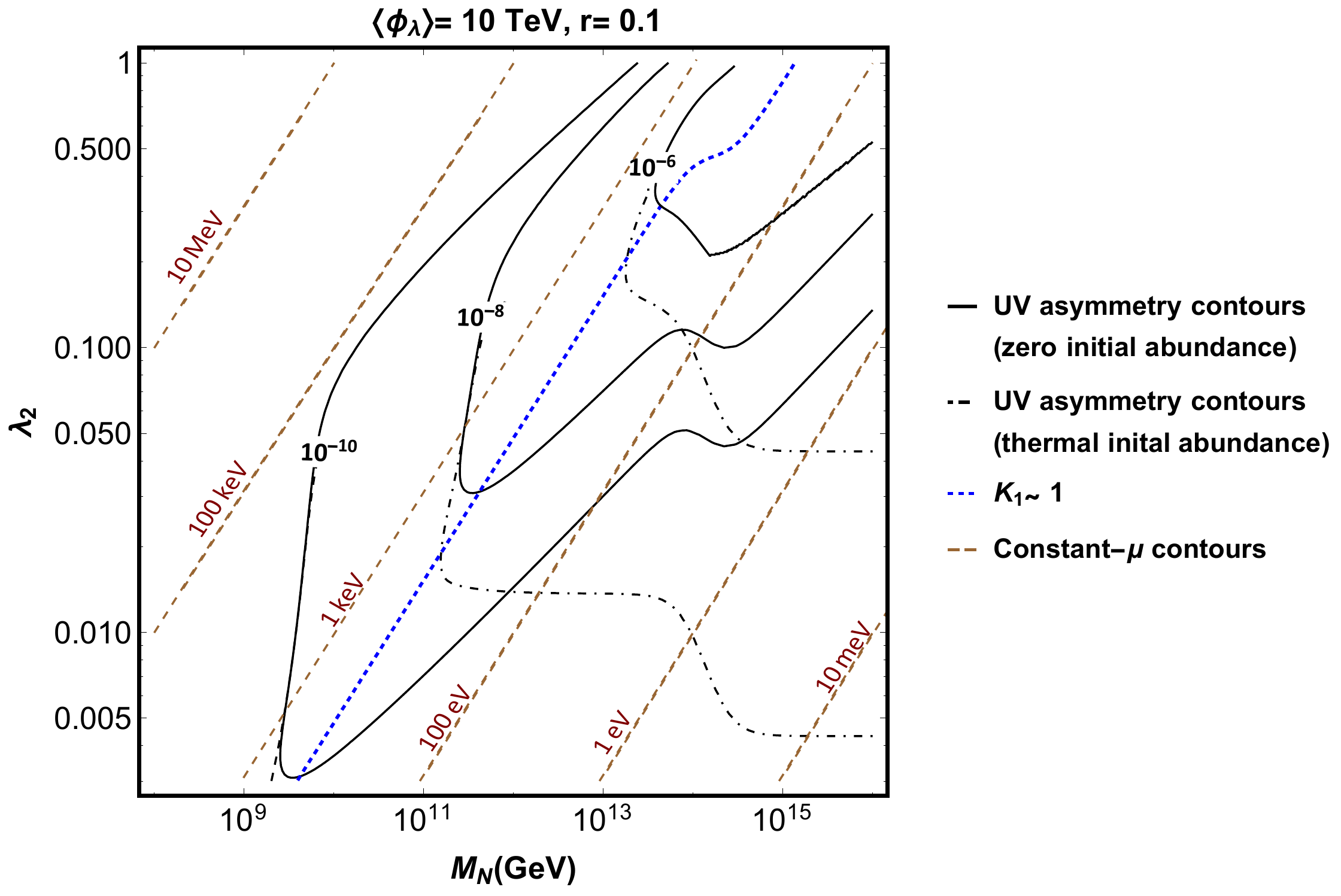}
\caption{Contours of  UV asymmetry and $\mu$ generated in the high scale module in the $(M_{N},\lambda_2)$ plane for $\langle \Phi_{\lambda} \rangle =10$ TeV and $r=0.1$. Solid curves are contours of UV asymmetry assuming zero initial abundance of $N$ and dot-dashed curves are contours of UV asymmetry with the assumption of thermal initial abundance for $N$. The brown dashed lines are contours of constant $\mu$. The blue dotted line sets the boundary between strong washout regime (to the left of the line) and weak washout regime (to the right of the line) for the high scale module. \label{fig:UVPlot}}
\end{figure}

The high scale module in eq.~(\ref{eq:two_modules}) has four parameters: $\lambda_2$, $r$, $M_N$ and $\langle\Phi_{\lambda}\rangle$.
Besides determining the generated UV asymmetry $Y_\Delta(z_{\scalebox{0.5}{UV},f})$,  they also control the TeV scale mass $\mu$ as \hyperlink{eq:T2}{T2} in table~\ref{tab:summary_formulae}.
Fixing two of the four parameters, we can plot contours of constant $\mu$ and $Y_\Delta(z_{\scalebox{0.5}{UV},f})$ in the 
plane defined by the remaining two.
For instance, fixing $\langle \Phi_{\lambda} \rangle$ and $r$, 
we obtain contours of constant $\mu$ and $Y_\Delta(z_{\scalebox{0.5}{UV},f})$ in the $(M_N, \lambda_2)$ plane. 
These contours are shown in figure \ref{fig:UVPlot}. 
As seen from \hyperlink{eq:T2}{T2} in table~\ref{tab:summary_formulae}, constant $\mu$ simply gives straight lines (brown, dashed).
Moving onto UV asymmetry, 
the blue dotted curve in the plot separates the regions of strong and weak UV washout. The transition seen in this curve around $M_N \sim 10^{14}$ GeV is due to the change in $g_\star$ as discussed in section~\ref{subsec:Initial-conditions}.
In the region to the left of the blue dotted curve, washout in the UV is strong and the UV asymmetry is a function mainly of $M_N$ with only a logarithmic dependence on $\lambda_2$ (\hyperlink{eq:T5}{T5} in table \ref{tab:summary_formulae}), as long as the washout from off-shell scattering is negligible (i.e. as long as $\lambda_2$ is not too large). So, in this region UV asymmetry contours (solid, black lines) are almost vertical lines until $\lambda_2$ becomes so large that washout from off-shell scattering becomes important. For such high $\lambda_2$ values, the UV asymmetry is very sensitive to $K_{N_2}^{\rm scatt}$ and contours of constant asymmetry roughly follow constant $K_{N_2}^{\rm scatt}\propto \frac{\lambda_2^4}{M_N}$ lines. 
In the weak washout region (to the right of the blue dotted curve) and with zero initial abundance for $N$, the UV asymmetry is $\propto \frac{\epsilon_1 K_1^2}{g_*} \propto \frac{\lambda_2^6}{g_* M_N^2}$. Therefore, as long as $g_*$ does not vary significantly, solid curves in this region follow constant $\frac{\lambda_2^3}{M_N}$ lines. The dot-dashed black curves are contours of constant UV asymmetry assuming thermal initial abundance for $N$. They differ from the solid curves only in the weak washout regime where the UV asymmetry is $\propto \frac{\lambda_2^2}{g_*}$ and follow constant $\lambda_2$ lines if $g_*$ is constant. 
The transition in both cases seen around $M_N\sim 10^{14}$ GeV is due to the change in $g_*$ mentioned earlier. 

%%%%%%%%%%%%%%%%%%%%%

\subsubsection{Combining high scale and TeV scale modules}

Here, we will combine the results of sections \ref{sec:param_TeV} and \ref{sec:param_UV} in order to get a better picture of the allowed parameter space. 
For given values of $\left(y,m_{ \Psi } \right)$, we can find from figure \ref{fig:TeVPlot} the required values of $\mu$ 
(for the right SM neutrino mass)
and
UV lepton asymmetry. Then, we can ``match'' these values to those generated by the high scale module from figure \ref{fig:UVPlot}: for a given $\langle \Phi_\lambda \rangle$ and $r$, we can determine the necessary $\left( M_N, \lambda_2 \right)$.
To make the parametric dependence more explicit, 
we
fix (at a time) values of two parameters out of ($M_N, \langle \Phi_\lambda \rangle$ and $r$) and present viable 
contours for various values of the 3rd parameter in the 2D $y - m_\Psi$ plane.
We show some of the curves in figures \ref{fig: WOregions}
and \ref{fig:Combined}: more details are given below.
%
%

%%%%%%%%%%%%%%

\begin{figure}[h]
\centering
\includegraphics[width=120mm]{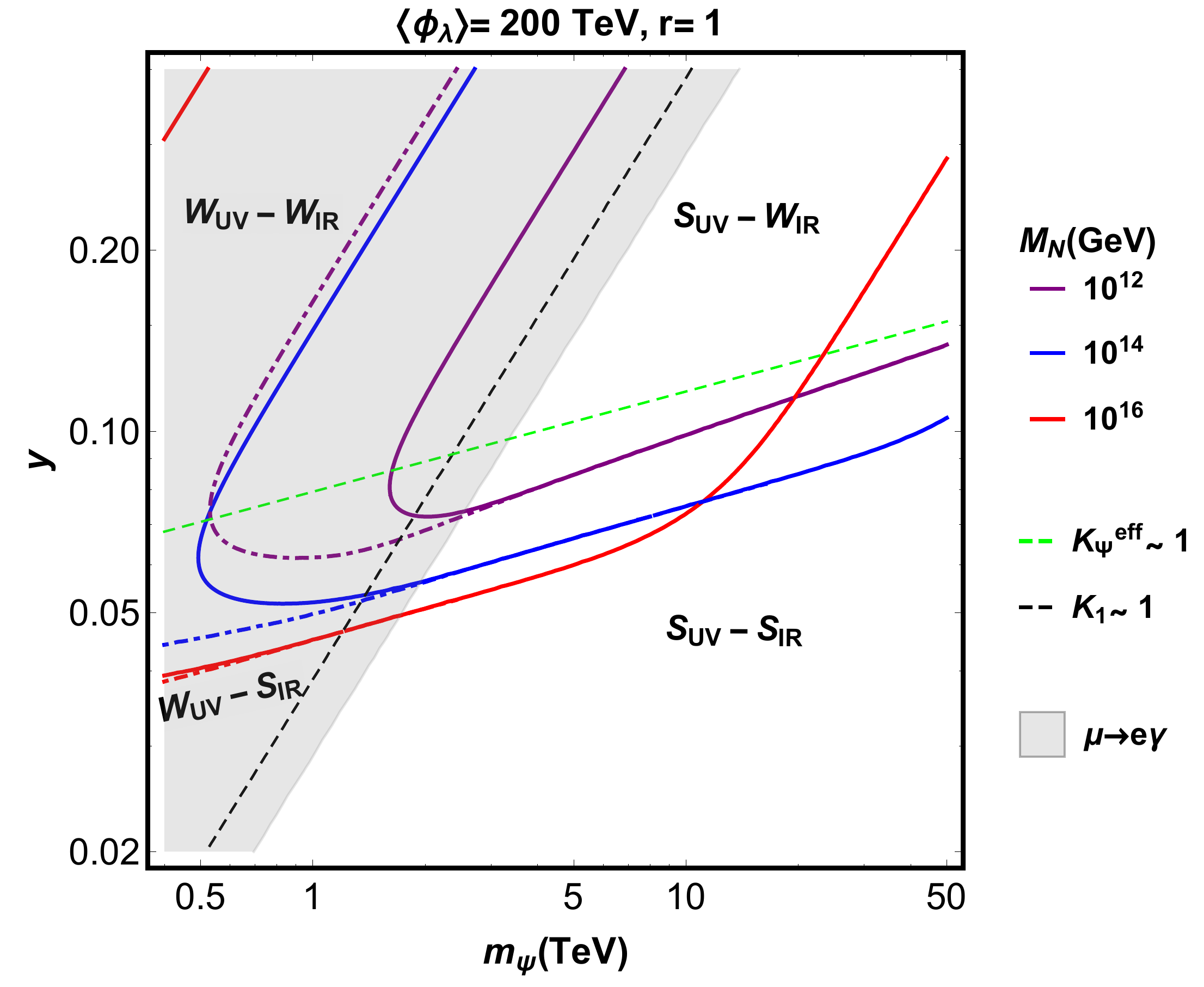}
\caption{Interplay of high-scale washout and asymmetry generation with TeV scale washout. Solid (dot-dashed) curves are contours on which observed baryon asymmetry and SM neutrino masses are produced for fixed $r$ and $\langle \Phi_\lambda \rangle$ and different choices of $M_N$, assuming zero (thermal) initial abundance for $N$ . The dashed green line sets the boundary between the weak washout and strong washout regimes in the IR (around the TeV scale), and the dashed black line is the boundary of weak and strong washout regimes in UV. The gray shaded region is constrained by $\mu \rightarrow e \gamma$.\label{fig: WOregions}}
\end{figure}

\begin{figure}[t!]
\centering
\includegraphics[width=75mm]{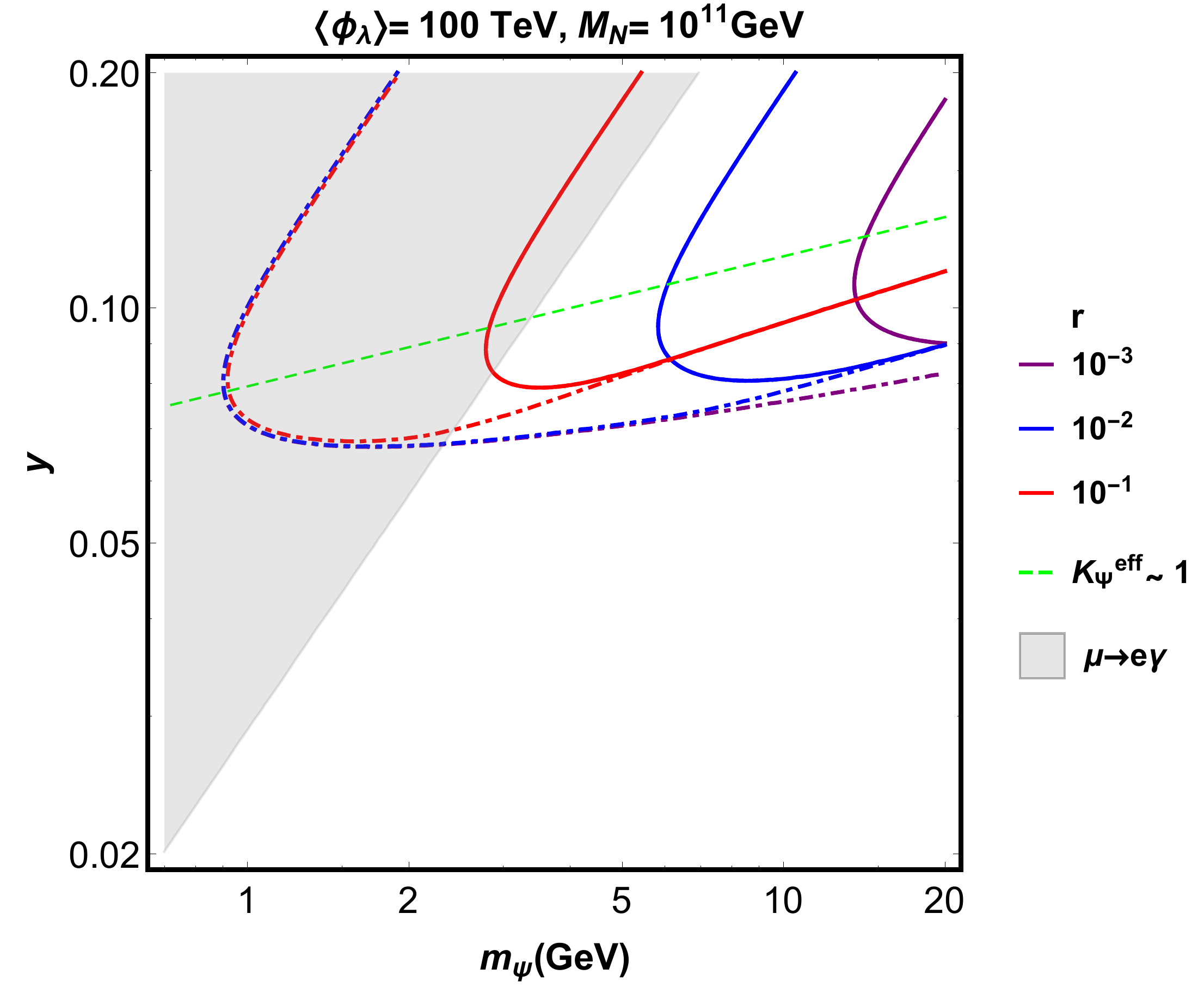} \includegraphics[width=75mm]{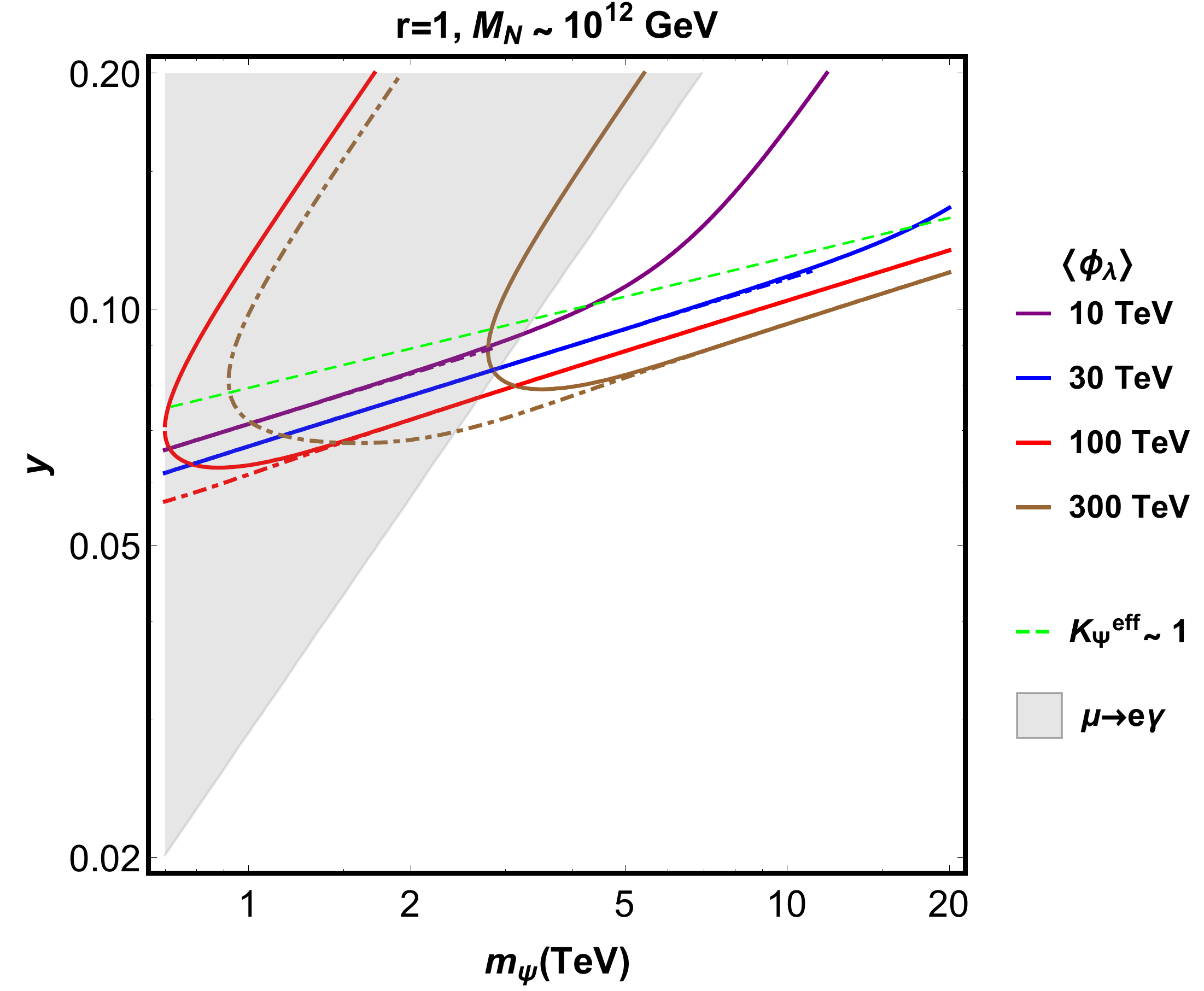}
\caption{ Solid (dot-dashed) curves are contours in $(m_{\Psi},y)$ plane on which the observed neutrino mass and baryon asymmetry is produced assuming zero (thermal) initial abundance for $N$, similar to figure~\ref{fig: WOregions}. In the left panel,  we vary $r$ and keep $M_N$ and $\langle \Phi_{\lambda}\rangle$ fixed, while the right panel shows the results with varying $\langle \Phi_{\lambda}\rangle$ while keeping $M_N$ and $r$ fixed. The gray shaded region is constrained by $\mu \rightarrow e \gamma$.
\label{fig:Combined}}
\end{figure}

%%%%%%%%%%%%%%

%
First of all, we mention some of the {\em general} ingredients going into
these plots.
In order to generate them,
we have assumed anarchic, non-degenerate Majorana singlet masses. We have estimated the CP asymmetry, UV washout factor and effective IR washout factor as shown in table \ref{tab:summary_formulae} and have used the analytic approximate expressions of appendix \ref{app:approximate_solutions}. Whenever washout from off-shell scattering becomes important (for regions in parameter space where $K_{N_2}^{\rm scatt}>0.1$), we take it into account by calculating the efficiency factor numerically.
The
solid curves in figures \ref{fig: WOregions}
and \ref{fig:Combined} are produced under the assumption of {\em zero} initial $N_1$ abundance and 
dot-dashed curves are obtained assuming thermal initial $N_1$ abundance. 
These differ only in the weak UV washout regime
since in the strong UV washout regime any asymmetry generated at the early time is efficiently erased 
and the final result only depends on the equilibrium abundance of $N_1$ at a time when the inverse decays freeze out.
As already noted above,
on each curve, we have fixed the SM neutrino mass (to $m_\nu = 0.05$ eV) and 
the final baryon asymmetry matches the observed one.
Finally, 
the region constrained by lepton flavor violating process $\mu \to e \gamma$ (\hyperlink{eq:T11}{T11} in table \ref{tab:summary_formulae}) is shaded in gray, see the upper left corner of each plot. As discussed in section~\ref{review}, such constraint can be further relaxed with flavor symmetries.

We now discuss in more detail some of the {\em specfic} features in these plots.
Interestingly, there is an important {\em interplay} between the asymmetry generation/washout effects in high scale and washouts in TeV scale modules. 
In order to see this, consider first figure \ref{fig: WOregions}, where 
we fix $r$ and $\langle \Phi_{ \lambda } \rangle$ and show working contours for several values of $M_N$: we observe
that it is divided into four regions by two dashed lines: 
the green dashed line denotes the $K_\Psi^{\rm eff} \sim 1$ boundary 
(of strong/weak washout in the IR)
and the black dashed line 
is for $K_1 \sim 1$ 
(i.e., boundary of strong/weak washout in the UV).
From table~\ref{tab:summary_formulae}, we see that $K_\Psi^{\rm eff} \sim m_\Psi / y^6$ and $K_1 \sim m_\Psi^2 / y^2$. The region above (below) the green dashed line has $K_\Psi^{\rm eff} < 1$ ($K_\Psi^{\rm eff} >1$), is identified as the weak (strong) IR washout region and labelled by $W_{\rm IR}$ ($S_{\rm IR}$).
The dashed black line, on the other hand, separates the high scale strong ($S_{\rm UV}$) and weak ($W_{\rm UV}$) washout regimes. The high scale washout is strong ($K_1>1$) below this line and weak ($K_1<1$) above it. 
One can identify four regimes from these combination: $S_{\rm UV}-S_{\rm IR}$, $S_{\rm UV}-W_{\rm IR}$, $W_{\rm UV}-W_{\rm IR}$ and $W_{\rm UV}-S_{\rm IR}$:

\begin{enumerate}[(i)]
\item $S_{\rm UV}-S_{\rm IR}$: 
In this regime, the high scale asymmetry generation happens in the strong washout regime 
and the UV asymmetry generated is determined primarily by $M_N$ and has only a very weak dependence on other parameters. 
Fixing all other parameters besides $\{m_\Psi, y \}$, from  
\hyperlink{eq:T5}{T5} in table \ref{tab:summary_formulae}, 
we see that the UV asymmetry has only logarithmic (mild) UV dependence on $K_1$. On the other hand, the final asymmetry 
is exponentially sensitive to TeV-scale washout, i.e. $K_\Psi^{\rm eff} \sim m_\Psi / y^6$ and hence a constant 
final asymmetry will lie along the constant $K_\Psi^{\rm eff}$ lines i.e. parallel to 
$K_\Psi^{\rm eff} \sim 1$ line. 
Eventually no curves will appear simply because TeV-washout becomes 
so strong that it is not possible to render the observed size of asymmetry for any choice of $M_N$.

Furthermore, one may notice from the red curve in figure~\ref{fig: WOregions} that its behavior differs from the others for larger $m_\Psi$. This may be understood by recalling that for larger $M_{N}$ washout in the UV from scattering by off-shell exchange of $N_2$ becomes larger (parametrized by $K_{N_2}^{\rm scatt}$ in \hyperlink{eq:T5}{T5} of table~\ref{tab:summary_formulae}) and at some point it becomes a significant factor 
in determining the final asymmetry. In this regime, the final asymmetry will follow a constant $K_{N_2}^{\rm scatt}$ line. Moreover, $K_{N_2}^{\rm scatt} \sim m_\Psi^4 / y^4$ and so the asymmetry curve appears to be parallel to constant $K_1$ line.

\item $S_{\rm UV}-W_{\rm IR}$: 
In this region washout at the TeV scale is negligible and the final asymmetry is set by the high scale parameters 
as \hyperlink{eq:T5}{T5} in table \ref{tab:summary_formulae}.
So the curves in this region follow a constant $K_1$ curve except
for some choice of parameters when washout due to scattering from off-shell $N_2$ exchange becomes relevant, 
in which case the curves are determined by a constant $K_{N_2}^{\rm scatt}$ line that coincides with a constant $K_1$ line.

\item $W_{\rm UV}-W_{\rm IR}$: 
In this region, washout at the TeV scale is negligible and the final asymmetry will be mainly dictated by the UV asymmetry. The asymmetry generated at the high scale, as shown in \hyperlink{eq:T6}{T6} or \hyperlink{eq:T7}{T7} in table \ref{tab:summary_formulae},
will be proportional to powers of $m_\Psi/y$ and they will lie on constant $K_1$ lines 
(for both zero and thermal initial $N_1$ abundance).

\item $W_{\rm UV}-S_{\rm IR}$: The generation of asymmetry at the high scale occurs in the weak washout regime and the washout at the TeV scale is strong. The curves in this region interpolate between strong-strong and weak-weak regions, starting from a constant $K^{\rm eff}_\Psi$ line near the $S_{\rm UV}-S_{\rm IR}$ region and ending roughly on constant $K_1$ lines near the  $W_{\rm UV}-W_{\rm IR}$ region.
 
\end{enumerate}

Useful complementary information can be found in figure \ref{fig:Combined}, where we show two plots with fixed ($M_N$, $\langle \Phi_\lambda\rangle$)
instead,
while varying $r$ and with fixed ($M_N$, $r$), for several choices of $\langle \Phi_\lambda\rangle$, respectively.
Analyses similar to that done for figure \ref{fig: WOregions} can be
performed here also, but for brevity, we will not repeat it.
As seen from \hyperlink{eq:T4}{T4} in table~\ref{tab:summary_formulae}, $K_1 \propto \frac{r^2}{\langle \Phi_\lambda \rangle^2}$ and thus as we change either $r$ or $\langle \Phi_\lambda \rangle$ (as we do in figure~\ref{fig: WOregions}), the $K_1 \sim1$ boundary will also change. To avoid too much complication in plots, therefore, we decided not to show $K_1 \sim 1$ lines for each case. 
For a  
discussion of 
other phenomenology of this model (such as collider and cosmological signals), 
see ref. \cite{short}.

\subsection{Selected benchmark points}
\label{subsec:benchmark_points}

In this section we are going to focus on some representative benchmark points. We categorize the possibilities based on $M_N$, the size of mass of heavy Majorana singlet. Specifically we present the choice of parameters that make leptogenesis possible for $M_N>10^{15}$ GeV and $M_N<10^{9}$ GeV, i.e., outside the usual range of Majorana singlet mass in type I seesaw leptogenesis scenarios.

\subsubsection{Super-heavy singlet: $\gtrsim 10^{ 15 }$ GeV}

We start with singlet masses close to the upper bound on the reheating temperature from BICEP, i.e., $M_N \sim 10^{ 16 }$ GeV \cite{Ade:2015tva}.\footnote{The constraint is on the Hubble scale during inflation, which with the assumption of instantaneous reheating, can be translated into a bound on the reheating temperature.}
In type I seesaw, leptogenesis fails in this regime because it suffers from too large washout due to off-shell scattering mediated by the Majorana singlet.
This washout has a rate proportional to $\lambda^4$ and as one increases $M_N$, one also has to increase $\lambda$ in order to generate sufficient SM neutrino mass (in the type I seesaw $m_\nu \sim \lambda^2 v^2/M_N$). When $\lambda > 1$, the washout becomes strong and efficiently erases the asymmetry.
On the other hand, in our hybrid model, we can keep $\lambda$ small enough to suppress the washout from scattering 
and adjust other parameters to obtain the SM neutrino mass 
even for such large values of $M_N$.
Indeed, taking into account the above considerations,  it was shown in \cite{short} that the upper bound on $M_N$ in hybrid model is given by 
\bea\label{upperB}
M_N \lesssim \left( \frac{16\pi^3 \sqrt{g_\ast} \; v^4}{M_{\rm Pl} \; m_\nu^2} \right)\times\left( \frac{y\langle \Phi_\lambda \rangle}{\kappa\langle \Phi_\kappa \rangle} \right)^4\sim10^{14}~{\rm GeV}\times\left( \frac{y\langle \Phi_\lambda \rangle}{\kappa\langle \Phi_\kappa \rangle} \right)^4,
\label{eq:upper_bound_M_N}
\eea
where the first factor is the bound for standard type I seesaw due to strong washout from scattering as discussed earlier. We see that the second factor, which can be interpreted as a TeV-modulation effect, buys us extra freedom and allows to relax the usual upper bound of $10^{14} \sim 10^{15}$ GeV.
Initial conditions for leptogenesis in this high temperature regime are discussed in section \ref{subsec:Initial-conditions}.

In order to illustrate successful leptogenesis for $M_{N_1} \gtrsim 10^{15}$ GeV, as a benchmark point we choose $M_{N_1}=10^{16}$ GeV and $M_{N_2}=3\times10^{16}$ GeV (shown in table \ref{tab:benchmark}).
 A choice of $\lambda \sim 0.5$  allows for $N_1$ decays and inverse decays to be in equilibrium around $T=M_{N_1}$ while keeping the off-shell scattering mediated by $N$'s out of equilibrium. With this choice of parameters, high scale leptogenesis happens in the strong washout regime ($K_1\approx 10$), and it generates a UV baryon asymmetry of $\sim 10^{-5}$. The washout at the TeV scale is then needed to dilute it down to the observed baryon asymmetry of $\sim 10^{-10}$. For this to happen we must have $K_{\Psi}^{\rm eff}\approx 27$. This value of $K_{\Psi}^{\rm eff}$ can be obtained by a choice of $m_{\Psi}\sim  3$ TeV and $y\sim 0.05$, which is also consistent with the $\mu \rightarrow e \gamma$ bound. To get the observed neutrino mass with the chosen parameters, we then take $\langle \Phi_{\lambda} \rangle \sim 400$ TeV.\footnote{As discussed in section~\ref{sec:4_extension_big_picture} and appendix~\ref{app:warped_seesaw}, such a value for
 $\langle \Phi_{ \lambda } \rangle$, i.e., $\gg$ TeV, can be ``effectively'' obtained with{\em out} any hierarchies in the {\em fundamental} parameters
 in the warped/composite UV completion of the hybrid model.}

Note that since we have not introduced any significant hierarchies 
in any of the mass or Yukawa matrices, we obtain an anarchic SM neutrino mass matrix in the SM flavor basis as is
sufficient
to fit to the observed neutrino masses and mixing angles.

\begin{table} 
\centering
\setlength{\arrayrulewidth}{0.3 mm}
\setlength{\tabcolsep}{8pt}
\renewcommand{\arraystretch}{1.2}
 
\begin{tabular}{ |P{0.9cm}|P{0.9cm}|P{1.2cm}|P{0.9cm}|P{1.2cm}|P{0.9cm}|P{0.8cm}| P{0.8cm}| P{0.8cm}| P{0.8cm}|  }
\hline
$\, M_{N_1}$ (GeV)& $\lambda_2$ & $r=\frac{\lambda_1}{\lambda_2}$ & $\frac{M_{N_2}}{M_{N_1}}$ & $\langle \Phi_\lambda \rangle$ (TeV) & $\, m_\Psi$ (TeV) & $y$ &$K_1$ & $K_\Psi^{\rm eff}$ & $\, \, \mu$ (keV)\\
\hline
$10^{16}$  & 0.5  & 1 & 3 & 400 & 3 & 0.05&10 & 27 &5\\
\hline
$10^{11}$ &0.04 & 0.3  & 3 &20 & 8 & 0.1 & 30&0.2&6\\
\hline
$10^{7}$ &  $0.01$ &$ 0.001$& 20& 4& 20& 0.2& 0.6& 0.03 &15\\
\hline
\end{tabular}
\caption{ Three different benchmark points consistent with neutrino mass data and leptogenesis, organized by  Majorana singlet mass scale.  \label{tab:benchmark}}
\end{table}

\subsubsection{Going below Davidson-Ibarra bound of $\sim 10^9$ GeV} \label{sec:LowMN}

At the other extreme we consider Majorana singlet masses below $10^9$ GeV. In standard type I seesaw model, one can not have successful leptogenesis for singlet masses below $~ 10^9$ GeV unless one employs flavor effects or resonant leptogenesis which requires hierarchical parameters and/or new ingredients
as discussed in section~\ref{subsec:others}.
This lower limit on the Majorana singlet mass is known as Davidson-Ibarra bound \cite{Davidson:2008bu}.
However, as shown in \cite{short}, the lower bound in hybrid seesaw is relaxed
\bea
M_N\gtrsim{10^{-7}}\frac{8\pi v^2}{m_\nu}
\left( \frac{y \left<\Phi_\lambda \right>}{\kappa \left<\Phi_\kappa \right>} \right)^2\sim{10^9~{\rm GeV}}
\times\left( \frac{y \left<\Phi_\lambda \right>}{\kappa \left<\Phi_\kappa \right>} \right)^2,
\label{M_N_lower_generic}
\eea
where the first factor is the Davidson-Ibarra lower bound for the case of standard type-I seesaw and the second factor is due to the TeV-modulation. Therefore, in our model we can have successful leptogenesis with Majorana singlet masses $\ll 10^9$ GeV , even if we ignore flavor effects and without any degeneracy between singlet masses. However, there exists another rather generic lower bound $M_N \gtrsim 10^{5}$ GeV. This is derived by the 
simultaneous requirements of large enough CP violation and small enough $\Delta = 2$ washout due to scattering. In order to see this more explicitly, we note that to suppress potentially dangerous washout by the $\Delta =2$ scattering from the off-shell $N_2$ mediation, we need to impose its rate to be smaller than the Hubble rate at $T=M_{N_{1}}$. Using eq.~(\ref{eq:CP_approx}) and (\ref{eq:scatt_rate}) this gives us the following condition:
%
%%%
\begin{eqnarray}
M_{N_{1}} & \gtrsim & \frac{4}{\pi}\frac{\epsilon_{1}^{2}}{\sin^{2}\left(\phi_{12}\right)}\frac{M_{{\rm Pl}}}{1.66\sqrt{g_{\star}}}\nonumber \\
 & = & 8.5\times10^{4} \frac{1}{\sin^{2}\left(\phi_{12}\right)}\left(\frac{\epsilon_{1}}{10^{-7}}\right)^{2}\sqrt{\frac{121.25}{g_{\star}}}\,{\rm GeV}.
 \label{eq:lowerbound_M_washout}
\end{eqnarray}
%%%
Hence we see that the requirement that the scatterings $\Phi_{\lambda}\Psi_{a}\leftrightarrow (\Phi_{\lambda} {\Psi}_{a})^*$ be out of equilibrium set a lower bound on the mass of $N_1$~\cite{Racker:2013lua}. The value $\epsilon\sim 10^{-7}$ is chosen because this is the minimum value of $\epsilon$ to get successful leptogenesis.

To achieve leptogenesis for the lowest value $M_{N_{1}}\sim 10^{5}$ GeV, we however need a
hierarchy in $\lambda_{1a}/\lambda_{2a}$. 
This can be seen by checking the allowed range of $\lambda_{1a}$ and $\lambda_{2a}$. Since the value of $\epsilon_1$  is already saturated to its minimum value for $M_{N_{1}}\sim 10^{5}$ GeV, we can not afford additional washout effects. As we saw in eq.~(\ref{eq:eta_N_1}), the maximum efficiency of the washout can be achieved if $K_1\lesssim 1$, and this leads to
\begin{eqnarray}
\left(\lambda\lambda^{\dagger}\right)_{11} & \lesssim & 1.66\sqrt{g_{\star}}16\pi\frac{M_{N_{1}}}{M_{{\rm Pl}}}\nonumber \\
 & = & 7.5\times10^{-12}\sqrt{\frac{121.25}{g_{\star}}}\left(\frac{M_{N_{1}}}{10^{5}\,{\rm GeV}}\right).
\end{eqnarray}
On the other hand, to have $\left|\epsilon_{1}\right|\gtrsim10^{-7}$,
from eq.~\eqref{eq:CP_approx}, we require
%%%%%%%
\begin{eqnarray}
\left(\lambda\lambda^{\dagger}\right)_{22} & \gtrsim & 8\pi10^{-7}\frac{M_{N_{2}}}{M_{N_{1}}}\frac{1}{\left|\sin\left(\phi_{12}\right)\right|}\nonumber \\
 & = & 2.5\times10^{-5}\left(\frac{M_{N_{2}}}{M_{N_{1}}}\frac{1}{10}\right)\frac{1}{\left|\sin\left(\phi_{12}\right)\right|}.
\end{eqnarray}
The above estimation shows that to achieve leptogenesis for $M_{N_{1}}\sim10^{5}$ GeV, we need $\lambda_{1a}/\lambda_{2a}\sim 5\times10^{-4}$ (i.e. a small value of $r$),\footnote{As discussed in \cite{short}, with the choice of {\em anarchic} parameters instead, 
we can only go down to $M_N \sim 10^{11}$ GeV.\label{fn:anarchy}} but not among different generations of $\Psi$'s.

The relaxation of the lower bound on the singlet Majorana masses, thus lowering the required reheating temperature of the Universe, 
may alleviate the gravitino overproduction problem \cite{Pagels:1981ke}  of SUSY models.  Namely, for gravitino masses (SUSY breaking scale) $\sim$ TeV (which is the ``natural'' range), we typically need reheating temperatures below ~$\sim 10^9$ GeV in order to avoid BBN bounds from excessive late decays of (very weakly-coupled) gravitinos \cite{Moroi:1993mb}. In the usual type I seesaw model, this might be in tension with leptogenesis.

We now present a specific choice of parameters consistent with leptogenesis for Majorana singlet mass $M_{N_1}\sim 10^7$ GeV (see table \ref{tab:benchmark}). We choose $\lambda_1 \sim 10^{-5}$ such that we get $K_1\sim 1$ in order to optimize the efficiency factor $\eta$. As already mentioned, we need to allow for a hierarchy between the Yukawa couplings of $N_1$ and $N_2$, which we choose to be $r=\frac{\lambda_1}{\lambda_2}\sim 10^{-3}$ corresponding to $\lambda_2 \sim 0.01$. This provides an enhancemnet in the UV asymmetry by a factor of $\frac{1}{r^2}\sim10^6$, compared to the anarchic ($r=1$) case (see \hyperlink{eq:T2}{T5} in table~\ref{tab:summary_formulae}), which is enough to account for the observed asymmetry. 
Note that with $\lambda_2 \gg \lambda_1$, the decay rate of $N_{2}$ is much larger than the Hubble rate, and so washout from inverse decay of $N_{2}$ can be potentially dangerous for leptogenesis. However, since at temperatures below the mass of $N_2$, the rate for this inverse decay is Boltzmann suppressed, a small hierarchy between $M_{N_2}$ and $M_{N_1}$ is sufficient for $N_{2}$ inverse decay to be out of equilibrium at $T\approx M_{N_1}$.
The condition we demand is
\begin{equation}
e^{-\frac{M_{N_2}}{M_{N_1}}} < \frac{\Gamma_{N_1}}{\Gamma_{N_2}} \sim r^2.
\end{equation}
We choose $M_{N_2} \approx 20 M_{N_1}$ as our benchmark value which gives a Boltzmann suppression of $e^{-20} \sim 10^{-9} \ll r^2$. With the chosen values for $M_{N_2}$ and $\lambda_2$, washout from off-shell scattering, mediated by $N_2$, is out of equilibrium when asymmetry generation happens.
 A choice of $m_\Psi \sim 20$ TeV and $y \sim 0.2$ would result in weak washout at the TeV scale ($K_\Psi^{\rm eff}\sim 0.03$) and is consistent with the $\mu \rightarrow e \gamma$ bound. We then can pick $\langle \Phi_{\lambda} \rangle = 3$ TeV to obtain the right SM neutrino mass scale. 

One might worry that with the hierarchies introduced it may not be possible to obtain a relatively anarchic SM neutrino mass matrix. We should however note that even though we are discriminating different $N$ generations (labeled by $i, j, ...$), we are not introducing any hierarchies distinguishing different $\Psi$ families (labeled by $a, b, ...$) or SM lepton flavors (labeled by $\alpha, \beta, ...$), and this results in an anarchic SM neutrino mass matrix in the flavor basis. Still, in the limit of $r \rightarrow 0$ the rank of the $\lambda$ matrix is reduced by one. This in turn reduces the rank of $\mu$ and SM neutrino mass matrices.  So in order to have a realistic neutrino mass matrix in scenarios with small $r$, we need to consider at least three generations of $N_i$. We take $N_3$ to have Yukawa couplings comparable to those of $N_2$. Choosing $M_{N_3}$ larger than $M_{N_2}$ by a factor of  a few ensures that contributions of $N_3$ to the CP asymmetry and to off-shell scattering washout 
are subdominant compared to those of $N_2$. Note that this scenario with small $r \sim 10 ^{-3}$ and three generations of $N$ results in one of the SM neutrino mass eigenvalues being much smaller than the other two, by a factor of $\sim r^2 \sim 10^{-6}$. Such a small mass for the lightest neutrino mass eigenstate is of course consistent with the current neutrino data.

\subsubsection{Intermediate scales: $\sim 10^9-10^{ 15 }$ GeV}

The region of parameter space with $M_N\sim 10^9-10^{ 15 }$ GeV works in our model as well as in the usual type I case. An example of a working point is presented, for $M_{N}\sim 10^{11}$ GeV, in table \ref{tab:benchmark}. No hierarchies in the Yukawa or mass matrices nor small Yukawa coupling are needed for this case.\footnote{As mentioned in
footnote~\ref{fn:anarchy}, this is the smallest value of $M_N$ which works with anarchy.} For the presented benchmark point, the asymmetry in the UV is generated in strong washout regime ($K_1 \sim 30$)  and the washout at the TeV scale is negligible ($K^{\rm eff}_\Psi \sim 0.2$).

%%%%%%%%%
\section{Conclusion and outlook}

\label{conclude}

%%%%%%%%%%%%%%%%%%%%%%%%
%
%
%%%%%%%%%%%%%%%%%%%%%%%%

The seesaw mechanism has been very successful in explaining the extreme smallness of the SM neutrino masses. At the same time the Majorana nature of SM neutrino, i.e., lepton-number violation, raises the highly attractive possibility of baryogenesis via leptogenesis. 
Part of our focus in this paper was devoted to a specific realization, the inverse seesaw. An attractive feature of this scenario is that the new TeV fermion singlets introduced to generate the small neutrino masses have unsuppressed couplings to the SM and are thus accessible at colliders.\footnote{These singlets could also be charged under new gauge symmetries broken at the TeV scale, giving additional production channels.} Unfortunately, this set-up also has two drawbacks. First, to obtain a small $m_\nu$ one needs to introduce a tiny Majorana mass term (for one chirality of the singlet) $\mu\sim1$ keV. Even though such a choice is technically natural, its value has no fundamental explanation and appears as a new scale of nature. %
Secondly, as we investigated in great detail in the present paper (a program which we started in \cite{short}), it is difficult to achieve successful leptogenesis from decays of the TeV-mass singlets, especially if we stick to the philosophy that the Yukawa couplings are not small {\em and} there is no particular flavor structure.

An indication that leptogenesis is hard to achieve in the inverse seesaw already appeared in \cite{Deppisch:2010fr}. A parametric estimate of the final asymmetry first appeared in our earlier paper~\cite{short} for $\mu$-terms and singlet mass being {\em anarchic} i.e., roughly of the same order, but not degenerate. There we showed that the final asymmetry --- in the case of strong washout --- is basically of the order of the mass of singlet in Planck scale units, and thus {\em in}dependent of the size of the Yukawa coupling and $\mu$. This demonstrates that the effect is too small for singlet masses in the TeV ballpark. 

In the present paper we have substantially elaborated upon this result in several aspects. First, we thoroughly studied the case of weak washout. This scenario turns out to involve various subtleties, but does {\em not} alter the conclusion regarding {\em un}successful leptogenesis. In addition, we generalized our earlier analysis to include scenarios with a degeneracy among different generations of singlets (which might require flavor symmetries to be at play); we demonstrated that this can indeed enhance the asymmetry compared to the above estimate, but still only barely reaching the observed value. Our approach is mostly based on analytical approximations, as opposed to earlier fully numerical studies. We think our approach makes the physics more transparent and represents a useful reference for future work.

We then moved on to another scenario with small lepton number violation and TeV scale singlets, the linear seesaw model. Here the singlets are {\em purely} Dirac and the lepton-number breaking needed for generating the SM neutrino (Majorana) mass arises from a small Yukawa coupling of one chirality (with the Yukawa coupling of the other being unsuppressed as in the inverse seesaw case). We showed that this model also encounters a similar fate in so far as leptogenesis is concerned. Interestingly, however, we found that a {\em merger} of the above two models (linear plus inverse) can indeed achieve successful leptogenesis. Unfortunately, this requires Yukawa couplings that are so small that collider and low-energy signatures are not relevant.

All of these problems are evaded by the hybrid seesaw scenario we proposed in \cite{short}. This picture combines the original inverse seesaw model with a high-scale module and simultaneously (1) motivates why $\mu$ is small and (2) realizes a successful leptogenesis with couplings of order unity.  

In the hybrid seesaw we add a {\em super}-heavy singlet (mass $M_N\gg$ TeV) which is {\em Majorana} in nature and {\em replace} the $\mu$-term by a mass-term mixing the TeV fermions with this new super-heavy singlet. In turn, this mass mixing originates from the VEV of a scalar field of size $O ( \hbox{TeV} )$, with the associated Yukawa coupling also being unsuppressed. Integrating out the super-heavy (Majorana) singlet then generates an effective Majorana mass (i.e., $\mu$-) term for the TeV mass singlets, which is super-small due to a {\em high}-scale seesaw
structure (akin to how the SM neutrino mass itself is rendered small in the {\em conventional} seesaw case). This provides an explanation of the smallness of $\mu$: it is the ratio of the two fundamental scales in the problem (the TeV and $M_N$).

Crucially, the super-heavy singlet does not couple directly to the SM Higgs-lepton sector. Hence, there is no contribution to the {\em SM} neutrino mass operator in the effective field theory just below the super-heavy singlet mass: it is clearly {\em not} quite the conventional high-scale seesaw for SM neutrino mass. Indeed, as in the original {\em inverse} seesaw model, the SM neutrino mass operator appears only at the {\em TeV} scale after the exchange of the particles in the inverse seesaw module. In other words, lepton-number {\em is} broken at a super-high scale by the mass of the super-heavy singlet, but this symmetry breaking has to be ``communicated'' to the SM Higgs-lepton via TeV-mass singlets. In this sense the neutrino masses emerge from a {\em hybrid} seesaw structure here, i.e., a combination of high-scale and inverse seesaw.

At the same time, decays of super-heavy singlet into the inverse seesaw fermions (and the physical scalar associated with their mass mixing) can create an asymmetry in the latter's abundance. Subsequently, the original asymmetry in the TeV-mass singlets gets transferred to the SM leptons, and ultimately quarks. Again, just like in the generation of the SM neutrino mass, it is the TeV-mass singlets which carry the message of high-scale lepton-number breaking (here in the form of asymmetry) to the SM leptons. In addition, while decays of TeV-mass singlets generate negligible lepton asymmetry (as seen when studying leptogenesis in the ordinary inverse seesaw), low scale physics can potentially lead to significant washout of the ``primordial'' asymmetry. As a result, we have shown that there is a subtle interplay between UV physics (i.e., related to the super-heavy singlet) and IR dynamics (the $\O (\hbox{TeV} )$-mass singlets). For certain choices of parameters, the UV asymmetry is already of roughly the observed size, in which case the IR washout has to be rather weak. Alternatively, the UV asymmetry could even be too large, and then diluted appropriately by strong washout in the IR. This interplay, merely outlined in~\cite{short}, has been studied in great detail in the present paper.

Due to the above structure, the hybrid model has a much wider spectrum of allowed couplings and masses compared to high-scale scenarios. As already mentioned in the previous paper, one can for example extend the range of super-heavy singlet masses, going above $\sim 10^{15}$ GeV and below $\sim 10^9$ GeV (the Davidson-Ibarra bound). The lowest mass regime was not investigated in detail in~\cite{short} because going to such low scales requires hierarchies in the couplings (which we assumed to be absent there). In the present paper, however, using the accurate formulas derived here and allowing modest hierarchies among couplings and masses, we have worked out in detail several benchmark scenarios. In particular, a careful discussion of the {\em lower} bound on the heavy singlet mass (which is, as already mentioned, lower than that in the standard seesaw and might thus ameliorate the SUSY gravitino problem) is presented. Overall, we believe the hybrid seesaw represents a {\em new} paradigm for leptogenesis.

In this paper we have also demonstrated that the specific coupling structure that characterizes the hybrid seesaw can be the result of a gauge symmetry. The gauged model we presented however should only be viewed as a partial UV completion of the hybrid scenario because it does not motivate why the inverse seesaw module, including the new scalars, lies at the TeV scale. A full explanation of the hybrid seesaw structure is given by warped extra dimensions \cite{short, Agashe:2015izu}, a framework that also addresses the hierarchy problem.

In fact, the hybrid seesaw scenario analyzed in the present paper may be viewed as an effective description of the warped/composite seesaw of \cite{Agashe:2015izu}. However, it remains to be seen whether that specific UV completion of the hybrid seesaw realizes leptogenesis precisely as discussed in this paper. There are in fact a few qualitative differences that might play an important role.
First, in warped extra dimensions/composite Higgs the inverse seesaw module is replaced by an entire {\em tower} of resonances (Kaluza-Klein states). What is the impact of these new degrees of freedom on leptogenesis? 
Relatedly, the TeV scale is associated to a {\em phase transition} at $\sim$ TeV temperatures. Note that in the models we discussed here there is also a phase transition when the vacuum of the scalars $\Phi_{\lambda,\kappa}$ is formed, but that is expected to be a smooth transition with no large entropy production. In warped extra dimensions, on the other hand, the phase transition is expected to be strongly first order \cite{Creminelli:2001th}. What implications does this have on leptogenesis? 
Finally, at temperatures relevant to the heavy singlet decay, the geometry of the extra dimension is qualitatively different from that at zero temperature \cite{Creminelli:2001th}. Does this affect leptogenesis in any way? We defer a discussion of these interesting questions to a follow-up paper.

\section*{Acknowledgements}

We would like to thank Bhupal Dev, Andre de Gouvea, Pedro Machado, Rabindra Mohapatra, Qaisar Shafi, Geraldine Servant, Jessica Turner and Yue Zhang for discussions.
We would also like to thank the authors of reference \cite{Dolan:2018qpy} for clarifications about their results.
The work of KA, PD, and ME was supported in part by NSF Grant No.~PHY-1620074 and the Maryland Center for Fundamental Physics. 
KA and PD were 
also supported by the Fermilab Distinguished Scholars Program and 
SH by NSF Grant No.~PHY-1719877. 
CSF acknowledges the hospitality of Fermilab theory group during his visit supported by FAPESP grant 2017/02747-7 when part of the work was carried out.
LV is supported by the Swiss National Science Foundation under the Sinergia network CRSII2-16081.

\appendix

\section{Leptogenesis in TeV scale linear seesaw}\label{app:sec3}
In section~\ref{inverse_lepto}, we discussed leptogenesis in TeV scale inverse seesaw model in detail.
Now we move on to study another well-motivated seesaw model, the linear seesaw. The Lagrangian is
\bea\label{eq:L_y'}
-{\mathcal L}_{\rm LSS}&=& y_{a\al}\Psi_a^cH\ell_\al +(m_\Psi)_{ab}\Psi_a\Psi^c_b+ y'_{a\al}\Psi_a H\ell_\al+{\rm h.c.},
\eea
where $\al$ is the SM lepton flavor index and $a,b=1,2$ denotes the generations of $\Psi,\Psi^c$.
By redefining the fields, the $(m_{\Psi})_{ab}$ matrix can be made real and diagonal whereas $y_{a\al}$ and $y'_{a\al}$ are complex. In the mass basis, we get the same form as in eq.~(\ref{eq:diagonalization}), but the parameters are changed to
\bea\label{eq:y'_Mandh}
m_1=m_2= m_{\Psi_1}~&;&~h_{1\al}= \frac{y_{1\al}+ y'_{1\al}}{\sqrt{2}}~,~h_{2\al}= \frac{i(y_{1\al} -y'_{1\al})}{\sqrt{2}}\nn
m_3=m_4= m_{\Psi_2}~&;&~h_{3\al}= \frac{y_{2\al}+ y'_{2\al}}{\sqrt{2}}~,~h_{4\al}= \frac{i(y_{2\al} -y'_{2\al})}{\sqrt{2}}.
\eea
The condition for small lepton number breaking in this case is $\vep'_{a\al}\ll 1$, where we define $\vep^{\prime}_{a\al}\equiv y'_{a\al}/y_{a\al}$.
It is clear that taking $\vep'_{1\al} \to 0$ limit, $h_{1\al}=ih_{2\al}$ and thus $(\tilde \Psi_1, \tilde \Psi_2)$ become a Dirac pair. \\
In this section, we only focus on the effect from the Lagrangian in eq.~(\ref{eq:L_y'}). Loops will generate a mass spitting between singlets within the same generation, which will effectively generate a $\mu$ term as in the ISS models. Though the size of this $\mu$ term is loop suppressed, it might change the parametric dependance of the final baryon asymmetry. We have studied the models with both $y'$ and $\mu$ turned on in section \ref{sec:ISS+LSS}.\\
\noindent\textbf{CP asymmetry}\\
According to the definition of the CP asymmetry parameter for $\tilde\Psi_i$, denoted as $\ep_i$ in eq.~(\ref{eq:epsilon}), we have 
\bea\label{eq:ep'12}
\ep_1+\ep_2&=&\frac{1}{8\pi (hh^\dagger)_{22}}\left(\sum_{i\neq 1}\textrm{Im}[(hh^\dagger)^2_{1i}]f_{1i}+\sum_{j\neq 2}\textrm{Im}[(hh^\dagger)^2_{2j}]f_{2j} \right)\nonumber\\
&~&+\frac{1}{8\pi}\frac{(hh^\dagger)_{22}-(hh^\dagger)_{11}}{ (hh^\dagger)_{11}(hh^\dagger)_{22}}\left(\sum_{i\neq 1}\textrm{Im}[(hh^\dagger)^2_{1i}]f_{1i} \right),
\eea
 Notice that $f_{ij}=f^{\rm v}_{ij}+f^{\rm self}_{ij}$, where $f^{\rm v}_{ij}$ and $f^{\rm self}_{ij}$ are defined in eqs.~(\ref{eq:fv}) and (\ref{eq:fself}) respectively. lt is easy to show that in pure linear seesaw [eq.~(\ref{eq:y'_Mandh})] with singlets in different generations being non-degenerate (i.e., $|m_{\Psi_2}-m_{\Psi_1}|>\Gamma_i$), 
\bea
\left.
\begin{array}{c}
f^{\rm v}_{12}=f^{\rm v}_{21} \\
 f^{\rm self}_{12}=f^{\rm self}_{21}=0 
\end{array}
\right\}~&\Rightarrow&~f_{12}=f_{21}\nn
\left.
\begin{array}{c}
f^{\rm v}_{13}=f^{\rm v}_{14}= f^{\rm v}_{23}= f^{\rm v}_{24} \\
f^{\rm self}_{13}\approx f^{\rm self}_{14}\approx f^{\rm self}_{23}\approx f^{\rm self}_{24}
\end{array}
\right\}~&\Rightarrow&~f_{13}\approx f_{14}\approx f_{23}\approx f_{24}
\eea
Using such relation and  $\textrm{Im}[(hh^\dagger)^2_{12}]=\textrm{Im}[(hh^\dagger)^2_{21}]=0$, eq.~(\ref{eq:ep'12}) reduces to
\bea\label{eq:ep'12_simp}
\ep_1+\ep_2&\approx&\frac{1}{8\pi (hh^\dagger)_{22}}\left(\textrm{Im}[(hh^\dagger)^2_{13}]+\textrm{Im}[(hh^\dagger)^2_{14}]+\textrm{Im}[(hh^\dagger)^2_{23}]+\textrm{Im}[(hh^\dagger)^2_{24}]\right)f_{13}\nn
&~&-\frac{1}{8\pi}\frac{(y'y^\dagger)_{11}+(yy^{\prime\dagger})_{11}}{ (hh^\dagger)_{11}(hh^\dagger)_{22}}\left(\textrm{Im}[(hh^\dagger)^2_{13}]+\textrm{Im}[(hh^\dagger)^2_{14}]\right)f_{13}.
\eea
It is worth mentioning that, if we only consider one generation of singlets, meaning only two degenerate Majorana states $\tilde\Psi_{1,2}$ left, $\ep_1+\ep_2$ will vanish due to the absence of the CP phase.\\
Furthermore,  according to eq.~(\ref{eq:y'_Mandh}), we can find that 
\bea
(hh^\dagger)^2_{13}&=&\frac{1}{4}\left[(yy^\dagger)_{12}+(y'y^\dagger)_{12}+(yy^{\prime\dagger})_{12}+(y'y^{\prime\dagger})_{12}\right]^2\nn
(hh^\dagger)^2_{14}&=&-\frac{1}{4}\left[(yy^\dagger)_{12}+(y'y^\dagger)_{12}-(yy^{\prime\dagger})_{12}-(y'y^{\prime\dagger})_{12}\right]^2\nn
(hh^\dagger)^2_{23}&=&-\frac{1}{4}\left[(yy^\dagger)_{12}-(y'y^\dagger)_{12}+(yy^{\prime\dagger})_{12}-(y'y^{\prime\dagger})_{12}\right]^2\nn
(hh^\dagger)^2_{24}&=&\frac{1}{4}\left[(yy^\dagger)_{12}-(y'y^\dagger)_{12}-(yy^{\prime\dagger})_{12}+(y'y^{\prime\dagger})_{12}\right]^2,
\eea
and the sum
\bea\label{eq:sum_hs}
(hh^\dagger)^2_{13}+(hh^\dagger)^2_{14}&=&(yy^\dagger)_{12}(yy^{\prime\dagger})_{12}+\mathcal O(y^{\prime2})\nn
(hh^\dagger)^2_{13}+(hh^\dagger)^2_{14}+(hh^\dagger)^2_{23}+(hh^\dagger)^2_{24}&=&2\left[(yy^\dagger)_{12}(y'y^{\prime\dagger})_{12}+(y'y^\dagger)_{12}(yy^{\prime\dagger})_{12}\right].\nn
\eea
Plugging eq.~(\ref{eq:sum_hs}) into eq.~(\ref{eq:ep'12_simp}), one could obtain
\bea
\ep_1+\ep_2\approx \frac{\textrm{Im}\left[(yy^\dagger)_{12}(y'y^{\prime\dagger})_{12}+(y'y^\dagger)_{12}(yy^{\prime\dagger})_{12}-2(yy^\dagger)_{12}(yy^{\prime\dagger})_{12}|yy^{\prime\dagger} |_{11}/(yy^\dagger)_{11}\right]}{2\pi(yy^\dagger)_{11}}f_{13},
\eea
to the second order in $y'$.
Assuming no hierarchy among $y_{a \al}$($y'_{a \al}$) and $m_{\Psi_a}$ are not degenerate, namely $f_{13}$ is $O(1)$ factor, the CP asymmetry can be schematically written as
\bea\label{eq:ep'}
\ep\equiv\ep_1+\ep_2\sim \frac{\textrm {Im}[(y'y^{\prime\dagger})(yy^\dagger)]}{(yy^\dagger)}\sim \frac{\Gamma}{m_\Psi}\vep^{\prime 2}  ~~~(\vep'\ll 1),
\eea
where  $\vep'$ is the schematic notation for $\vep'_{a\al}$ . 

According to eq.~(\ref{eq:epsilon}), one can also find that 
\bea
\ep_1\approx-\ep_2 \sim \frac{\Gamma}{m_\Psi}\vep^{\prime },
\eea
which is first order in $\vep'$, while the sum $\ep$ is second order in $\vep'$ [see eq.~(\ref{eq:ep'})]. As argued in appendix~\ref{app:BE_ISS_LSS}, we shall use the sum $\ep$ instead of $\ep_1$ or $\ep_2$ in the estimation of final asymmetry.

\noindent\textbf{Washouts and baryon asymmetry}\\
Now we want to evaluate the effective washouts in linear seesaw. Follow the method in ref. \cite{Blanchet:2009kk}, one can calculate
\bea\label{eq:K'_tot}
K^{ \rm eff }\sim \frac{\Gamma}{H}\mathcal \vep^{\prime 2}.
\eea
Using the formula for baryon asymmetry [eq.~(\ref{eq:YB})] and the efficiency factor $\eta\lesssim 1/K^{ \rm eff }$, the baryon asymmetry in linear seesaw is
\bea
Y_{\Delta B}\lesssim 10^{-3}\frac{\ep}{K^{\rm eff }}\lesssim 10^{-3}\sqrt{g_*}\frac{m_\Psi}{M_{\rm Pl}},
\eea
which is remarkably the same as the result in inverse seesaw [eq.~(\ref{eq:general-YB})]. Leptogenesis in linear seesaw is also summarized in table~\ref{tab:results}.
%
%
%
%%%
\section{Boltzmann equations and analytical approximate solutions\label{app:approximate_solutions}}

We start with a brief review of the general BEs in the next section before proceeding to derive analytical approximate solutions used in this work in section \ref{app:approx_sols}.

\subsection{Generalities}\label{app:BE_gen}
A general BE describing the evolution of $n_X$ in time $t$ can be written as~\cite{Kolb:1979qa}
%%%
\bea
\frac{dn_X}{dt} + 3 H n_X 
= - \sum_{b,...,i,j,...} [Xb ... \leftrightarrow ij ...]
\label{eq:BE_general}
\eea
%%%
where the Hubble rate is $H = 1.66\sqrt{g_\star} \frac{T^2}{M_{\rm Pl}}$ with $g_\star$ the total number of relativistic degrees of freedom (of the Universe) and $M_{\rm Pl} = 1.22 \times 10^{19}$ GeV the Planck mass and
%%%
\bea
[Xb ... \leftrightarrow ...] 
&=& \Lambda^{ij...}_{Xb...}
\left[
|{\cal A}(Xb...\to ij...)|^2 
f_X f_b ... 
(1 + \eta_i f_i)(1 + \eta_j f_j) ... 
\right.
\nonumber \\
& & 
\left.
- |{\cal A}(ij...\to Xb...)|^2 
f_i f_j ... 
(1 + \eta_X f_X)(1 + \eta_b f_b) ...
\right],
\eea
%%%
where
%%%
\bea
\Lambda^{ij...}_{Xb...} 
&\equiv& 
\int d\Pi_X d\Pi_b ... d\Pi_i d\Pi_j ...
(2\pi)^4 \delta^{(4)}(p_X + p_b + ... - p_i -p_j -...)
\nonumber \\
d\Pi_x 
&\equiv&
\frac{d^3p_x}{(2\pi)^3 2 E_x}.
\eea
%%%
In the above, $f_x$ is general phase space distribution with $\eta_x = 1(-1)$ if $x$ is a boson (fermion) and $|{\cal A}(ab...\to ij...)|^2$ is the squared amplitude summed over initial and final spin states and gauge multiplicities. 

In our study, we will consider $X$ as the only massive particle with mass $M_X$ while all other particles are massless. In this scenario, it is convenient to trade $t$ for $z \equiv \frac{M_X}{T}$ and number density $n_x$ for \emph{abundance} $Y_x = \frac{n_x}{s}$ where $s = \frac{2\pi^2}{45}g_\star T^3$ is the entropic density. In this case, eq.~(\ref{eq:BE_general}) can be written, during radiation-dominated epoch, as
%%%
\bea
sHz\frac{dY_X}{dz} 
= - \sum_{b,...,i,j,...} [Xb ... \leftrightarrow ij ...].
\label{eq:BE_general2}
\eea
%%%

For massless particles, we assume kinetic equilibrium with phase space distribution $f_x = (e^{\frac{E_x - \mu_x}{T}} - \eta_x)^{-1}$ where $\mu_x$ is the chemical potential for $x$. For real scalar, we have $\mu_x = 0$; otherwise, we assume the chemical potential of the antiparticle $x^*$ is given by $\mu_{x^*} = - \mu_x$. Given that $n_x = \int \frac{d^3 p}{(2\pi)^3} f_{x}$, at leading order in $\mu_x/T$, we have the relation
%%%
\bea
\frac{2\mu_x}{T} & = & 
\frac{Y_{\Delta x}}
{\zeta_x g_x Y^{\rm eq}},
\eea
%%%
where $Y_{\Delta x} \equiv \frac{n_x - n_{\bar x}}{s}$, $g_x$ is the number of degrees of freedom of $x$ and $\zeta_x = 1(2)$ for relativistic fermion (boson) and 
%%%
\bea
Y^{\rm eq} = \frac{15}{8\pi^2 g_\star}.
\label{eq:Y_eq}
\eea
%%%

For the massive particle $X$, we approximate $f_X \approx \frac{Y_X}{Y_X^{\rm eq}} f_X^{\rm eq}$ 
where $f_x^{\rm eq} = (e^{E_x/T} - \eta_x)^{-1}$ is the equilibrium phase space distribution and $Y_x^{\rm eq} = n_x^{\rm eq}/s$ denote the equilibrium abundance of $x$.

As shown in detail in appendix A of refs.~\cite{Fong:2011yx}, with the above approximations and expanding the right-hand side of eq.~(\ref{eq:BE_general2}) up to first order in $\mu_x/T$, the BEs can be written in terms of $Y_X$, $Y_X^{\rm eq}$, $Y_{\Delta x}$, $\zeta_x g_x Y^{\rm eq}$ and (equilibrium) thermal averaged reaction densities
%%%
\bea
\gamma(ab... \leftrightarrow ij...)
& \equiv &
\Lambda^{ij...}_{ab...}
\left[
|{\cal A}(ab...\leftrightarrow ij...)|^2 
f_a^{\rm eq} f_b^{\rm eq} ... 
(1 + \eta_i f_i^{\rm eq})(1 + \eta_j f_j^{\rm eq}) ...
\right].
\label{eq:thermal_reaction_densities}
\eea
%%%
%where $f_x^{\rm eq} = (e^{E_x/T} - \eta_x)^{-1}$. 
Notice that for the time reversal process, the only difference is in the squared amplitude while the phase space distribution combination remains the same i.e. $f_i^{\rm eq} f_j^{\rm eq} ... (1 + \eta_a f_a^{\rm eq}) (1 + \eta_b f_b^{\rm eq}) ... = f_a^{\rm eq} f_b^{\rm eq} ... 
(1 + \eta_i f_i^{\rm eq})(1 + \eta_j f_j^{\rm eq}) ...$ due to energy conservation. Finally, as discussed in section~\ref{subsec:generalities}, once the approximate $U(1)$ charges are identified, all particle asymmetries $Y_{\Delta x}$ can be expressed in term of these charges as in eq.~(\ref{eq:particle_asymmetry}).

Assuming Maxwell-Boltzmann distribution and ignoring the Fermi-Dirac/Bose-Einstein factor $1 + \eta_x f_x^{\rm eq}$, the (inverse) decay process $X \leftrightarrow ij$ can be written as
%%%
\bea
\gamma(X \leftrightarrow ij) = n_X^{\rm eq} \Gamma(X \leftrightarrow ij) \frac{{\cal K}_1(z)}{{\cal K}_2(z)},
\label{eq:decay_rec_den}
\eea
%%%
where $\Gamma(X \to ij)$ is the decay width for $X \to ij$ , $\Gamma(ij \to X) = \Gamma(X^* \to i^* j^*)$ should be interpreted as the CP conjugate process, and $\frac{{\cal K}_1(z)}{{\cal K}_2(z)}$ is the thermal averaged time dilation factor with $K_n(z)$ the modified Bessel function of second kind of $n$-th order. The equilibrium number density of $X$ is
%%%
\bea
n_X^{\rm eq} = \frac{g_X}{2\pi^2} z^2 {\cal K}_2(z),
\label{eq:X_therm_den}
\eea
%%%
where $g_X$ is the number degrees of freedom of $X$.

Under the same approximations as above, for the scatterings $ij \leftrightarrow kl$, we have
%%%
\bea
\gamma(ij \leftrightarrow kl) 
= n_i^{\rm eq} n_j^{\rm eq} \left< \sigma(ij \leftrightarrow kl) \right>
= n_i^{\rm eq} \Gamma(ij \leftrightarrow kl),
\label{eq:scatt_rec_den}
\eea
%%%
where $\left< \sigma(ij \leftrightarrow kl) \right>$ is the thermal averaged cross section, $n_i^{\rm eq} = n_j^{\rm eq} = \frac{T^3}{\pi^2}$ and we have defined the scattering rate as\footnote{The number of degrees of freedom for initial and final states have been absorbed into the cross section $\sigma(ij \leftrightarrow kl)$. }
%%%
\bea
\Gamma(ij \leftrightarrow kl) 
\equiv n_j^{\rm eq}\left< \sigma(ij \leftrightarrow kl) \right>
= \frac{\gamma(ij \to kl)}{n_i^{\rm eq}}. 
\label{eq:scatt_rate_gen}
\eea
%%%

Finally, one can define the CP parameter for the decay $X \to ij$ as
%%%
\bea
\epsilon(X \to ij) = 
\frac{\gamma(X \to ij) - \gamma(X^* \to i^* j^*)}{\gamma_X},
\label{eq:CP_parameter0}
\eea
%%%
where we have defined the total decay reaction density as
%%% 
\bea
\gamma_X \equiv \sum_{ij}
\left[\gamma(X \to ij) + \gamma(X^* \to i^* j^*)\right].
\label{eq:total_decay_rec_den}
\eea
%%%
With the Maxwell-Boltzmann approximation, eq.~(\ref{eq:CP_parameter0}) can be written only in term of decay widths $\Gamma(X \to ij)$ and $\Gamma(X^* \to i^*j^*)$.

%%%%%%%%%%%%%%
\subsection{Analytical approximate solutions}\label{app:approx_sols}

One can construct the BEs according to the procedure discussed in the previous section. In this section, we will derive analytical approximate solutions to the set of BEs used in sections \ref{sec:Formalism} and \ref{results}. The BEs we consider involves only decays and inverse decays of a heavy particle $X$ of mass $M_{X}$, which captures the dominant generation and washout of the asymmetry in $\Delta$:
%%%
\begin{eqnarray}
\frac{dY_{X}}{dz} & = & -D\left(z\right)\left(\frac{Y_{X}}{Y_{X}^{{\rm eq}}}-1\right),\label{eq:BE_YX_app}\\
\frac{dY_{\Delta}}{dz} & = & \epsilon D\left(z\right)\left(\frac{Y_{X}}{Y_{X}^{{\rm eq}}}-1\right)-\frac{1}{2}cD\left(z\right)\frac{Y_{\Delta}}{Y^{{\rm eq}}},\label{eq:BE_Yq_app}
\end{eqnarray}
%%%
where $z = \frac{M_{X}}{T}$, $\epsilon$ is the CP parameter\footnote{In eq.~(\ref{eq:BE_Yq_app}) one may notice that there is overall sign difference for the term $\propto \epsilon$ compared to equations appearing in section~\ref{subsec:quantitative}. This sign depends on the precise definition of the aymmetry parameter $\Delta$ and equations with one sign $\epsilon$ are related to equations with opposite sign $\epsilon$ by a simple change $\epsilon \to - \epsilon$. Physically, $\epsilon \to - \epsilon$ just changes the notion of particle $\leftrightarrow$ anti-particle. } from decay of $X$ defined in eq.~(\ref{eq:CP_parameter0}), $Y^{{\rm eq}}=\frac{15}{8\pi^{2}g_{\star}}$ [see eq.~(\ref{eq:Y_eq})] and $Y_{X}^{{\rm eq}}=\frac{45 g_X}{4\pi^{4}g_{\star}}z^{2}{\cal K}_{2}\left(z\right)$ [see eq.~(\ref{eq:X_therm_den})].
%.
In the (inverse decay) washout term [the second term
of eq.~(\ref{eq:BE_Yq_app})], the coefficient $c$ can account for the following two effects. Firstly, it can capture the relevant spectator effects~\cite{Buchmuller:2001sr,Nardi:2005hs}. In hybrid-genesis discussed in section \ref{sec:Formalism}, the spectator effects are captured by $c=c_{W_1}$ defined in eq.~(\ref{eq:defc}) or $c=c_{W_2}$ defined in eq.~(\ref{eq:BE_washout}) with their values given in appendix \ref{app:Spectator-effects}.
Moreover, choosing $c=K^{\rm eff}/K\ll1$ can also represent the reduced washout due to approximate symmetry in small lepton number violating models. For example, one can set $c\sim (\mu/\Gamma)^2$ in ISS (section~\ref{inverse_lepto}) or $c\sim (y'/y)^2$ in LSS (appendix~\ref{app:sec3}) to study the BEs for leptogenesis in these models. The total decay reaction density eq.~(\ref{eq:total_decay_rec_den}) compared to the Hubble expansion rate $H$ is denoted as 
%%%
\begin{eqnarray}
D\left(z\right) & \equiv & \frac{\gamma_{X}}{sHz}=Kz\frac{{\cal K}_{1}\left(z\right)}{{\cal K}_{2}\left(z\right)}Y_{X}^{{\rm eq}}=\frac{1}{2}KY_{X}^{{\rm eq}}(0)z^{3}{\cal K}_{1}\left(z\right),\label{eq:decay_term}
\end{eqnarray}
%%%
where the \emph{washout parameter} is defined as
%%%
\begin{eqnarray}
K & \equiv &\frac{\Gamma_{X}}{H(T=M_{X})},\label{eq:washout_param}
\end{eqnarray}
%%%
with $\Gamma_{X}$ the total decay width of $X$.

Let's first study the Boltzmann equation for $Y_{X}$ as in eq. (\ref{eq:BE_YX_app}). Assuming  $Y_{X}\left(z_{i}\right)=0$,
we can define $z_{{\rm eq}}$ as the temperature in which
%%%
\begin{eqnarray}
Y_{X}\left(z_{{\rm eq}}\right) & = & Y_{X}^{{\rm eq}}\left(z_{{\rm eq}}\right).\label{eq:zeq_def}
\end{eqnarray}
%%%

For $z<z_{{\rm eq}}$, we can approximate $\frac{dY_{X}}{dz}\approx D\left(z\right)$
and obtain
%%%
\begin{eqnarray}
Y_{X}\left(z\right) \approx\int_{z_{i}}^{z}dz'D\left(z'\right)=\frac{1}{2}Y_{X}^{{\rm eq}}(0)Kf\left(z_{i},z\right),
\end{eqnarray}
%%%
where we have set $Y_X(z_i) = 0$ and defined
%%%
\begin{eqnarray}
f\left(z_{i},z\right) & \equiv & \int_{z_{i}}^{z}dz'z'^{3}{\cal K}_{1}\left(z'\right).\label{eq:f_function}
\end{eqnarray}
%%%

Taking high initial temperature $z_{i}\to0$, let us consider the
following two cases. For $K\ll1$, $X$ reaches its equilibrium abundance
at late time $z_{{\rm eq}}\gg1$ and we can approximate\footnote{We approximate the result with the identity $f\left(0,\infty\right)=\frac{3\pi}{2}$.}
%%%
\begin{eqnarray}
Y_{X}\left(z_{\rm eq}\right) & \approx & \frac{3\pi}{4}Y_{X}^{{\rm eq}}(0)K\equiv Y_{a}.\label{eq:YK_a}
\end{eqnarray}
%%%
On the other hand, for $K\gg1$, $X$ reaches its equilibrium abundance
at early time $z_{{\rm eq}}\ll1$ and we can approximate\footnote{We approximate the result with $f\left(0,z\ll1\right)=\frac{z^{3}}{3}$. }
%%%
\begin{eqnarray}
Y_{X}\left(z_{\rm eq}\right) & \approx & \frac{1}{6}Y_{X}^{{\rm eq}}(0)Kz_{{\rm eq}}^{3}\equiv Y_{b}.\label{eq:YK_b}
\end{eqnarray}
%%%
According to the definition of $Y_{X}\left(z_{{\rm eq}}\right) = Y_{X}^{{\rm eq}}\left(z_{{\rm eq}}\right)\approx Y_{X}^{{\rm eq}}\left(0\right)$, we have $z_{{\rm eq}}\approx\left(6/K\right)^{1/3}$.

Next, we will look at the Boltzmann equation for $Y_{\Delta}$ as in eq.~\eqref{eq:BE_Yq_app}. It is convenient to parametrize the asymmetry
generated in $Y_{\Delta}$ by
%%%
\begin{eqnarray}
Y_{\Delta}(z) & \equiv & \epsilon\,\eta(z) Y_{X}^{{\rm eq}}(0),\label{eq:eta_parametrization}
\end{eqnarray}
%%%
where $\eta\equiv\eta(\infty)$ is known as the \emph{efficiency} factor, which shows the effect from washout.
$\eta\leq 1$ by definition and  we will get $\eta=1$ when there is thermal initial abundance of $Y_{X}$ and no washout.
Substituting eq. (\ref{eq:eta_parametrization}) into eq.
(\ref{eq:BE_Yq_app}), we have
%%%
\begin{eqnarray}
\frac{d\eta(z)}{dz} & = & \frac{D\left(z\right)}{Y_{X}^{{\rm eq}}(0)}\left(\frac{Y_{X}}{Y_{X}^{{\rm eq}}}-1\right)-\frac{1}{2}c\frac{D\left(z\right)}{Y^{{\rm eq}}}\eta.\label{eq:BE_Yeta}
\end{eqnarray}
%%%
Notice that the equation above is independent of $\epsilon$. This
simplification arises because we have considered zero temperature
CP parameter which is independent of temperature and the problem boils
down to solving for $\eta(z)$. The formal solution for the equation
above is
%%%
\begin{eqnarray}
\eta\left(z\right) & = & \eta\left(z_{i}\right)e^{-\frac{c}{2Y^{{\rm eq}}}\int_{z_{i}}^{z}dz'D\left(z'\right)}+\frac{1}{Y_{X}^{{\rm eq}}(0)}\int_{z_{i}}^{z}dz'D\left(z'\right)\left(\frac{Y_{X}}{Y_{X}^{{\rm eq}}}-1\right)e^{-\frac{c}{2Y^{{\rm eq}}}\int_{z'}^{z}dz''D\left(z''\right)}\nonumber \\
 & = & \eta\left(z_{i}\right)e^{-\frac{c}{2Y^{{\rm eq}}}\int_{z_{i}}^{z}dz'D\left(z'\right)}-\frac{1}{Y_{X}^{{\rm eq}}(0)}\int_{z_{i}}^{z}dz'\frac{dY_{X}}{dz^{'}}e^{-\frac{c}{2Y^{{\rm eq}}}\int_{z'}^{z}dz''D\left(z''\right)}.\label{eq:eta_solution_formal}
\end{eqnarray}
%%%
In the following, we will assume no initial asymmetry and set $\eta\left(z_{i}\right)=0$. (After
all, our aim is to generate an asymmetry dynamically.)

For $z\leq z_{{\rm eq}}$, we define $\eta^{-}\left(z\right)$:
%%%
\begin{eqnarray}
\eta^{-}\left(z\right)\equiv\eta\left(z\right) & \approx & -\frac{1}{Y_{X}^{{\rm eq}}(0)}\int_{z_{i}}^{z}dz'D\left(z'\right)e^{-\frac{c}{2Y^{{\rm eq}}}\int_{z'}^{z}dz''D\left(z''\right)}\nonumber \\
 & = & -\frac{2Y^{{\rm eq}}}{cY_{X}^{{\rm eq}}(0)}\left[1-e^{-\frac{c}{2Y^{{\rm eq}}}\int_{z_{i}}^{z}dz'D\left(z'\right)}\right]\nonumber \\
 & = & -\frac{2}{R\, c}\left[1-e^{-\frac{c}{2Y^{{\rm eq}}}Y_{X}\left(z\right)}\right]\label{eq:eta_minus},
\end{eqnarray}
%%%
where we use the approximation: $\frac{dY_{X}}{dz}\approx D\left(z\right)$ and define
%%%
\begin{eqnarray}
R & \equiv & \frac{Y_{X}^{{\rm eq}}(0)}{Y^{{\rm eq}}}.\label{eq:Rc}
\end{eqnarray}
%%%

For $z>z_{{\rm eq}}$, we have
\begin{eqnarray}
\eta\left(z\right) & = & \eta^{-}\left(z_{\rm eq}\right)e^{-\frac{c}{2Y^{{\rm eq}}}\int_{z_{\rm eq}}^{z}dz'D\left(z'\right)}
+
\eta^+(z),\label{eq:eta_solution_final}
\end{eqnarray}
where
\begin{eqnarray}
\eta^{+}\left(z\right) & = & -\frac{1}{Y_{X}^{{\rm eq}}(0)}\int_{z_{\rm eq}}^{z}dz'\frac{dY_{X}}{dz^{'}}e^{-\frac{c}{2Y^{{\rm eq}}}\int_{z'}^{z}dz''D\left(z''\right)}.\label{eq:eta_plus}
\end{eqnarray}
The first term of eq.~(\ref{eq:eta_solution_final}) is the contribution when $X$ is being populated while the the second is the contribution when $X$ decays.
Next we will discuss the solutions in the following regimes. 
%%%%%%%%%%%%%%%%%%%%%%%%
\vspace{0.5cm}
\subsubsection{Weak washout regime ($c K \ll 1$) with $Y_X(z_i)=0$}
\label{app:weak_washout}
As shown in eq.~(\ref{eq:BE_Yq_app}), the washout of the asymmetry is controlled by $c K$ while the generation is controlled by $K$. In the weak washout regime: $c K\ll1$, there is still freedom to choose $K \ll1 $ or $K \gg 1$ because $c\ll 1$ in ISS or LSS models.  The region $ c K \ll 1$ and $K \gg 1$ is not possible in the standard leptogenesis with type I seesaw due to $c$ being order unity. Now we shall discuss these two cases in the weak washout regime.

\noindent\textbf{Case I: $ c K \ll 1$ and $K \ll 1$}\\
Since  $z_{{\rm eq}}\gg1$ when $K\ll1$, we can neglect the washout for $z>z_{{\rm eq}}$ in eq.~\eqref{eq:eta_solution_final}.
Hence, we have
%%%
\begin{eqnarray}
\eta\left(z\right) & \approx & \eta^{-}\left(z_{\rm eq}\right)
-\frac{1}{Y_{X}^{{\rm eq}}(0)}\int_{z_{{\rm eq}}}^{z}dz'\frac{dY_{X}}{dz'}, \nonumber \\
& = & -\frac{2}{R\, c}\left[1-e^{-\frac{c}{2Y^{{\rm eq}}}Y_a}\right]
+
\frac{1}{Y_{X}^{{\rm eq}}(0)}\left[Y_a-Y_{X}\left(z\right)\right],
\label{eq:eta_solution_weak1}
\end{eqnarray}
%%%
where we have used eq.~\eqref{eq:YK_a}.

For the final efficiency, we take $z\to\infty$ where $Y_X(\infty) = 0$ and obtain
%%%
\begin{eqnarray}
\eta_{K \ll 1}^{w}\left(K,c\right) 
 & \approx & -\frac{2}{R\,c}\left[1-e^{-\frac{c}{2Y^{{\rm eq}}}Y_a}\right]
 +\frac{1}{Y_{X}^{{\rm eq}}(0)}Y_a\nonumber \\
 & \approx & -\frac{2}{R\,c}\left[\frac{c}{2Y^{{\rm eq}}}Y_a-\frac{1}{2}\left(\frac{c}{2Y^{{\rm eq}}}Y_a\right)^{2}\right]+\frac{1}{Y_{X}^{{\rm eq}}(0)}Y_a\nonumber \\
 & = & \frac{1}{Y_{X}^{{\rm eq}}(0)}\frac{c}{4Y^{{\rm eq}}}Y_a^{2}\nonumber \\
 & = & \frac{9\pi^{2}}{64}R\, c\,K^{2}.
 \label{eq:eff_weak_a}
\end{eqnarray}
%%%
In the above, we have expanded the exponent in $cK\ll1$ up to second order. If we choose $c\sim 1$, this gives the standard result $\eta_{K \ll 1}^{w}\sim K^2$ in the weak washout regime. While in the models with small lepton number breaking ($c=K^{\rm eff}/K\ll1$), we would have $\eta_{K \ll 1}^{w}\sim K^{\rm eff} K$.

\noindent\textbf{Case II: $ c K \ll 1$ and $K \gg 1$}\\
Since $z_{{\rm eq}}\ll1$ when $K\gg1$, we cannot neglect the washout for $z>z_{{\rm eq}}$ in eq.~\eqref{eq:eta_solution_final}. 
In this case, we have
%%%
%%%
\begin{eqnarray}\label{eq:Kgg1_approx_1}
 \eta^{-}\left(z_{\rm eq}\right)
e^{-\frac{c}{2Y^{{\rm eq}}}\int_{z_{{\rm eq}}}^{z}dz'D\left(z'\right)} & \approx & -\frac{2}{R\,c}\left[1-e^{-\frac{c}{2Y^{{\rm eq}}}Y_b}\right]e^{-\frac{c}{2Y^{{\rm eq}}}\int_{z_{{\rm eq}}}^{z}dz'D\left(z'\right)}\nonumber \\
 & \approx & -\frac{2}{R\,c}\left[1-e^{-\frac{1}{2}R\,c}\right]e^{-\frac{c}{2Y^{{\rm eq}}}\int_{z_{{\rm eq}}}^{z}dz'D\left(z'\right)},
 \eea
 and
 \bea\label{eq:Kgg1_approx_2}
\eta^{+}\left(z\right)
 & \approx & -\frac{1}{Y_{X}^{{\rm eq}}(0)}\int_{z_{{\rm eq}}}^{z}dz'\frac{dY_{X}^{{\rm eq}}}{dz'}e^{-\frac{c}{2Y^{{\rm eq}}}\int_{z'}^{z}dz''D\left(z''\right)}\nonumber \\
 & = & \frac{1}{Y_{X}^{{\rm eq}}(0)K}\int_{z_{{\rm eq}}}^{z}dz'\frac{1}{z'}D\left(z'\right)e^{-\frac{c}{2Y^{{\rm eq}}}\int_{z'}^{z}dz''D\left(z''\right)},
\end{eqnarray}
where we have used the following approximation:
%%%
\begin{eqnarray}
\frac{dY_{X}}{dz'} &\approx& \frac{dY_{X}^{{\rm eq}}}{dz'}=-Y_{X}^{{\rm eq}}\frac{{\cal K}_{1}\left(z\right)}{{\cal K}_{2}\left(z\right)}=-\frac{1}{Kz}D\left(z\right).
\end{eqnarray}
%%%
Writing the integrand as $e^{-g(z',z)}$, the dominant contribution for $\eta^+\left(z\right)$ comes from a
region around $z_B$ where $g(z',z)$ has a minimum.
Following the approximation of ref.~\cite{Buchmuller:2004nz} by replacing
the exponent of the integrand $D\left(z\right)$ by $\overline{D}\left(z\right)=\frac{\bar{z}}{z}D\left(z\right)$
with $\bar{z}=\min\left(z,z_{B}\right)$, we have
%%%
\begin{eqnarray}
\eta^+\left(z\right) & \approx & \frac{1}{Y_{X}^{{\rm eq}}(0)K\bar{z}}\int_{z_{{\rm eq}}}^{z}dz'\overline{D}\left(z'\right)e^{-\frac{c}{2Y^{{\rm eq}}}\int_{z'}^{z}dz''\overline{D}\left(z''\right)}\nonumber \\
 & = & \frac{2}{\bar{z}R\,c K}\left[1-e^{-\frac{c}{2Y^{{\rm eq}}}\int_{z_{{\rm eq}}}^{z}dz'\overline{D}\left(z'\right)}\right].\label{eq:eta_plus_large_K}
\end{eqnarray}
%%%
In the above, $z_{B}$ is well approximated by \cite{Buchmuller:2004nz} to be
%%%
\begin{eqnarray}
z_{B}\left(K,c\right) & \approx & 1+\frac{1}{2}\ln\left[1+\frac{\pi K^{2}R^2\,c^{2}}{1024}\left(\ln\frac{3125\pi K^{2}R^2\,c^{2}}{1024}\right)^{5}\right],\label{eq:zB}
\end{eqnarray}
%%%
for all $K$. For $K\gg1$, we can approximate $z_{{\rm eq}}\approx 0$
and integrate eq.~\eqref{eq:eta_plus_large_K}\footnote{We use the identity $\int_{0}^{\infty}dzz{}^{2}{\cal K}_{1}\left(z\right)=2$. }
%%%
\begin{eqnarray}
\eta^+\left(\infty\right) & \approx & \frac{2}{ z_B R\,cK}\left[1-e^{-\frac{1}{4}z_{B}KR\,c\int_{0}^{\infty}dz'z'^{2}{\cal K}_{1}\left(z'\right)}\right]
\nonumber \\
& = &
\frac{2}{z_{B}R\,cK}\left[1-e^{-\frac{1}{2}z_{B}R\,cK}\right].\label{eq:eta_plus_large_K_z}
\end{eqnarray}
%%%
Plugging eqs.~(\ref{eq:Kgg1_approx_1}) and~\eqref{eq:Kgg1_approx_1} into eq.~(\ref{eq:eta_solution_final}), the final efficiency ($z\to\infty$) is
%%%
\begin{eqnarray}
\eta_{K\gg 1}^{w}\left(K,c\right) 
 & \approx & -\frac{2}{R\,c}\left[1-e^{-\frac{1}{2}R\,c}\right]e^{-\frac{c}{2Y^{{\rm eq}}}Y_a}+\frac{2}{z_{B}R\,cK}\left[1-e^{-\frac{1}{2}z_{B}R\,cK}\right]\nonumber \\
 & \approx & \frac{1}{4}R\,cK\left(\frac{3\pi}{2}-z_{B}\right),\label{eq:eff_weak_b}
\end{eqnarray}
where we have kept only the leading term in $cK\ll1$. Notice that the result above depends on $cK$ instead of $K^2$ as in the usual weak washout regime. To our knowledge, this is a \emph{novel} which has not been presented elsewhere.

%%%%%%%%%%%%%%%%%%%%%%%%

\subsubsection{Strong washout regime ($c K \gg 1$)} \label{app:strong_washout}

In the strong washout regime $cK\gg1$, we will also have $K\gg 1$ due to $c$ being at most order unity. In this case, the term involving $\eta^{-}(z_{\rm eq})$ in eq.~(\ref{eq:eta_solution_final}) is negligible because it suffers a strong exponential washout $e^{-\frac{c}{2Y^{{\rm eq}}}\int_{z_{{\rm eq}}}^{z}dz'D\left(z'\right)}$. 
For $\eta^{+}\left(\infty\right)$, we can make use of eq.~\eqref{eq:eta_plus_large_K_z} which is valid for $K\gg1$. According to eq.~(\ref{eq:eta_solution_final}), the final efficiency is simply\footnote{For improved approximation, we can also include contribution from
$\eta_{}$ which gives $-\frac{2}{R\,c}\left[1-e^{-\frac{1}{2}R\,c}\right]e^{-\frac{3\pi}{8}R\,cK}$. }
%%%
\begin{eqnarray}
\eta^{s}\left(K,c\right) & \approx & \eta^+\left(\infty\right)\approx\frac{2}{z_{B}R\,cK}\left[1-e^{-\frac{1}{2}z_{B}R\,cK}\right].\label{eq:eff_strong}
\end{eqnarray}
%%%

\subsubsection{Regimes with thermal initial abundance of $X$} \label{app:thermal_initial_X}
Now we move on to study the regimes with thermal initial abundance of $X$, i.e. $Y_{X}\left(z_{i}\right)=Y_{X}^{{\rm eq}}\left(z_{i}\right)$.
According to the definition in eq.~(\ref{eq:zeq_def}), we have $z_{\rm eq}=z_i$. This means $\eta(z)=\eta^+(z)$ in eq.~(\ref{eq:eta_solution_final}).
Hence for the case of thermal initial
abundance of $X$, a approximate solution good for all $K$ is 
%%%
\begin{eqnarray}
\eta^{{\rm th}}\left(K,c\right)=  \eta^+\left(\infty\right)\approx\frac{2}{z_{B}R\,cK}\left[1-e^{-\frac{1}{2}z_{B}R\,cK}\right].\label{eq:eff_thermal}
\end{eqnarray}
%%%
We can check several limits:
\bea
\eta^{{\rm th}}\left(K,c\right)\approx\left\{
\begin{array}{ll}
1-\frac{1}{4}z_B R\, cK~~~~~~(c K\ll 1)\\
\frac{2}{z_{B}R\,cK}~~~~~~~~~~~~~~(c K \gg 1 )
\end{array}
\right ..
\eea
In the weak washout regime $cK\ll1$, $\eta^{{\rm th}}\left(K,c\right)\approx 1$, meaning there is almost no washout effect as expected. While in the strong washout regime, it coincides with eq.~(\ref{eq:eff_strong}) because the efficiency factor is not sensitive to the initial condition in this region.
%%%%%%%%%%%%%%%%%%%%%%%%

\subsubsection{For all regimes}

For the case of thermal initial abundance of $X$, we can use eq.
(\ref{eq:eff_thermal}) for all $K$ and $c$. For the case of zero
initial abundance of $X$, following ref.~\cite{Buchmuller:2004nz},
we can interpolate $\eta$ for all $K$ and $c$ wtih 
%%%
\begin{eqnarray}
\eta\left(K,c\right) & \approx & \eta^{-}\left(K,c\right)+\eta^{+}\left(K,c\right),
\end{eqnarray}
where
%%%
\begin{eqnarray}
\eta^{-}\left(K,c\right) & = & -\frac{2}{R\,c}e^{-\frac{3\pi}{8}R\,cK}\left\{ \exp\left[\frac{\frac{3\pi}{8}K}{\left(1+\sqrt{\frac{3\pi}{4}K}\right)^{2}}R\,c\right]-1\right\} ,\label{eq:eta_ap_minus}\\
\eta^{+}\left(K,c\right) & = & \frac{2}{z_{B}R\,cK}\left\{ 1-\exp\left[-\frac{\frac{3\pi}{8}K}{\left(1+\sqrt{\frac{3\pi}{4}K}\right)^{2}}z_{B}R\,cK\right]\right\} .\label{eq:eta_ap_plus}
\end{eqnarray}
%%%
The equations above reproduce the approximate solutions eqs.~(\ref{eq:eff_weak_a}), (\ref{eq:eff_weak_b}) and (\ref{eq:eff_strong})
in their respective regimes.

%%%%%%%%%%%%%%%%%%%%%%%%

\section{Spectator effects\label{app:Spectator-effects}}

Here we shall discuss the relevant spectator effects in analyzing BEs at different temperature regimes in section~\ref{sec:Formalism}.

In the thermal bath, through fast scatterings, asymmetries will also be induced in other particles not directly involved in asymmetry generation (they are known as spectators). Although the effects remain generally less than order of one~\cite{Buchmuller:2001sr,Nardi:2005hs}, they are included for completeness. Such effects from spectators are encoded in $c_{\Psi}, c_{\Phi_{\lambda}}$ in BEs [see eq.~(\ref{eq:BE_Yq})], which are defined as the ratio of $Y_{\Delta_{\Psi}}/Y_{\Delta}$ or $Y_{\Phi_{\lambda}}/Y_{\Delta}$ respectively. $c_{\Psi}, c_{\Phi_{\lambda}}$ can be calculated using the charge matrix in eq.~(\ref{eq:particle_asymmetry}), which depends on the effective $U(1)$ symmetries [see section~\ref{subsec:generalities}] present at the relevant temperature regime. In the following, we will briefly discuss the interactions which are in or out of thermal equilibrium and the conserved charges in different temperature regimes. 

In our hybrid seesaw model, we always assume singlet particles $(\Psi,\Psi^c,\Phi_\lambda,\Phi_\kappa)$ are in equilibrium when the genesis happens due to large couplings within the sector [see section \ref{subsec:Initial-conditions}]. Whether the SM particles are in equilibrium or not depends on the temperature. For $T\gtrsim10^{15}$ GeV, as discussed in section~\ref{subsec:Initial-conditions}, the SM particles cannot be in equilibrium via the SM interactions. Since we have $y \Psi^c H \ell$ interaction in our hybrid seesaw model [see eq.~(\ref{eq:hybrid_model})] with unsuppressed coupling $y$, we assume the SM lepton doublets and Higgs are in equilibrium but not other SM particles in this temperature regime. For $T\lesssim10^{15}$ GeV, the SM gauge interactions are in equilibrium.
For $T\gtrsim10^{12}$ GeV, EW sphaleron processes as well as all charged lepton Yukawa interactions are out of thermal equilibrium. In addition,
the first and second family quark Yukawa interactions are out
of thermal equilibrium while the third family quark Yukawa interactions
are in thermal equilibrium.\footnote{For $T\gtrsim 10^{13}$ GeV, QCD sphaleron processes and bottom Yukawa interactions are also out of thermal equilibrium.} For $T\lesssim10^{12}$ GeV , the EW sphalerons get into thermal equilibrium. 
For $T\lesssim10^{11}$ GeV, the $\tau$ and charm Yukawa interactions get into thermal equilibrium. For $T\lesssim10^{9}$ GeV, $\mu$ Yukawa interactions are in thermal equilibrium. For $T\lesssim10^{7}$ GeV, down Yukawa interactions are in thermal equilibrium and finally for $T\lesssim10^{4}$ GeV, electron Yukawa interactions get into thermal equilibrium as well.

\begin{table}
\begin{centering}
\renewcommand{\arraystretch}{1.5}
\setlength{\arrayrulewidth}{0.3 mm}
\begin{tabular}{|c|c|c|c|c|}
\hline 
 \multirow{2}{*}{Temperature regimes} & \multicolumn{3}{c|}{Fully-symmetric model }&Non-symmetric model \tabularnewline
 \cline{2-5}
& $c_{\Psi}$ & $c_{\Phi_{\lambda}}$ & $c_{W1}$ & $c_{W1}$\tabularnewline
\hline 
\hline 
$10^{15}\,{\rm GeV}\lesssim T$ & $\frac{5}{7}$ & 1 & $\frac{31}{42}$&$\frac{1}{6}$\tabularnewline
\hline 
$10^{13}\,{\rm GeV}\lesssim T\lesssim10^{15}$ GeV & $\frac{2}{3}$ & 1 & $\frac{13}{18}$&$\frac{1}{9}$\tabularnewline
\hline 
$10^{12}\,{\rm GeV}\lesssim T\lesssim10^{13}$ GeV & $\frac{35}{53}$ & 1 & $\frac{229}{318}$&$\frac{4}{39}$\tabularnewline
\hline 
$10^{11}\,{\rm GeV}\lesssim T\lesssim10^{12}$ GeV & $\frac{15}{23}$ & $1$ & $\frac{33}{46}$& $\frac{1}{11}$\tabularnewline
\hline 
$10^{9}\,{\rm GeV}\lesssim T\lesssim10^{11}$ GeV & $\frac{42}{65}$ & $1$ & $\frac{93}{130}$& $\frac{5}{61}$\tabularnewline
\hline 
$10^{7}\,{\rm GeV}\lesssim T\lesssim10^{9}$ GeV & $\frac{573}{887}$ & $1$ & $\frac{1269}{1774}$& $\frac{17}{208}$\tabularnewline
\hline 
$10^{4}\,{\rm GeV}\lesssim T\lesssim10^{7}$ GeV & $\frac{363}{565}$ & $1$ & $\frac{807}{1130}$&$\frac{10}{131}$\tabularnewline
\hline 
\end{tabular}
\par\end{centering}
\caption{ We list the values of $c_{\Psi}$, $c_{\Phi_{\lambda}}$
and $c_{W1}=\frac{1}{3}c_{\Psi}+\frac{1}{2}c_{\Phi_{\lambda}}$ in
different temperature regimes in the \textbf{fully-symmetric model}, as well as the values of $c_{W1}=\frac{1}{3}c_{\Psi}$ in the \textbf{non-symmetric model.} For simplicity, in our numerical estimations, we will fix $c_{W1}=0.7(0.1)$ for all temperature regimes in the \textbf{fully-symmetric (non-symmetric) model} . \label{tab:cW1-sym}}
\end{table}

Knowing the relevant interactions at a given temperature regime, we can figure out the conservation of the charges. Since EW sphalerons are not in equilibrium  for $T\gtrsim10^{12}$ GeV, the baryon number $B$ is an effective symmetry. Therefore, we impose baryon number conservation, i.e., $Y_{\Delta B} = 0$. We do not impose such conservation for $T\lesssim10^{12}$ GeV because EW sphalerons processes break baryon number symmetry. Notice that our definition of $Y_\Delta$ changes from $-Y_{\Delta L'}$ for $T\gtrsim10^{12}$ GeV  to $Y_{\Delta (B-L')}$ for $T\lesssim 10^{12}$ GeV [see eq.~(\ref{eq:Delta_hybrid_model})], which will lead to small changes in the spectator effects.
Since we have two realizations of the hybrid seesaw model, namely \textbf{fully-symmetric model} and \textbf{non-symmetric model} (defined in section~\ref{subsubsec:Symmetry-hybrid-model}), we will discuss them separately here:
%%%
\begin{itemize}
\item\textbf{Fully-symmetric model}\\
In this type of model, on top of hypercharge conservation, we also impose $U(1)_{B-L}$ and $U(1)_{\lambda - B}$ conservation in all temperature regimes. In table \ref{tab:cW1-sym}, we list the values of $c_{\Psi}$, $c_{\Phi_{\lambda}}$ and $c_{W1}=\frac{1}{3}c_{\Psi}+\frac{1}{2}c_{\Phi_{\lambda}}$ [introduced in eq.~(\ref{eq:defc})] in different temperature regimes. 

\item \textbf{Non-symmetric model}\\
In this type of model,  we only impose hypercharge conservation due to the absence other global symmetries. Moreover, since the scalar $\Phi_{\lambda}$ in this model does not carry any charge, so $c_{\Phi_{\lambda}}=0$ and we will get $c_{W1}=\frac{1}{3}c_{\Psi}$ according to eq.~\eqref{eq:defc}. 
In table~\ref{tab:cW1-sym}, we list the values of $c_{W1}=\frac{1}{3}c_{\Psi}$ in different temperature regimes. 
\end{itemize}
%%%
As can be seen in table~\ref{tab:cW1-sym}, although the exact values varies at different temperatures, we can find that in all temperature regimes $c_{W1}\approx 0.7$ in the \textbf{fully-symmetric model} and  $c_{W1}\approx 0.1$  in the \textbf{non-symmetric model}. Therefore, we will fix $c_{W1}=0.7 (0.1)$ in the estimation of BEs in the \textbf{fully-symmetric (non-symmetric) model} for simplicity.

%%%%%%%%%%%%
\section{Comments on Boltzmann equations for ISS and LSS\label{app:BE_ISS_LSS}}

When we study leptogenesis in ISS and LSS models in section~\ref{inverse_lepto}, we have used the sum of CP parameter of each particle within the pseudo-Dirac pair to estimate the size of total CP asymmetry as well as the final lepton or baryon asymmetry. In this section, we will justify this argument using a general parametric estimation. 

Consider Boltzmann equation for the asymmetry in $\Delta (B-L)$ in inverse or linear seesaw,

 \begin{equation} \label{eq:BE_ISS_1}
\frac{dY_{\Delta (B-L)}}{dz}  = \sum_i \epsilon_i \frac{dY_{\tilde{\Psi}_i}}{dz} - \frac{1}{2}W(z) Y_{\Delta (B-L)},
\end{equation}
where $\tilde{\Psi}_i$ are the mass eigenstates of the singlet fermions with $m_i \leq m_{i+1}$, $\epsilon_i$ is the CP asymmetry parameter for $\tilde{\Psi}_i$ decays and  $z=\frac{m_1}{T}$.  For simplicity let's focus on the asymmetry generated from decays of the lightest pseudo-Dirac pair only, i.e. consider the sum in eq. (\ref{eq:BE_ISS_1}) to be only over $i=1,2$. The qualitative conclusion will not change when we include more generations of $\tilde\Psi_i$. The washout is controlled by $W(z)$ and we assume the dominant washout comes from the inverse decay (on-shell part of $\Delta (B-L)=2$ scattering process). Correctly including the interference among $\tilde{\Psi}_{1,2}$, one would get (see ref. \cite{Blanchet:2009kk})
\bea
W(z)\approx\frac{1}{2} K^{\rm eff} \frac{Y^{\rm eq}_{\tilde{\Psi}}(0)}{Y^{\rm eq}}z^3 {\cal K}_1(z),
\eea
where $K^{\rm eff}=K_1 \delta_1^2 $ for ISS and $K^{\rm eff}=K_1 \vep_1^{\prime2} $ for with $\delta\sim \mu/\Gamma$, $\vep'\sim y'/y$ and $K_i$ defined in eq.~(\ref{eq:K}). This is the same washout factor in eq.~(\ref{eq:BE_Yq_app}) with $c=K^{\rm eff}/K_1$.
The formal solution to the differential equation eq.~(\ref{eq:BE_ISS_1}) is 
\bea
Y_{\Delta (B-L)}(z)=\int_{0}^{z}dz' \sum_i \epsilon_i \frac{dY_{\tilde{\Psi}_i}}{dz'} e^{- \frac{1}{2}\int^{z}_{z'}dz'' W(z'')},
\eea
assuming $Y_{\Delta (B-L)}(0)=0$, meaning no initial asymmetry. Since $W(z)$ is the same for all  $\tilde{\Psi}_i$, we can simply put the sum out of the integration:
\bea\label{eq:general_sum_Y_delta}
Y_{\Delta (B-L)}(\infty)&=&\sum_i \int_{0}^{\infty}dz' \epsilon_i \frac{dY_{\tilde{\Psi}_i}}{dz'} e^{-\frac{1}{2}\int^{\infty}_{z'}dz'' W(z'')}\\\nonumber
&=&\sum_i \epsilon_i \eta_i Y^{\rm eq}_{\tilde{\Psi}}(0),
\eea
where $ \eta_i\equiv \eta_i(\infty)$ and $\eta(z)$ is defined in eq.~(\ref{eq:BE_Yeta}) with $c=K^{\rm eff}/K_1$ . This means we could treat the generation of asymmetry and washout separately for each $\tilde \Psi_i$ and the total effect is the sum of the result of each one. Now eq.~(\ref{eq:general_sum_Y_delta}) can be rewritten as
\bea\label{eq:ep_factorization}
\frac{Y_{\Delta (B-L)}(\infty)}{Y^{\rm eq}_{\tilde{\Psi}}(0)}=\left[(\ep_1+\ep_2)\eta_1+\ep_2(\eta_2-\eta_1)\right].
\eea
As discussed in section~\ref{inverse_lepto} and appendix~\ref{app:sec3}, we have $\epsilon_1 \approx -\epsilon_2 =\mathcal O( \vep y^2)$, where $\vep\sim \mu/m_\Psi$ and we assume $y^2\gg \mu/\Gamma$, in ISS and $\epsilon_1 \approx -\epsilon_2=\mathcal O(  \vep' y^2)$ in LSS. The sum $\epsilon_1 +\epsilon_2$, however, is second order in $\vep(\vep')$:
$\ep_1+ \epsilon_2 =\mathcal O(\vep^2 /y^2)$ in ISS and $\epsilon_1 + \epsilon_2=\mathcal O( \vep^{\prime 2}y^2)$ in LSS. Since the mass splitting and the difference in Yukawa couplings within the pseudo-Dirac pair are controlled by $\vep $ or $\vep'$, the difference of the $\eta_i$ should go to zero as $\vep$ or $\vep'\to 0$ . Taking $\eta_1-\eta_2\propto \vep(\vep')\eta_1$ as an conservative estimation, we would find the first term in eq.~(\ref{eq:ep_factorization}) is $\mathcal O(\vep^2 /y^2)\eta_1$ in ISS or $\mathcal O(\vep^{\prime 2 } y^2) \eta_1$ in LSS. Whereas the second term is at most $O(\vep^2 y^2) \eta_1$ in ISS or $\mathcal O(\vep^{\prime 2 } y^2) \eta_1$ in LSS. This means the second term in eq.~(\ref{eq:ep_factorization}) is parametrically smaller or at most the same order as the first term. Since our estimation in this paper is only order of magnitude, it is appropriate to keep only the term with $\epsilon_1 + \epsilon_2$, meaning eq.~(\ref{eq:ep_factorization}) approximates to 
 \begin{equation} \label{eq:BE_simplified}
\frac{Y_{\Delta (B-L)}(\infty)}{Y^{\rm eq}_{\tilde{\Psi}}(0)} \sim(\ep_1+\ep_2)\eta_1.
\end{equation}
This allows us to simply treat the contribution to the $Y_{\Delta (B-L) }$ from a pseudo-Dirac pair as if only one of the particle (say $\tilde{\Psi}_1$) decays with the effective CP asymmetry parameter being $\epsilon_1+\ep_2$.

We shall justify the above argument with analytic approximations of $\eta_2-\eta_1$ in both ISS and LSS models. We first consider the LSS case where $m_2=m_1$. There is a unified definition of $z$ for each $\tilde\Psi_i$ because $z\equiv\frac{m_1}{T}=\frac{m_2}{T}$. Therefore we could simply use the results for $\eta_i$ derived in appendix~\ref{app:approximate_solutions}:
\bea
\eta_i\sim\left\{
\begin{array}{ll}
1/(K^{\rm eff} z_B)~~~~~~(K^{\rm eff}\gg 1)\\
K^{\rm eff}~~~~~~~~~~~~~~~(K^{\rm eff}\ll 1 \&K_i \gg 1 \textrm{with zero initial}\; \tilde\Psi_i )\\
 K^{\rm eff}K_i ~~~~~~~~~~~~(K_i \ll 1 \textrm{with zero initial}\; \tilde\Psi_i )\\
 1~~~~~~~~~~~~~~~~~~~(K^{\rm eff}\ll 1  \textrm{with thermal initial}\; \tilde\Psi_i )
\end{array}
\right .,
\eea
where $z_B$ is defined in eq.~(\ref{eq:zB}) and it only depends on $K^{\rm eff}$. Since only in $K_i \ll 1$ region (with zero initial $\tilde\Psi_i$) $\eta_i$ depends on $K_i$ and knowing that $K_2\approx K_1(1-4\vep'_1)$ [derived using eq.~(\ref{eq:y'_Mandh})] to the first order in $\vep'_1$, we can conclude that 
\bea
\frac{Y_{\Delta (B-L)}(\infty)}{Y^{\rm eq}_{\tilde{\Psi}}(0)}\approx\left\{
\begin{array}{ll}
\left[(\ep_1+\ep_2)\eta_1 -4\ep_2\vep'_1\eta_1  \right]~~~~(K_i \ll 1 )\\
(\ep_1+\ep_2)\eta_1 ~~~~~~~~~~~~~~~~~~~~(\textrm{others})
\end{array}
\right ..
\eea
Since $(\ep_1+\ep_2)=O(\vep^{\prime2} y^2) \sim \ep_2\vep'_1=O(\vep^{\prime 2} y^2)$, it matches our estimation in eq.~(\ref{eq:BE_simplified}).

Now we move on to study the case of ISS, which is bit subtler due to mass difference $m_2=m_1(1+\vep_1)$. We can not simply use the same expression in appendix~\ref{app:approximate_solutions} to get $\eta_2$. Instead, we should consider
\bea\label{eq:Y_psi_2}
\eta_2=\frac{1}{Y^{\rm eq}_{\tilde{\Psi}}(0)}\int_{0}^{\infty}dz' \frac{dY_{\tilde{\Psi}_2}(z_2')}{dz_2'}e^{-\frac{1}{2}\int^{\infty}_{z'}dz'' W(z'')},
\eea
where $z'_2=z'(1+\vep_1)$.
In the strong washout region ($K^{\rm eff}\gg 1$), as discussed in appendix~\ref{app:strong_washout}, we can treat $Y_{\tilde{\Psi}_2}\approx Y^{\rm eq}_{\tilde{\Psi}}$ in all relevant regions. Therefore the part involving $z_2'$ in eq.~(\ref{eq:Y_psi_2}) can be approximated  to be
\bea\label{eq:approx_z_2'_eq}
\frac{dY_{\tilde{\Psi}_2}(z_2')}{dz_2'}&\approx& \frac{dY^{\rm eq}_{\tilde{\Psi}}(z_2')}{dz_2'}\nonumber\\
&\approx&\frac{dY^{\rm eq}_{\tilde{\Psi}}(z')}{dz'}+\vep_1 z'\frac{d^2Y^{\rm eq}_{\tilde{\Psi}}(z')}{dz^{\prime2}},
\eea
to the first order in $\vep_1$.
Plugging the first part of the second line of eq.~(\ref{eq:approx_z_2'_eq}) to eq.~(\ref{eq:Y_psi_2}) will simply get $\eta_1$ and thus we can write $\eta_2\approx \eta_1+\delta \eta$
with
\bea\label{eq:delta_eta}
\delta\eta=\frac{\vep_1}{Y^{\rm eq}_{\tilde{\Psi}}(0)}\int_{0}^{\infty}dz'  z'\frac{d^2Y^{\rm eq}_{\tilde{\Psi}}(z')}{dz^{\prime2}}e^{-\frac{1}{2}\int^{\infty}_{z'}dz'' W(z'')}.
\eea
 For $K^{\rm eff}\gg 1$, due to the exponential washout controlled by $W(z)$, the integration in the region $z\gtrsim z_B\gg 1$ dominates. This allows us  to keep only $z'\gg 1$ region of the integrand:
\bea\label{eq:approx_W}
\frac{1}{Y^{\rm eq}_{\tilde{\Psi}}(0)} z'\frac{d^2Y^{\rm eq}_{\tilde{\Psi}}(z')}{dz^{\prime2}}\overset{z'\gg 1}{\approx } - \frac{1}{K^{\rm eff}R}W(z'),
\eea
where $R\equiv \frac{Y_{\tilde\Psi}^{{\rm eq}}(0)}{Y^{{\rm eq}}}$ and we have used the properties of $\mathcal K_1(z')$ function:
\bea
z'\frac{d(z^{\prime 2} \mathcal K_1(z'))}{dz'}\overset{z'\gg 1}{\approx } -z^{\prime3} \mathcal K_1(z').
\eea
Combining eq.~(\ref{eq:delta_eta}) and eq.~(\ref{eq:approx_W}), one will find
\bea
\delta\eta&\approx&-\frac{\vep_1}{K^{\rm eff}R}\int_{0}^{\infty}dz' W(z') e^{-\frac{1}{2}\int^{\infty}_{z'}dz'' W(z'')}\\\nonumber
&=&\frac{2\vep_1}{K^{\rm eff}R}\left(1- e^{-\frac{1}{2}\int^{z}_{0}dz'' W(z'')}\right)\\\nonumber
&=&\frac{2\vep_1}{K^{\rm eff}R}\left(1- e^{-\frac{3\pi}{8} K^{\rm eff}R}\right).
\eea
Using the expression for $\eta_1$ in eq.~(\ref{eq:eff_strong}) and $K^{\rm eff}\gg 1$, we can get
\bea
\delta\eta\approx z_B \vep_1 \eta_1.
\eea
Therefore, the asymmetry in ISS in the strong washout region can be summarized as 
\bea
\frac{Y_{\Delta (B-L)}(\infty)}{Y^{\rm eq}_{\tilde{\Psi}}(z=0)}\approx\left[(\ep_1+\ep_2)\eta_1 +\ep_2\vep_1z_B\eta_1  \right]~~~~~~~(K^{\rm eff}\gg 1 ).
\eea
Since $(\ep_1+\ep_2)=O(\vep^2 /y^2) \gg \ep_2\vep_1z_B=O(\vep^2 y^2\ln K^{\rm eff}) $, this will reduce to eq.~(\ref{eq:BE_simplified}). We also checked numerically that the results for weak washout regions are consistent with eq.~(\ref{eq:BE_simplified}).

%%%%%
\section{Gauge model\label{app:gauge_model}}

The structure of the hybrid seesaw introduced in eq.~(\ref{eq:hybrid_model}) can be obtained introducing appropriate gauge symmetries and additional fields. In this appendix we present a minimal model that reduces to our hybrid scenario after the additional fields ($\chi,S$ in table~\ref{tab:gauge_model}) have been integrated out or decoupled. 
In this model, $U(1)_{B-L}$ global symmetry in table~\ref{tab:global_charges} is promoted to be a gauge symmetry while $U(1)_{\lambda-B}$ arises as an accidental global symmetry.

We assume the full model has gauge group $G_{\rm SM}\times U(1)_{B-L}\times U(1)_{X}$, where $G_{\rm SM}\equiv SU(3)_{c}\times SU(2)_{L}\times U(1)_{Y}$. While the new gauge symmetry $U(1)_{B-L}$ is different from the usual $(B-L)$ symmetry of the SM, we decided to use this name since SM particles are charged as baryon ($B$) minus lepton ($L$) number symmetry.
The fields beyond the SM are singlets under $G_{\rm SM}$. Their charges under the two new $U(1)$ gauge groups are specified in table~\ref{tab:gauge_model}. It is easy to check that the gauge symmetry is anomaly-free~\cite{Batra:2005rh}. Notice that in our case we have two $N_i$ and, as a result, the lightest SM neutrino will be massless. Scenarios with three $N_i$ (or more) can be constructed, but at the cost of introducing new fermions.

%%%%%%%%%%%%%%%%%%%%%
\begin{table}[t]
\begin{center}
\begin{tabular}{c|ccc} 
\rule{0pt}{1.2em}%
 & $U(1)_{B-L}$ & $U(1)_X$ & spin\\
 \hline
 $\Psi^c_{a=1,2,3}$ &  $+1$ & $0$ &  $1/2$  \\
 $\Psi_{a=1,2,3}$ &  $0$ & $\alpha$ & $1/2$ \\
 $N_{i=1,2}$ &  $0$ & $-4\alpha$ & $1/2$\\
 $\Phi_\kappa$ & $-1$ & $-\alpha$ & $0$\\
 $\Phi_\lambda$ & $0$ & $3\alpha$ & $0$\\
 \hline
$\chi$ & $0$ & $5\alpha$ & $1/2$\\
%$S$ & $0$ & $-10\alpha$ & $0$\\
$S$ & $0$ & $8\alpha$ & $0$\\
  \end{tabular}
\end{center}
\caption{Beyond the SM fields and their charges under new gauge symmetries $U(1)_{B-L} \times U(1)_X$. Here $\al$ is some arbitrary real number.}
\label{tab:gauge_model}
\end{table}
%%%%%%%%%%%%%%%%%%%%%%%

Other than the kinetic terms, the only renormalizable couplings allowed by the symmetries are
\begin{eqnarray}\label{gauge}
-{\cal L}_{\rm Yukawa}&=&y_{a\alpha}\Psi^c_aH\ell_\alpha+\kappa_{ab}\Psi_a^c\Phi_\kappa\Psi_b+\lambda_{ai}\Psi_a \Phi_\lambda N_i+c_{ij}SN_iN_j +\textrm{h.c.}
%+y_\chi\chi\chi S.
\end{eqnarray}
In addition to the fields of the hybrid seesaw [eq.(\ref{eq:hybrid_model})] we have added a Weyl fermion $\chi$ and one complex scalar $S$. As mentioned earlier, the former is necessary to obtain a gauge anomaly-free $U(1)_X$. The scalar $S$ is assumed to acquire large VEV, thus generating the large Majorana masses for $N_i$. The hybrid model is effectively recovered once $S$ gets a VEV $M_N\propto\langle S\rangle$ and its radial mode gets integrated out. In particular, note that no number-changing interaction between $\Phi_{\lambda,\kappa},S$ is allowed by gauge invariance at the renormalizable level. The lowest dimensional operator in the scalar potential that breaks the $U(1)_{\lambda-B}$ symmetries in table~\ref{tab:global_charges} (after $S$ gets a VEV) arises at dimension 11, $\sim \frac{\Phi_\lambda^8 \left( S^* \right)^3}{M_{\rm Pl}^7}$, % and contains $8\, \Phi_\lambda$ and $3\,S^*$. 
and as a result, its effect is negligible at low energies. Further number changing operators may exist, e.g. $\frac{\Phi_\lambda^8 \left( S^* \right)^3}{M_{\rm Pl}^7} \left( \frac{\Phi_\kappa^\dagger \Phi_\kappa}{M_{\rm Pl}^2} \right) \to \frac{\langle S \rangle^3}{M_{\rm Pl}^3} \frac{\Phi_\lambda^8 \Phi_\kappa^\dagger \Phi_\kappa}{M_{\rm Pl}^6}$. Unlike previous dimension 11 operator, this operator can generate number changing processes among $\Phi_\lambda$'s and $\Phi_\kappa$'s within the EFT of hybrid model, and in principle can washout asymmetry as discussed in section~\ref{subsubsec:Survival of the asymmetry at intermediate temperatures}. However, being dimension 13 or higher, those effects can safely be ignored. 

The global symmetry $U(1)_{\lambda-B}$ is spontaneously broken by $\langle \Phi_\lambda \rangle \sim$ TeV, generating Nambu-Goldstone boson (NGB). Due to explicit breakings by higher-dimensional operators, this NGB will acquire mass. This fact, together with resulting phenomenological implications were discussed in \cite{short}.

The additional fermion $\chi$ is very light and stable on cosmological scales. Its mass dominantly arises from $\chi\chi(S^*\Phi_\lambda)^2/M_{\rm Pl}^3$\footnote{This operator also breaks $U(1)_{\lambda-B}$ and induces $U(1)_{\lambda-B}$-violating decay of $\Phi_\lambda$. However, such decay is not harmful for the genesis if either (i) corresponding decay rate is slow (and it is: at any temperature $T \lesssim M_{\rm Pl}$ the decay is inactive) or (ii) (even if $\Phi_\lambda$ decay were rapid) $\chi$ does not interact with SM sector strongly that it does not transfer asymmetry (from $\Phi_\lambda$ decay) to the SM.}, and is thus of order $m_\chi\sim (M_N/M_{\rm Pl})^2({\rm TeV}^2/M_{\rm Pl})\ll10^{-3}$ eV. However, because it is also very weakly-coupled, its presence is still allowed by all experimental data. After the gauge boson associated with $U(1)_{X}$ becomes massive, the dominant interaction involving $\chi$ is $ \bar\chi\chi \bar\Psi \Psi/\langle S\rangle ^2$. This decouples around $T\sim (\langle S\rangle^4/M_{\rm Pl})^{1/3}\sim(M_N^4/M_{\rm Pl})^{1/3}$, which is always much higher than the QCD phase transition in our model. This ensures that $\chi$ behaves like dark radiation and contributes negligibly to $\Delta N_{\rm eff}$ at BBN and CMB \cite{Brust:2013xpv}. Other constraints on the hybrid seesaw model are discussed in \cite{short}.

\section{Warped/Composite Higgs seesaw}
\label{app:warped_seesaw}

Here, we briefly review the seesaw mechanism in the context of SM fields propagating in a warped extra dimension,
a scenario dual to the SM Higgs being a composite of new strong dynamics, in particular, a (broken) CFT 
(with rest of the SM particles being {\em partially} composite).
We will show that this framework {\em naturally} leads to the structure of the hybrid model introduced in section~\ref{sec:4_extension_big_picture} [see eq.~(\ref{eq:hybrid_model})].

Mainly for simplicity's sake, we provide a description from the CFT viewpoint and for warped extra-dimensional discussion and for more details, we refer to \cite{Agashe:2015izu}. 
An implementation of the seesaw mechanism (to begin with, high-scale version) in the composite Higgs framework may be represented by the following Lagrangian:
\bea
\mathcal{L} = \mathcal{L}_{\rm CFT} + \lambda \overline{N_R} \mathcal{O}_N + \frac{1}{2} M_N^{\rm bare} N_R^2
\eea
where $N_R$ is an elementary (external to CFT) right-handed fermion and $\mathcal{O}_N$ is a CFT operator that mixes with $N_R$ with coupling $\lambda$\footnote{in analogy with
similar effect for the SM charged fermions and gauge bosons.} and hence interpolates left-handed composite fermionic states. We take the bare Majorana mass of $N_R$ to be its natural size $M_N^{\rm bare} \lesssim M_{\rm Pl}$. We assume that CFT sector preserves lepton number, and the only source of lepton-number violation is the Majorana mass term $M_N^{\rm bare}$ present in the sector external to the CFT. It turns out that the observed neutrino mass can be reproduced when the operator $\overline{N_R} \mathcal{O}_N$ is relevant, i.e. the scaling dimension of $\mathcal{O}_N$ is $\left[ \mathcal{O}_N \right] < 5/2$. In this case, the theory flows to a new IR fixed point where the operator $\overline{N_R} \mathcal{O}_N$ becomes marginal so that $\Big[ N_R^2 \Big] > 3$.
For the case of $M_N^{ \rm bare } < M_{ \rm Pl }$, renormalization group (RG) flow then drives the mass term to a significantly smaller value until the singlet fermion $N_R$ gets integrated out at its physical mass,\footnote{Of course, the singlet field $N_R$ will mix with composite fermion states and hence it is not quite mass eigenstate. Still, the composite state that mixes with $N_R$ will have a mass of $\sim M_N^{\rm phy}$. The resulting mass eigenstate, therefore, will have a mass $\sim M_N^{\rm phy}$.} which can be estimated to be %\shc{Giving this formula may be too much. I am giving since one of the point I tried to make is how $M_N \ll M_{\rm Pl}$ happens naturally in composite seesaw. Still, if you think it is too much, we can remove it.}
\bea
M_N^{\rm phy} \sim M_N^{\rm bare} \left( \frac{M_N^{\rm bare}}{M_{\rm Pl}} \right)^{\frac{1}{2\left[\mathcal{O}_N\right]-4}-1}.
\eea
Integrating out $N_R$ at this $M_N^{\rm phy}$ scale generates 
\bea
\Delta \mathcal{L}_{\rm CFT} =\lambda \overline{N_R} \mathcal{O}_N + \frac{1}{2} M_N^{\rm phy} N_R^2 
& \rightarrow & \frac{\lambda^2}{M_N^{\rm phy}} \mathcal{O}_N^2.  
\eea
As is clear from the appearance of lepton-number breaking spurion $M_N^{\rm phy}$, the CFT operator $\mathcal{O}_N^2$ is a lepton-number violating perturbation to the CFT sector. Integrating out $N_R$, therefore, effectively transfers lepton-number breaking into the CFT sector. One notices that this is like generating Weinberg operator in type I seesaw 
and a rather precise match may be seen when $\mathcal{O}$ is (roughly) identified with $H \ell$. 

RG running the theory further down to the TeV scale where strongly coupled sector confines we get
\bea\label{eq:warped_seesaw_O_square}
\Delta \mathcal{L}_{\rm CFT} \sim \frac{\lambda^2}{M_N^{\rm phy}} \left( \frac{{\rm TeV}}{M_N^{\rm phy}} \right)^{2 \left[ \mathcal{O}_N \right] - 5} \mathcal{O}_N^2 \sim \frac{\lambda^2}{M_N^{\rm bare}} \left( \frac{{\rm TeV}}{M_{\rm Pl}} \right)^{2 \left[ \mathcal{O}_N \right] - 5} \mathcal{O}_N^2,
\eea
where 
{$\left[ \mathcal{O}_N \right]$ denotes the scaling dimension of $\mathcal{O}_N$}
and we used (hence assumed accordingly) the large-$N$ approximation for the scaling dimension of $\mathcal{O}_N^2$. Using this lepton-number violating spurion and dimensional analysis, we can make a quick estimation for the SM neutrino mass. Including a factor of $\sim \left( {\rm TeV} / M_{\rm Pl} \right)^{2 \left[ \mathcal{O}_L \right] - 5}$\footnote{Here, we 
assume for concreteness that $\left[ \mathcal{O}_L \right] > 5/2$.} to account for the square of coupling of SM lepton doublet ($\ell$) 
to the CFT (where Higgs arises as a composite state) via linear coupling $\lambda_L \overline{\ell} \mathcal{O}_L$
(with $\mathcal{O}_L$ being a CFT operator with SM lepton quantum numbers), we get
\bea
m_\nu \sim \frac{\lambda_{L}^2\lambda^2v^2}{M_N^{\rm bare}} \left( \frac{\rm TeV}{M_{\rm Pl}} \right)^{2 \left( \left[ \mathcal{O}_N \right] + \left[ \mathcal{O}_L \right] - 5 \right)},
\label{eq:m_nu_CFT_spurion}
\eea
where we assume the couplings among composite states are $O(1)$.

There is another way of viewing this result, which may provide the connection with inverse seesaw 
(both minimal model of section~\ref{review} and its UV completion in section~\ref{sec:4_extension_big_picture}) clearer. 
For this note that, when the CFT sector confines at TeV scale, each operator $\mathcal{O}_N$, when acted on the vacuum, creates a tower of left-handed composite fermions, which can be identified with $\Psi$ of the inverse seesaw model. They combine with the right-chirality states ($\Psi^c$) generated by another CFT operator to form composite Dirac fermions, with mass starting at TeV and with TeV mass gap between adjacent states, i.e, we have $m_{ \Psi } \sim$ TeV.
The ``Weinberg''-type operator in eq.~(\ref{eq:warped_seesaw_O_square}) then can be viewed as generating a small Majorana mass terms for left-handed fermion (called $\mu$ in previous sections), i.e.,
\bea
\mu & \sim & \frac{\lambda^2 \; \hbox{TeV}^2 }{M_N^{\rm bare}} \left( \frac{{\rm TeV}}{M_{\rm Pl}} \right)^{2 \left[ \mathcal{O}_N \right] - 5}
\label{eq:mu_CFT}
\eea
Together with the $\sim$ TeV Dirac mass, this makes these composite fermions pseudo-Dirac. Finally, the composite singlet has a coupling to SM Higgs and composites with quantum numbers of SM lepton (interpolated by $\mathcal{O}_L$).  
We then obtain a coupling between composite singlet ($\Psi^c$), SM Higgs and lepton via mixing of elementary lepton with latter composites (including a different RG factor, i.e., determined by scaling dimension of $\mathcal{O}_L$)\footnote{For the case of $\left[ \mathcal{O}_L \right] > 5/2$ assumed here, the corresponding mixing is {\em ir}relevant. Note also that a similar factor was used in the spurion/dimensional analysis estimate in eq.~(\ref{eq:m_nu_CFT_spurion}) above.}:
\bea
y & \sim & \lambda_L \left( \frac{ \rm TeV }{ M_{\rm Pl} } \right)^{ \left[ \mathcal{O}_L \right] - 5/2 }
\label{y_CFT}.
\eea
%
%Here (for simplicity and naturalness), the composite Yukawa coupling is taken to be $g_{\rm NP}=O(1)$.
%
It is then straightforward to check that the exchange of $\Psi-\Psi^c$ [with above coupling in eq.~(\ref{y_CFT}) and $\mu$-term in eq.~(\ref{eq:mu_CFT}) plugged into eq.~(\ref{neutrino_mass})] re-produces the mass for the SM neutrino obtained in 
eq.~(\ref{eq:m_nu_CFT_spurion}) using simply spurion/dimensional analysis (i.e., we see here the ``anatomy'' of the earlier estimate). Namely, the underlying dynamics for the neutrino mass generation is precisely that of inverse seesaw. Crucially, notice [as per eq.~(\ref{eq:mu_CFT})] that in this composite Higgs seesaw framework, the required small Majorana mass terms is generated dynamically via type I seesaw mechanism. This solves the naturalness problem for $\mu$ term in 4D inverse seesaw model. 

Moreover, as should be evident by now, the effective model of eq.~(\ref{eq:hybrid_model}) in section~\ref{sec:4_extension_big_picture} shares many features with 5D warped/4D composite model and therefore may be viewed as a ``toy'' version of the full 5D model; in fact, quite remarkably, the warped/composite model makes up for the deficiencies of the toy model (as follows).
First of all, elementary $N_R$ above (with mass term $M^{ \rm bare }_N$) is roughly analogous to that $N$ in eq.~(\ref{eq:hybrid_model}).
In the toy model, $N$ mixes with $\Psi$ (once $\Phi_{ \lambda}$ acquires VEV), which matches on to elementary $N_R$ mixing with composite singlet interpolated by $\mathcal{O}_N$ (via the $\lambda \overline{ N_R } \mathcal{O}_N$ coupling).
Note that a set of new scalar fields we had to introduce in eq.~(\ref{eq:hybrid_model}) (for $N - \Psi$ and $\Psi- \Psi^c$ mass terms) is not needed here, since the scale TeV involved in both these mass terms is dynamically generated by confinement, instead of by VEVs of elementary scalars. 
Nonetheless, there could be a {\em composite} scalar associated with ``fluctuations'' of the confinement scale (dilaton),
which can play the role of the {\em physical} scalar $\Phi_{ \lambda }$  which is of course crucial for leptogenesis. Moreover, its mass is naturally $\sim$ TeV, i.e., the compositeness scale. So, the issue (a) mentioned in section~\ref{sec:4_extension_big_picture} in the context of the toy model, i.e., hierarchy problem for scalars, is absent. Moving on, as we have already briefly mentioned, due to the fact that theory consists of a weakly coupled (external) sector and a CFT sector, the absence of any other interaction terms but the linear coupling $\lambda N \mathcal{O}_N$ is completely natural; in particular, a direct coupling of $N_R$ (elementary) to SM Higgs (composite) and SM lepton is forbidden. 
Similarly, (bare) Majorana mass terms for $\Psi$, $\Psi^c$ are not allowed since CFT sector (by itself) preserves lepton-number. Thus, the issue (b) of the toy model, i.e., the particular structure of the Lagrangian, is solved.

Finally, regarding the neutrino mass, the TeV-modulation factor mentioned in section~\ref{sec:4_extension_big_picture}, i.e. the 2nd factor in eq.~(\ref{eq:neutrino_mass_in_hybrid_model}), corresponds to an RG running in composite seesaw. Matching eq.~(\ref{eq:m_nu_CFT_spurion})  with eq.~(\ref{eq:neutrino_mass_in_hybrid_model}) using eq.~(\ref{y_CFT}) and identifying $M_N$ with $M_N^{\rm bare }$, we get
\bea
\left( \frac{\langle \Phi_\lambda \rangle}{ \kappa \langle \Phi_\kappa \rangle} \right)^2\leftrightarrow\left( \frac{\rm TeV}{M_{\rm Pl}} \right)^{2 \left[ \mathcal{O}_N \right] - 5},
\eea
and thus can naturally be much larger or smaller than 1.
Specifically, if this modulation factor is (much) larger than unity, then $M_N^{\rm bare}$ as large as $M_{ \rm Pl }$ 
can still give the required SM neutrino mass. 
In this way, we realize that the high-scale type I seesaw implemented in the composite Higgs framework (or warped extra-dimensional framework via AdS/CFT correspondence) seems to be a very 
natural theory for the SM neutrino mass. 
 Concerning {\em leptogenesis}, given that studying it
explicitly in the actual 5D theory, or its dual 4D CFT, can be appreciably more complicated (due in part to much more degrees of freedom to deal with and issues regarding thermal phase transition), understanding the relevant physics within the more effective framework of eq.~(\ref{eq:hybrid_model}) {\em to begin with} is highly-motivated.

 \end{document}